\documentclass{emulateapj}
\usepackage{apjfonts}
\usepackage{graphicx}
\usepackage{amsmath}

\newcommand{\be}{\begin{equation}}
\newcommand{\ee}{\end{equation}}
\newcommand{\beqa}{\begin{eqnarray}}
\newcommand{\eeqa}{\end{eqnarray}}

\def\Mo{{\rm M_\odot}}
\newcommand{\hMpc}{\ h^{-1}\rm{Mpc}}
\newcommand{\hMsun}{\ h^{-1}\rm{M}_{\odot}}

\newcommand{\hkpc}{\ h^{-1}\rm{kpc}}
\newcommand{\hpc}{\ h^{-1}\rm{pc}}
\def\Mpc{\ {\rm Mpc}}
\def\kpc{\ {\rm kpc}}
\def\pc{\ {\rm pc}}
\def\kms{{\ }{\rm km}\,{\rm s}^{-1}}
\def\LCDM{$\Lambda$CDM}
\def\degrees{^\circ}

\begin{document}
\submitted{The Astrophysical Journal, accepted}
\vspace{1mm}

\shortauthors{KAZANTZIDIS ET AL.}
\shorttitle{CDM Substructure and Galactic Disks I:}

\title{Cold Dark Matter Substructure and Galactic Disks I: \\
Morphological Signatures of Hierarchical Satellite Accretion}

\author{Stelios Kazantzidis,\altaffilmark{1,2}
        James S. Bullock,\altaffilmark{3}
        Andrew R. Zentner,\altaffilmark{4}\\
        Andrey V. Kravtsov,\altaffilmark{5}
        and Leonidas A. Moustakas\altaffilmark{6}}

\vspace{2mm}

\begin{abstract}
  We conduct a series of high-resolution, fully self-consistent dissipationless 
  $N$-body simulations to investigate the cumulative effect of substructure 
  impacts onto thin disk galaxies in the context of the {\LCDM} paradigm of 
  structure formation. Our simulation campaign is based on a hybrid approach 
  combining cosmological simulations and controlled numerical experiments. 
  Substructure mass functions, orbital distributions, internal structures, and 
  accretion times are culled directly from cosmological simulations of
  galaxy-sized cold dark matter (CDM) halos. We demonstrate that accretions of 
  massive subhalos onto the central regions of host halos, where the galactic
  disk resides, since $z \sim 1$ should be common occurrences. In contrast,
  extremely few satellites in present-day CDM halos are likely to have a
  significant impact on the disk structure. This is due to the fact that massive subhalos
  with small orbital pericenters that are most capable of strongly perturbing
  the disk become either tidally disrupted or suffer substantial mass loss
  prior to $z=0$. One host halo merger history is subsequently used to seed 
  controlled $N$-body experiments of repeated satellite encounters with an 
  initially-thin galactic disk. These simulations track the effects of six
  dark matter substructures, with initial masses in the range $\sim (0.7-2) 
  \times 10^{10}\,\Mo$ ($\sim 20-60\%$ of the disk mass), crossing the 
  disk in the past $\sim 8$~Gyr. We demonstrate that these accretion events
  produce several distinctive morphological signatures in the stellar disk 
  including: long-lived, low-surface brightness, ring-like features in the
  outskirts; significant flares; bars; and faint filamentary
  structures that (spuriously) resemble tidal streams in configuration space. 
  The final distribution of disk stars exhibits a complex vertical
  structure that is well-described by a standard ``thin-thick'' disk
  decomposition, where the ``thick'' disk component has emerged primarily as a
  result of the interaction with the most massive subhalo.
  Though our simulation campaign was not designed to elucidate the nature of 
  specific Galactic structures, we compare one of the resulting dynamically 
  cold ring-like features in our simulations to the Monoceros ring stellar structure 
  in the Milky Way (MW). The comparison shows quantitative agreement in both spatial distribution 
  and kinematics, suggesting that such observed complex stellar components 
  may arise naturally as disk stars are excited by encounters with 
  CDM substructure. We conclude that satellite-disk interactions of the kind expected in {\LCDM} 
  models can induce morphological features in galactic disks that are similar to those being 
  discovered in the MW, M31, and in other nearby and distant disk galaxies. 
  These results highlight the significant role of CDM substructure
  in setting the structure of disk galaxies and driving galaxy evolution. 
  Upcoming galactic structure surveys and astrometric satellites may be able
  to distinguish between competing cosmological models by testing whether the
  detailed structure of galactic disks is as excited as predicted by the CDM paradigm.
\end{abstract}


\altaffiltext{1}{Kavli Institute for Particle Astrophysics and Cosmology;
        and Department of Physics;
        and Stanford Linear Accelerator Center, Stanford University, 
        2575 Sand Hill Rd, Menlo Park, CA 94025 USA; {\tt stelios@slac.stanford.edu}.}
        \altaffiltext{2}{Present Address: Center for Cosmology and Astro-Particle Physics; 
        and Department of Physics; and Department of Astronomy, 
        The Ohio State University, Physics Research Building, 191 West
        Woodruff Avenue, Columbus, OH 43210 USA; {\tt stelios@mps.ohio-state.edu}.}
        \altaffiltext{3}{Center for Cosmology;
        and Department of Physics \& Astronomy, The University of California at Irvine, Irvine, 
        4168 Reines Hall, CA 92697 USA; {\tt bullock@uci.edu}.}
        \altaffiltext{4}{Department of Physics and Astronomy,
        University of Pittsburgh, 100 Allen Hall, 3941 O'Hara Street,
        Pittsburgh, PA 15260 USA; {\tt zentner@pitt.edu}.}
        \altaffiltext{5}{Kavli Institute for Cosmological Physics; 
        and Department of Astronomy \& Astrophysics; 
        and the Enrico Fermi Institute, The University of Chicago, 5640 South Ellis Avenue, Chicago, IL 60637 USA; 
        {\tt andrey@oddjob.uchicago.edu}.}
         \altaffiltext{6}{Jet Propulsion Laboratory, California Institute of Technology, MS 169-327, 4800
        Oak Grove Dr, Pasadena, CA 91109 USA; {\tt leonidas@jpl.nasa.gov}.}

\section{INTRODUCTION}
\label{section:introduction}

In the cold dark matter (CDM) paradigm of structure formation,
galaxy-sized dark matter halos form via the continuous accretion of
smaller systems \citep[e.g.,][]{Blumenthal_etal84}. 
Recently, an explosion of data has disclosed unexpected kinematic and
structural complexity in the Milky Way (MW), its neighbors, and distant 
galaxies. There is a growing body of evidence from local star-count surveys
that the MW has, in fact, experienced numerous accretion events.
Since the original discovery of the Sagittarius dwarf galaxy and its
associated tidal stream \citep{Ibata_etal94,Yanny_etal00,Ivezic_etal00,Ibata_etal01a,
Ibata_etal01b,Majewski_etal03}, at least $5$ additional, spatially-coherent
structures have been found in and around the MW
\citep{Newberg_etal02,Gilmore_etal02,Yanny_etal03,Rocha-Pinto_etal04,Martin_etal04,
Mdelgado_etal05,Grillmair06a,Grillmair06b,Grillmair_Dionatos06,
Rocha-Pinto_etal06,Belokurov_etal06,Juric_etal08}. These numbers are in general agreement
with predictions of dwarf galaxy accretion and disruption in the prevailing  
{\LCDM} cosmological model
\citep{Bullock_etal01,Bullock_Johnston05,Bell_etal07} 
and the character of the structures themselves are similar to those expected 
in simulations of satellite galaxy disruption \citep{Johnston_etal95}. 

Among the more intriguing stellar structures discovered in the MW is the
Monoceros ring \citep{Newberg_etal02}. This coherent,
low-metallicity feature spans $\sim 170^\circ$ degrees around the
Galaxy and lies near  the disk plane at a Galacto-centric distance of
$\sim 20\kpc$ \citep{Newberg_etal02,Yanny_etal03,Ibata_etal03,Rocha-Pinto_etal03,Conn_etal05}. 
It is unclear whether this  structure is yet another example of tidal
debris from an accreted dwarf galaxy
\citep{Yanny_etal03,Helmi_etal03,Martin_etal04,Penarrubia_etal05} 
or a stellar extension of the disk itself
\citep{Ibata_etal03,Helmi_etal03,Momany_etal04,Moitinho_etal06,Momany_etal06}.

In addition to revealing a complex assortment of structure in the
outer disk and halo, photometric surveys are allowing a more detailed
``tomography'' of the MW disk itself \citep{Newberg_etal06,Juric_etal08}.
Traditional ``thin-thick'' disk decompositions to these data yield thin disk
exponential scale heights of $\sim 250\pc$ and corresponding thick disk heights
typically $\sim 2-4$ times larger \citep{Siegel_etal02,Newberg_etal06,Juric_etal08}. 
However, small but significant differences in published scale height
determinations may reflect a more complicated disk structure
than uniform scale height models allow \citep{Newberg_etal06}. 

Ongoing studies in the Andromeda galaxy (M31) suggest that its 
history is at least as rich and complex as that of the MW 
\citep[e.g.,][and references therein]{Ferguson_etal05,Kalirai_etal06,Ferguson07,Ibata_etal07}.
At least one recent, high-metallicity accretion event is evidenced by
the ``giant stream'' structure \citep{Ibata_etal01c} and Spitzer MIPS imaging
has revealed a nearly circular, $\sim 10\kpc$ star-forming 
ring that may have been triggered by a past disk interaction
\citep{Gordon_etal06}. Particularly intriguing is the extended $\sim 40\kpc$
disk-like configuration of stars around  M31 \citep{Ibata_etal05}. 
This extended  ``disk''
component is clumpy, but possesses a relatively low velocity dispersion,
$\sim 30 \kms$. It lags behind the rotation velocity of the 
standard M31 disk by $\sim 40 \kms$
and contains intermediate age stars similar to those of the thick disk of the MW
\citep{Brown_etal06,Faria_etal07}. Recently, \citet{Penarrubia_etal06} have investigated
models where this structure is formed by a co-planar,
near-circular accretion event. Though observations of stellar halos and thick
disks in more distant galaxies are extremely challenging, deep imaging and
star-count investigations have also revealed that these faint structures are
likely ubiquitous \citep[e.g.,][]{Sackett_etal04,Dalcanton_Bernstein02,Zibetti_Ferguson04,
Dalcanton_etal05,Yoachim_Dalcanton06,Barth07,Dejong_etal07a}.

The over-abundance of substructure in galaxy-sized dark matter halos 
has received a lot of attention in the literature, particularly 
regarding the problem of the ``missing'' galactic satellites 
\citep{Klypin_etal99,Moore_etal99,Stoehr_etal02,Zentner_Bullock03,Kazantzidis_etal04b,
Kravtsov_etal04,Mayer_etal07,Strigari_etal07}. It is important to emphasize that the satellites
that will be most crucial for altering disk morphologies will be the 
few most massive systems of the entire population that are distinctly outside of the regime 
relevant to the substructure problem. 
Indeed, it can be shown that in the impulsive heating approximation 
the tidal effects of a subhalo population scale as 
$dE/dt \propto \int n(M_{\rm sub}) \, M_{\rm sub}^2 \, dM_{\rm sub}$, where 
$n(M_{\rm sub})$ and $M_{\rm sub}$ are the number density of subhalos and
subhalo mass, respectively \citep{White00}. Since cosmological simulations
predict that the mass function of CDM halos can be described by a power law, 
$n(M_{\rm sub})\propto M_{\rm sub}^{-\alpha}$, with
$\alpha\approx 1.8-1.9$ \citep{Ghigna_etal98,Gao_etal04}, the dynamical
effects on stellar disks will be dominated by the most massive substructures
expected to be luminous. Therefore, of concern are not the ``missing'' or ``dark'' subhalos, 
but rather the population LMC and SMC-sized objects that fell into the disk over the past
$\sim 8$~Gyr and that are now (presumably) dispersed throughout the Galaxy as part of its stellar 
halo \citep[e.g.,][]{Bullock_Johnston05,Abadi_etal06}.

In the present paper we examine the morphological signatures that halo substructure
induces in thin galactic disks in the context of the CDM model of structure
formation. Rather than focus on the stellar material liberated from accreted
dwarf galaxies, we focus on the effects of satellite halo impacts on disk galaxies 
themselves. Specifically, we bombard an initially-thin galaxy-sized stellar disk with dark
matter subhalos whose mass functions, orbital distributions, internal
structures, and accretion times have been extracted directly from cosmological 
simulations of galaxy-sized dark matter halos and study the 
ramifications of such encounters for disk structure.
We note that our technique was not designed to reproduce any specific galactic structures and it
is worth bearing this in mind when comparing our results with those of studies aimed
at producing specific features.

Disk galaxies have been notoriously difficult to form in {\LCDM} cosmological
simulations \citep[e.g.,][]{Navarro_etal95}.
This difficulty in conjunction with the prevalence of disks in the universe
might point to an issue of fundamental cosmological importance. For example,
the dark matter may not be cold or the power spectrum 
may not be hierarchical on small scales \citep[e.g.,][]{Sommer-Larsen_Dolgov01,Zentner_Bullock03,
Strigari_etal06}. However, another possibility is that current {\LCDM}
cosmological simulations lack the physics and/or resolution necessary to form
disks. Indeed, more recent, higher 
resolution studies with new astrophysical inputs have been more successful 
\citep[e.g.,][]{Weil_etal98,Thacker_Couchman01,Abadi_etal03,Sommer_Larsen_etal03,Governato_etal04,
Brook_etal04,Robertson_etal04, Governato_etal07}. Nonetheless, the problem of disk galaxy
formation within a hierarchical cosmology is far from being resolved, and any
successes are strongly dependent on poorly understood physical processes that occur
below the numerical resolution of contemporary simulations.

In addition, despite the continuing increase in dynamic range, limited
numerical resolution renders current cosmological simulations inadequate to
address fully problems that encompass the enormous range 
of dynamical scales involved in studies of satellite-disk interactions. Given
this fact and the uncertainties regarding {\it ab initio} formation of disk
galaxies, we are motivated to explore the interplay 
between CDM substructure and galactic disks using {\it controlled} $N$-body
simulations. Such investigations can have profound implications for our
current understanding of galaxy formation and evolution.

Our work is informed considerably by many past numerical investigations into the
resilience of galactic disks to infalling satellites 
\citep{Quinn_Goodman86,Quinn_etal93,Walker_etal96,Huang_Carlberg97,Sellwood_etal98,
Velazquez_White99,Font_etal01,Ardi_etal03,Gauthier_etal06,Hayashi_Chiba06,Read_etal08,Villalobos_Helmi08}
and we present a more direct comparison to these studies in a companion paper 
(Kazantzidis et al. 2007, Paper II). Successes notwithstanding, a significant
fraction of these earlier numerical investigations had the drawback of not
being fully self-consistent, modeling various components of the primary disk
galaxy and sometimes also the satellites as rigid potentials, and focusing 
on experiments with infalling systems on nearly circular orbits which are poor
approximations of the highly eccentric orbits that typify the trajectories of
CDM substructure. 

Moreover, most of the numerical work published to 
date considered the encounters of {\it single} satellites with larger
multi-component disk galaxies. Notable exceptions are the studies of \citet{Font_etal01} and
\citet{Gauthier_etal06}, who utilized the {\it present-day} properties of  
a large ensemble of dark matter subhalos in order to investigate the dynamical
effects of substructure on stellar disks. However, satellite populations
possess vastly different properties at earlier 
times \citep[e.g.,][]{Zentner_Bullock03,Gao_etal04,Zentner_etal05a} and
investigations based on the $z=0$ substructure do not account for previous
encounters of since disrupted systems with the disk. Elucidating the effects
of CDM substructure on galactic disks clearly requires a more realistic
treatment of the evolution of the subhalo populations.

Our {\it dissipationless} numerical experiments extend and expand upon those
of earlier studies in at least three major respects. We adopt for the first
time a model in which the subhalo
populations are truly representative of those accreted and possibly destroyed in 
the past, instead of the $z=0$  {\it surviving} substructure present in a CDM halo.
In particular, our method tracks the host halo accretion events since $z=1$,
and as discussed below, this results in a considerably larger number of 
important satellite-disk interactions than previously
considered. Second, we employ self-consistent satellite
models whose masses, internal structures, and orbital parameters are directly 
extracted from high-resolution cosmological $N$-body simulations of galaxy-sized 
CDM halos. This element of the modeling allow us to model the internal properties
and impact parameters of the infalling systems without the need for
arbitrary choices. 

Finally, the primary disk galaxy models are motivated by the prevailing {\LCDM}
paradigm of structure formation and are derived from explicit distribution 
functions. In addition, they are flexible enough to
permit detailed modeling of actual disk galaxies, including the MW and M31. 
In the present study we adopt a model that satisfies a broad range of observational 
constraints available for the MW galaxy. The self-consistency of the galaxy
models in synergy with the adopted high-resolution 
allows us to construct equilibrium $N$-body models of disk 
galaxies that are as thin as the {\it old}, thin stellar disk of the MW with an 
{\it exponential} scale height of $h_z \sim 0.25 \pm 0.05$~kpc 
\citep[e.g.,][]{Kent_etal91,Dehnen_Binney98,Widrow_Dubinski05}. It is worth noting that the 
great majority of past investigations \citep[e.g.,][]{Velazquez_White99,Font_etal01} considered 
$N$-body models of disk galaxies that were much thicker compared to the old,
thin disk of the MW. As we demonstrate in Paper II, 
owing to their larger vertical velocity dispersions, thicker disks are
relatively less affected by encounters with satellites compared to their thin counterparts.

We model the infalling systems as pure dark matter subhalos and do not
follow the evolution (or deposition) of the {\it stars} in the
accreted systems. Our focus, in this study, is on the evolution of
the stellar material in an initially-thin disk. Though our numerical
experiments employ a best-fit model for the present-day structure of the MW,  
it is worth emphasizing that the adopted simulation campaign is neither
designed to follow the evolution of nor to draw specific conclusions about the MW 
or any other particular system. Given the complex interplay of effects 
(e.g., gas cooling, star formation, chemical evolution) relevant to the 
formation and evolution of galactic disks, our collisionless simulations 
aim at investigating the most {\it generic qualitative features} of the
evolution of a disk galaxy subject to bombardment by CDM substructure.

We demonstrate that a thin disk component may survive, even strongly
perturbed, significant bombardment by halo substructure and show that a wealth
of morphological signatures are generated in stellar disks in response to these 
encounters. These include pronounced flares, bars, and a number of 
(sub-dominant) low-surface brightness ring-like features and filamentary structures
that arise in and above the disk plane. We also show that interactions with
infalling satellites cause stellar disks to develop a complex vertical
morphology that resembles the commonly adopted ``thin-thick'' disk profiles
used in the analysis of disk galaxies. The resulting morphological features 
resemble the complex faint structures that are seen in the MW, M31, and in 
other nearby and distant disk galaxies. Quantitative comparison between 
the properties of ring-like features in our simulations and those of the 
Monoceros ring in the MW suggests that such observed complex stellar
structures may arise naturally as disk stars are excited by halo 
substructure of the kind predicted by the {\LCDM} paradigm of structure 
formation. We reiterate that our simulations were not designed to reproduce such 
features. Rather, our campaign was formulated to mimic satellite infall in the 
prevailing model of structure formation and no parameters were tuned to produce these results. 
The fact that such features are excited in our calculations suggests that they are of a very 
general nature, and thus require little fine tuning. 
All of these findings highlight the substantial role of CDM substructure in setting the observable 
structure of disk galaxies and driving galaxy morphological evolution.

The outline of this paper is as follows. In Section~\ref{sec:methods}
we use cosmological simulations to study substructure
properties and accretion statistics in four galaxy-sized CDM halos, select 
target subhalos from one host halo merger history for subsequent re-simulation, 
and describe the controlled numerical experiments to model the impact of 
these accretion events on the structure of an initially-thin galactic disk. 
Section~\ref{sec:results} contains results from this simulation campaign
pertaining to the global morphological evolution that infalling CDM satellites 
can induce in galactic stellar disks. Implications and extensions of our
findings are discussed in Section~\ref{sec:discussion}. Finally, in Section~\ref{sec:summary},
we summarize our main results and conclusions. Note that in  the  remainder of the
paper we use the terms ``satellites,'' ``subhalos,'' and ``substructures'' 
interchangeably.

\section{Numerical Methods}
\label{sec:methods}

The main goal of the present study is to investigate the morphological 
signatures created by the gravitational interaction between a population
of {\it dark} subhalos and a thin galactic disk. We begin this section by 
describing cosmological numerical simulations of the 
formation of four galaxy-sized halos in the concordance {\LCDM} cosmological 
model. In \S~2.2 we analyze these simulations 
to quantify the frequency with which massive subhalos approach the 
very central regions of their hosts, where the galactic disk presumably 
lies. We identify one of these host halos, and in \S~2.3 we extract from it
the properties of individual substructures for subsequent controlled 
re-simulation. To be specific, we derive the orbital parameters,
masses, density profiles, and accretion times of selected satellites 
from the cosmological simulations and use these properties to initialize controlled 
numerical simulations of subhalo-disk encounters. This allows us to connect the 
cosmological evolution of a galaxy-sized CDM halo with the evolution of a thin, 
galaxy-sized disk. In \S~2.4 we introduce the primary galaxy models and the 
numerical methods we employ to construct them. Finally, in \S~2.5 we 
describe in detail our simulation techniques as well as the controlled 
numerical experiments of the adopted halo merger history.

\subsection{Hierarchical Cosmological Simulations}
\label{sec:cosmo_sim}

The cosmological halos we analyze come from two different  
simulations that were performed using the Adaptive Refinement Tree (ART) $N$-body
code \citep{Kravtsov_etal97,Kravtsov99}. Both simulations assume a
flat {\LCDM} cosmology with $\Omega_m=1-\Omega_{\Lambda}=0.3$,
$\Omega_b=0.043$, $h=0.7$  and $\sigma_8=0.9$. 

The first simulation followed the evolution of three galaxy-sized dark matter halos 
within a cubic volume of $25\hMpc$ on a side using the multiple mass resolution technique 
described in \citet{Klypin_etal01}. We denote this simulation ``L25'' and
the halos as ``G$_1$'', ``G$_2$'', and ``G$_3$''. The mass resolution in the vicinity of each 
system was $m_p \simeq 1.2 \times 10^6 \hMsun$, corresponding to $1024^3$ particles over the entire 
computational box, and the simulation achieved a maximum spatial resolution of $191 \hpc$. 
Various properties of the L25 halos and their substructure have been discussed previously in 
\citet{Klypin_etal01}, \citet{Kravtsov_etal04}, and \citet{Zentner_etal05b}.

The second simulation followed the evolution of a single galaxy-sized halo in a cubic 
volume of $20 \hMpc$ on a side using the same multiple mass technique. We refer to this simulation 
as ``L20'' and the halo as ``G$_4$''. The maximum spatial resolution for L20 was $153 \hpc$ and 
the mass per particle within the region of halo G$_4$ was $m_p \simeq 6.1 \times 10^5 \hMsun$, 
corresponding to $1024^3$ particles in the entire volume. The L20 halo has
been discussed previously in \citet{Prada_etal06}, and \citet{Gnedin_Kravtsov06}.  

We define virial masses, $M_{\rm vir}$, and radii, $r_{\rm vir}$, for each
host halo at $z=0$ according to a fixed overdensity $\Delta = 337$ with
respect to the mean density of the universe, $\bar{\rho}$, centered on the
particle with the highest local density. This implies that virial mass and
radius are related by $M_{\rm  vir} = 4\pi\Delta\bar{\rho}r_{\rm vir}^3/3$. 
Halos G$_1$ through G$_4$ have virial masses 
$M_{\rm vir} \simeq (1.5,1.1, 1.1, 1.4) \times 10^{12} \hMsun$, and virial radii  
$r_{\rm vir} \simeq (234,215,216,230) \hkpc$, and so they are of sizes appropriate 
to harbor typical disk galaxies such as the MW or M31. Halos G$_2$ and G$_3$ are 
neighbors located approximately $427 \hkpc$ from each other in a configuration that resembles 
that of the Local Group. Halo G$_1$ is relatively isolated
and is $\sim 2\Mpc$ away from the G$_2$-G$_3$ pair.
All three of these halos accrete only a small fraction of their final 
masses and experience no major mergers at $z \lesssim 1$ (a look-back time of
$\approx 8$~Gyr), and therefore may reasonably host a disk galaxy. Moreover,
their merger histories do not appear to be extreme compared to the expected spread in 
halo accretion histories measured in simulations with better statistics 
\citep[e.g.,][]{Wechsler_etal02}. We have chosen G$_1$ as our fiducial case for 
the satellite-disk interaction experiments described in \S~\ref{sub:control_sims}.

\begin{figure*}[t!]
\centerline{\epsfxsize=7.2in \epsffile{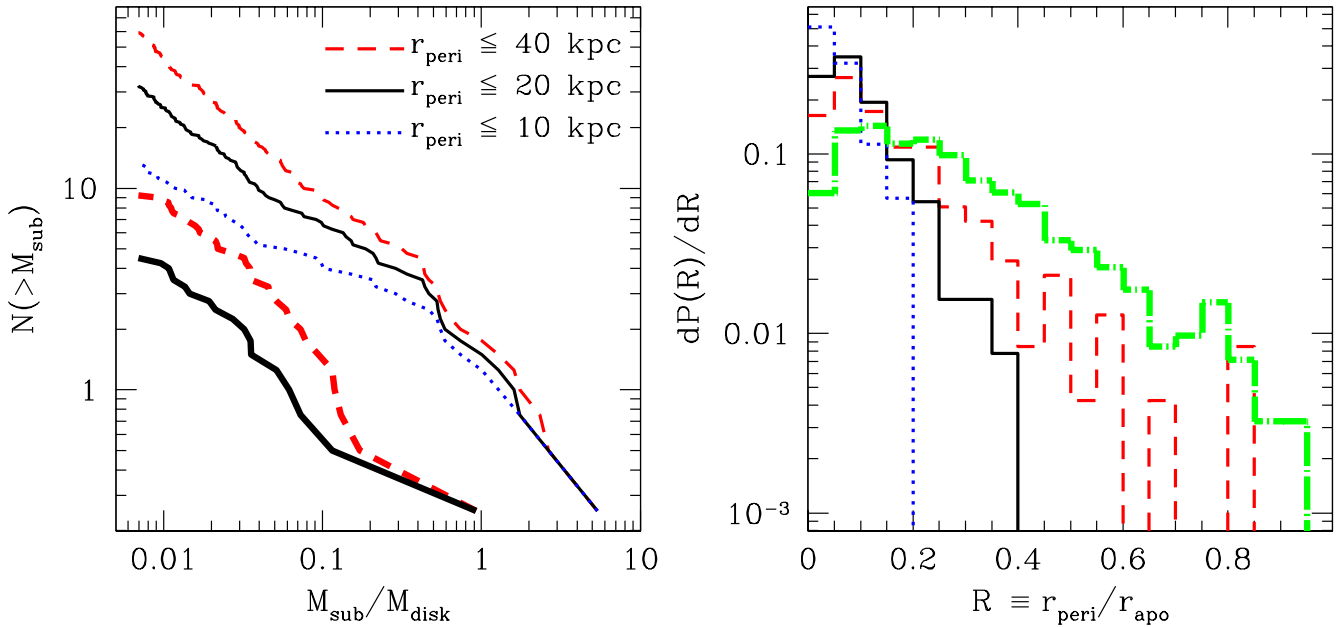}}
\caption{{\it Left:} Cumulative mass functions of subhalos that approach the central regions of 
their hosts averaged over four galaxy-sized halos formed in the {\LCDM} cosmology.
Upper {\it thin} lines present results for objects that have had pericenter
crossings within $10$ ({\it dotted}), $20$ ({\it solid}), and $40\kpc$ ({\it
  dashed}) since a redshift $z=1$. The masses are defined at the simulation
output time nearest to the first inward crossing of a sphere of radius
$50\kpc$ (in the rescaled units of the controlled simulations) and the
pericenters are computed from interpolated orbits between timesteps of the
simulation outputs. The lower {\it thick} lines correspond to the (surviving) subhalos identified at
$z=0$ with pericenters of $r_{\rm peri} \le 40\kpc$ ({\it dashed}) and $r_{\rm
  peri} \le 20\kpc$ ({\it solid}). These pericenters are also estimates based on
the orbit of a test particle in a static NFW potential whose properties match 
those of the host CDM halo at $z=0$. CDM models predict several close encounters of massive subhalos with
the galactic disk since $z=1$. In contrast, extremely few satellites in
present-day subhalo populations are
simultaneously massive enough and characterized 
by small orbital pericenters to have a profound effect on a galactic disk.
{\it Right:} The distribution of orbital pericentric-to-apocentric distance 
ratios, $R \equiv r_{\rm peri}/r_{\rm apo}$, for subhalos with $M_{\rm sub} > 0.05 M_{\rm disk}$. 
Line types for {\it thin} lines are as in the left panel. The {\it thick} dot-dashed line corresponds to 
the eccentricity distribution for {\it all} substructures at $z=0$. Objects that have penetrated
deep enough to encounter a disk are on more radial orbits 
compared to those that belong to the present-day subhalo population. Note that
in both panels, all units are relative to the re-scaled units of the
controlled satellite-disk encounter simulations of \S~\ref{sub:control_sims}.}
\label{fig1}
\end{figure*}

In order to construct mass accretion histories of the host halos 
and orbital trajectories of the accreted subhalos, we identified hosts
and substructures at a large number of simulation outputs up to a redshift 
of approximately $z \approx 10$. We used $96$ available simulation 
outputs in this interval for L25 and $148$ for L20.  
We identified halos using a variant of the Bound Density Maxima 
algorithm which has been discussed extensively in the literature\footnote{For completeness, we used 
$r_{\rm find}=5 \hkpc$ and $r_{\rm find}=2.5 \hkpc$ 
as the search radii in the L25 and L20 simulations, respectively.} \citep{Klypin_etal99}.  
Virial radii are meaningless for subhalos so their radii (and masses) 
are defined using an outer ``truncation'' radius, $r_{\rm trunc}$.
Here we adopt $r_{\rm trunc}$ as the radius at which the logarithmic slope of
the spherically-averaged density profile of the subhalo exceeds a critical
value, ${\rm  d}\ln \rho(r)/{\rm  d}\ln r \vert_{r_{\rm trunc}} = -0.5$. 
This criterion is based on the fact that field halo profiles are never
shallower than this value, and because this definition empirically yields a
good approximation to the subhalo tidal radius, $r_{\rm tid}$,
where the background density of the host halo equals the density of particles
bound to the subhalo. We note here that the spherical approximation 
for substructures is fairly well justified \citep{Moore_etal04,Kazantzidis_etal06} 
and that the outer profiles of subhalos typically fall off with 
${\rm d}\ln \rho/{\rm d}\ln r < -3$ so that the bound mass 
converges well before the formal truncation radius is reached.  

The procedure for tracking subhalo orbital trajectories was developed in 
\citet{Kravtsov_etal04} and we briefly summarize it here for completeness.  
The progenitor for each subhalo at a given timestep is selected using the
$25 \%$ most bound particles. We search a previous output timestep
for the halo with the highest common fraction of these particles and
declare this system to be the primary progenitor. Many examples of
such tracks for the L25 halos and a more complete description of our
methods and their tests are given in \citet{Kravtsov_etal04} and
\citet{Zentner_etal05b}.  

\subsection{Encounters of CDM Substructures with Galactic Disks}
\label{sec:sat_cen_encounter}

We begin by demonstrating that the $z=0$ properties of subhalo 
populations are strongly biased relative to the {\it history} 
of central encounters since a redshift $z=1$. 
Figure~\ref{fig1} shows mass functions and orbital eccentricity 
distributions of substructure populations within the halos G$_1$-G$_4$.
For this presentation, we have scaled the masses and orbital radii of all
subhalos ($M \rightarrow f_m M$ and $r \rightarrow f_r r$) in order to
normalize them to a single host halo mass and radius: $f_m = M_{\rm vir}/M_h$ and 
$f_r = r_{\rm vir}/R_h$. Here, $R_h = 244.5 \kpc$ and $M_h = 7.35 \times 10^{11} \Mo$ 
are the dark halo tidal radius and the dark halo mass interior to this radius of
the primary disk galaxy model, which we use in the controlled satellite-disk encounter 
simulations described in \S 2.5. The mass of the disk in this galaxy
model is $M_{\rm disk} = 3.53 \times 10^{10} \Mo \approx 0.048 M_h$,
and we use this number to characterize the importance of disk impacts in the
figures below. Detailed description of the primary disk galaxy model will be
given in \S 2.4.

The left panel of Figure~\ref{fig1} shows cumulative mass 
functions of substructures with different orbital pericenters that approach
the central regions of their hosts, where the galactic disk presumably lies.
The mass functions are averages over all four galaxy-sized host halos in the
cosmological simulations that we analyze. Satellite masses are measured at the epoch
closest to when the subhalo first crossed inward a (scaled) infall
radius of $r_{\rm inf}=50\kpc$ from the host halo center. We define this to be the
epoch that our controlled simulations initiate. The radius $r_{\rm inf}$ is
used to select infalling satellites that are likely
to have a substantial effect on the disk structure. Subhalos that remain on
the outskirts of the halo will not affect the disk as significantly. We refer
to satellite properties at the simulation output time nearest to the first inward
crossing of $r_{\rm inf}$ several times in what follows. The pericenters associated 
with these subhalos are computed from the 
the orbit of a test particle in a static NFW potential whose properties match 
those of the host CDM halo at the time of $r_{\rm inf}$.

Many of these satellites are no longer present at $z=0$ due
to disruption, but were present at some earlier epoch. We also note 
that these mass functions are constructed taking into account that a single 
distinct object may pass the host halo center multiple times since $z=1$.
This panel serves to illustrate that interactions of massive subhalos with
the center of the host potential where the galactic disk resides since
$z=1$ should be common occurrences in the standard {\LCDM}
cosmological model. Indeed, in the adopted cosmology, $\sim 5$ systems 
with masses $M_{\rm sub} \gtrsim 0.2M_{\rm disk}$ cross
through the central region ($r_{\rm peri} \le 20\kpc$) of a galaxy-sized
halo over a $\sim 8$~Gyr period.  

On the other hand, by examining the {\it present-day} substructure population 
of a galaxy-sized halo one would be led to believe that close encounters of
massive subhalos with the galactic disk are very rare: on average, $\sim 1$
system more massive than $0.1M_{\rm disk}$ is expected to be on a potentially
damaging orbit with $r_{\rm peri} \le 40\kpc$ at $z=0$. The difference between
the properties of subhalo populations that were accreted in the past
and those corresponding to the surviving systems is not surprising.
Subhalos on highly eccentric orbits that pass closest to the
host halo center are precisely those that are most likely to lose a
large percentage of their mass or become tidally disrupted after pericenter crossing
\citep[e.g.,][]{Zentner_Bullock03,Kravtsov_etal04,Gao_etal04,Zentner_etal05a}.
As orbiting substructures continuously lose mass, the fraction of
massive satellites that interact with disks and are likely to
have a significant impact on the disk structure declines with redshift
so that few remain by $z=0$.

The right panel of Figure~\ref{fig1} presents orbital eccentricity distributions
($R \equiv r_{\rm peri}/r_{\rm apo}$) for subhalos with $M_{\rm sub} > 0.05
M_{\rm disk}$ averaged over the same cosmological simulations. As in the left
panel, results are presented for systems that have had different pericenter
crossings since $z=1$. We also show the eccentricity distribution for {\it
  all} subhalos identified at $z=0$. This is consistent with previous studies 
reporting a median apocenter-to-pericenter ratio of $r_{\rm apo}/r_{\rm peri}
= 6/1$ for present-day substructure populations
\citep[e.g.,][]{Ghigna_etal98}. The plot illuminates another bias in 
the properties of different subhalo populations: objects that penetrate deep into
the center of the host potential are on fairly radial orbits, with $r_{\rm
  peri}/r_{\rm apo} \lesssim 0.2$, compared to the overall $z=0$ subhalo
population. It is also worth noting that circular orbits appear to be far from
the norm for systems that sink into the central parts of halos and interact
with galactic disks. 

While not demonstrated explicitly in the panel discussed above, we find that
the orbits of subhalos infalling since $z \sim 1$ are consistent with
isotropic infall from a radius of $50\kpc$.  It is known that infall from
larger radii is not isotropic \citep{Knebe_etal04,Zentner_etal05b}. However, 
given the limited statistical leverage of our sample and our neglect of 
baryonic components in shaping substructure orbits we are unable to make a more 
general statement regarding the directionality of satellite infall onto the disk.

\subsection{Selection Criteria and Properties of Simulated Cosmological Satellites}
\label{sub:simsatprop}

The primary aim of this work is to investigate the morphological signatures that arise
from a {\LCDM}-motivated satellite accretion history.
While in the previous section we have addressed the characteristics of entire 
subhalo populations within host halos G$_1$-G$_4$, here we describe 
the properties of the {\it particular} set of infalling CDM substructures
that we use as the basis for the controlled subhalo-disk encounter 
simulations below. We choose to focus on the satellite accretion history in host 
halo G$_1$ discussed in \S~2.1.

After computing the orbital tracks of all subhalos relative to the center of the
primary host halo G$_1$, we record all satellites that cross within a
(scaled) infall radius of $r_{\rm inf}=50\kpc$ from the host center
since $z=1$. For our re-simulation campaign, we focus on subhalos that
are a significant fraction of the disk mass, but not more massive than the
disk itself. Figure~\ref{fig1} illustrates that, on average, a galaxy-sized 
dark matter halo will have $\sim 1$ object more massive than $M_{\rm disk}$,
and $\sim 5$ objects more massive than $0.2 M_{\rm disk}$ cross within
the central $20\kpc$ since $z=1$. In the present study we select 
satellites with $0.2 M_{\rm disk} \lesssim M_{\rm sub} \lesssim M_{\rm
  disk}$. The choice of fairly massive subhalos is motivated by the fact 
that they dominate the tidal heating rate by substructure. 
Neglecting extremely massive satellites with $M_{\rm sub} > M_{\rm disk}$
is conservative in this case, because we aim to study the
more subtle features about a relatively well-preserved disk galaxy,
similar to the MW or M31. Such massive systems may well destroy or
considerably alter the morphology of the disk component.
In addition, very massive satellites would add substantial number of their
stars to the disk thereby significantly affecting its morphology and appearance.
We investigate the general robustness of thin disks to encounters with systems
more massive than the ones considered here in a forthcoming study. 

The final selection criterion is related to the orbital pericenters of satellites. 
To be specific, from the sample of selected subhalos, we choose systems which 
cross the central region of halo G$_1$ with $r_{\rm  peri} \lesssim 20 \kpc$. 
Small pericenters are essential in order for satellites to be 
regarded as potentially efficient perturbers. The imposed selection criteria
result in six accreted substructures, which we denote S1-S6, to simulate over a 
$\approx 8$~Gyr period.

One of the advantages of the present study is that the adopted subhalo models
are comparatively free from assumptions associated with their internal
structure. All previous numerical investigations of satellite-disk
interactions employed satellite models with poorly-motivated density profiles 
ranging from dense $\gamma$-models to low-concentration King profiles and truncated 
\citet[][hereafter NFW]{Navarro_etal96} models. The contrast between low- and
high-density satellites was primarily emphasized by \citet{Huang_Carlberg97}
who argued that the latter survive tidal stripping and are accreted onto the
disk, whereas the former are incorporated into outer halos. 
Our approach is to extract the {\it exact} density structures of cosmological
subhalos and use them to construct high-resolution, self-consistent satellite
models for the controlled simulations below.

Density profiles of selected cosmological satellites are computed at the
simulation output time nearest to when
the subhalo first passes within $r_{\rm inf}=50\kpc$ of the host halo center. 
Subhalos S1-S6 show no sign of a central density core and are quite 
cuspy down to the minimum resolved radius, as expected from previous work 
\citep{Kazantzidis_etal04b}. Individual infalling substructures are modeled using a  
multi-parameter $(\alpha, \beta, \gamma)$ density profile law 
\citep{Zhao96,Kravtsov_etal98}. A truncation in the density profile is also
introduced to mimic the effects of the host halo tidal field. This function
allows a range of inner, $\gamma$, and outer, $\beta$, slopes across a fit
scale radius, $r_s$, and includes a sharpness parameter,
$\alpha$, to control the nature of the transition between the two regimes. The 
normalization of the density profile is set by a characteristic inner density, $\rho_s$.
From a practical perspective, truncating the profiles sharply for $r > r_{\rm
  tid}$, where $r_{\rm tid}$ denotes the tidal radius of the satellite,
results in models that are out of equilibrium. Following \citet{Kazantzidis_etal04a},
we implement an exponential cutoff in the profile of each subhalo which sets in at the tidal 
radius and ensures that for $r > r_{\rm tid}$  the density of self-bound
satellite material falls off smoothly to very low values. This procedure
results in some additional bound mass beyond $r_{\rm tid}$, the precise amount
of which depends upon the adopted model parameters. In the models considered
here, the mass exterior to $r_{\rm tid}$ is typically only about $\sim 2\%$ of 
the bound mass for each subhalo. 

While it is cumbersome to present the detailed fits for each selected
cosmological satellite here, we report that they exhibit NFW behavior in their
central parts being very well approximated by 
a steep inner cusp of $\rho(r)\propto r^{-1}$. In contrast, the
NFW profile does not provide an accurate description of the internal structure
of subhalos at intermediate and large radii in agreement with previous 
numerical studies \citep{Ghigna_etal98,Kazantzidis_etal04b}. In all cases, the
asymptotic outer slopes are found to be much steeper than $-3$. This is attributable 
to the fact that tidal stripping from the host halo removes the outer parts of 
the orbiting satellites steepening their density profile. The best-fit values
vary from $\beta = -3.7$ to $\beta = -4.2$ suggesting that subhalos are better 
represented by the \citet{Hernquist90} profile in their outer parts in
agreement with \citet{Ghigna_etal98}.

We follow the procedure outlined in \citet{Kazantzidis_etal04a} to construct 
self-consistent, $N$-body realizations of satellite models. This method is
based on calculating the exact phase-space distribution function (DF) for a
given density profile through an Abel transform. For the purposes of our
study, we assume that the DF depends only on the binding energy per unit mass, $E$. 
Numerical models of infalling substructures are then readily generated by
randomly sampling the particle positions and velocities from this unique 
DF. A critical reader may note that satellites in the present study are 
modeled under the assumptions of spherical symmetry and an isotropic velocity
dispersion tensor even though cosmological halos exhibit strong departures
from spherical symmetry and velocity isotropy \citep[e.g.,][]{Cole_Lacey96}. 

While this is generally correct, it has been shown that mass loss processes
(tidal stripping and shocking) inside a host potential drive substructures
toward shapes that are nearly spherical in both configuration and velocity
space \citep{Moore_etal04,Kazantzidis_etal06}. Therefore, the assumptions of 
spherical symmetry and velocity isotropy for subhalos adopted here are 
well-grounded. Finally, the masses, positions and velocities of each selected
satellite are scaled in order to correspond to the 
same fractions of virial quantities $M_{\rm rvir}$, $r_{\rm rvir}$, and
$V_{\rm vir}$, respectively, in both the parent CDM halo and in the primary
galaxy model. 

\begin{figure}[t]
\centerline{\epsfxsize=3.5in \epsffile{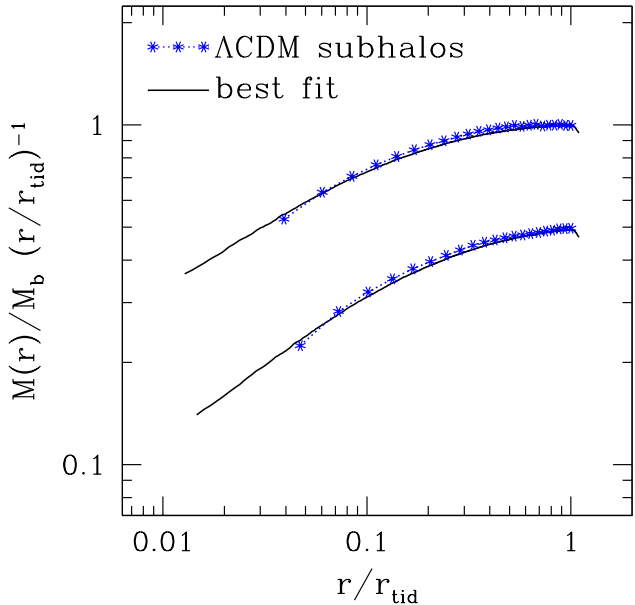}}
\caption{Cumulative mass profiles, $M(r)$, for representative 
cosmological satellites S1 and S2. {\it Stars} show profiles derived directly from the 
cosmological simulation of host halo G$_1$. {\it Solid} curves present fits to 
the mass distribution obtained according to the method described in the 
text (\S 2.3) and are plotted from the adopted force resolution 
($2\epsilon_{\rm sat}=300\pc$) outward. In order to emphasize the quality of 
the fit, we plot the quantity $M(r)/r$ on the vertical axis rather than $M(r)$.
Masses and radii are normalized to the bound mass, $M_b$, and tidal 
radius, $r_{\rm tid}$, of each subhalo, respectively. 
For clarity, the mass profile corresponding to the lower curve is vertically 
shifted downward by $0.5$ dex. The adopted procedure
ensures that the satellite models employed in this study are initialized
according to the density structure of subhalos found in high-resolution
cosmological $N$-body simulations.
\label{fig2}}
\end{figure}
%


\begin{table*}
\caption{Properties of the Satellite Models}
\begin{center}
\begin{tabular}{lccccccccccc}
\tableline\tableline\\
\multicolumn{1}{c}{}              &
\multicolumn{1}{c}{}              &
\multicolumn{1}{c}{$t_{\rm acc}$} &
\multicolumn{1}{c}{$d$}           &
\multicolumn{1}{c}{}              &
\multicolumn{1}{c}{$r_{\rm tid}$} &
\multicolumn{1}{c}{$V_{\rm max}$} &
\multicolumn{1}{c}{$r_{\rm max}$} &
\multicolumn{1}{c}{}              &
\multicolumn{1}{c}{}              &
\multicolumn{1}{c}{$\theta$}      & 
\multicolumn{1}{c}{}              
\\
\multicolumn{1}{c}{Model}&
\multicolumn{1}{c}{$z_{\rm acc}$} &
\multicolumn{1}{c}{($\rm Gyr$)}&
\multicolumn{1}{c}{($\rm kpc$)}&
\multicolumn{1}{c}{$M_{\rm sub}/M_{\rm disk}$} &
\multicolumn{1}{c}{(${\rm kpc}$)}&
\multicolumn{1}{c}{(${\rm km s^{-1}}$)}&
\multicolumn{1}{c}{(${\rm kpc}$)}&
\multicolumn{1}{c}{$r_{\rm peri}/R_d$} &
\multicolumn{1}{c}{$r_{\rm apo}/R_d$} &
\multicolumn{1}{c}{($\degrees$)}&
\multicolumn{1}{c}{$r_{\rm peri}^{*}/R_d$} 
\\
\multicolumn{1}{c}{(1)}&
\multicolumn{1}{c}{(2)}&
\multicolumn{1}{c}{(3)}&
\multicolumn{1}{c}{(4)}&
\multicolumn{1}{c}{(5)}&
\multicolumn{1}{c}{(6)}&
\multicolumn{1}{c}{(7)}&
\multicolumn{1}{c}{(8)}&
\multicolumn{1}{c}{(9)}&
\multicolumn{1}{c}{(10)}&
\multicolumn{1}{c}{(11)}&
\multicolumn{1}{c}{(12)}
\\
\\
\tableline
\tableline
\\
S1 & $ 0.96 $ & $ 7.6 $ & $ 45.2 $ & $ 0.33 $ & $ 24.8 $ & $ 42.4 $ & $ 6.9 $ & $ 2.6 $ & $ 17.7 $ & $93.3$  & $1.2$ \\
S2 & $ 0.89 $ & $ 7.3 $ & $ 40.7 $ & $ 0.57 $ & $ 21.5 $ & $ 59.8 $ & $ 8.1 $ & $ 2.6 $ & $ 15.7 $ & $86.6$  & $1.2$ \\     
S3 & $ 0.54 $ & $ 5.3 $ & $ 34.0 $ & $ 0.42 $ & $ 23.0 $ & $ 50.3 $ & $ 7.6 $ & $ 6.2 $ & $ 19.7 $ & $45.1$  & $3.5$ \\ 
S4 & $ 0.32 $ & $ 3.6 $ & $ 28.8 $ & $ 0.45 $ & $ 19.6 $ & $ 41.1 $ & $ 4.1 $ & $ 0.5 $ & $ 10.3 $ & $59.9$  & $1.4$ \\ 
S5 & $ 0.20 $ & $ 2.4 $ & $ 50.3 $ & $ 0.22 $ & $ 27.3 $ & $ 41.5 $ & $ 5.7 $ & $ 3.7 $ & $ 34.3 $ & $117.7$ & $2.6$ \\ 
S6 & $ 0.11 $ & $ 1.4 $ & $ 50.5 $ & $ 0.21 $ & $ 23.2 $ & $ 38.8 $ & $ 3.7 $ & $ 1.1 $ & $ 21.6 $ & $144.5$ & $1.9$ \\ 
\\
\tableline
\end{tabular}
\end{center}
{\sc Notes.}--- Columns (2)-(10) refer to the properties of cosmological
satellites recorded in the simulation output time closest to first crossing
within a sphere of radius $r_{\rm inf}=50\kpc$ from the center of host halo G$_1$.
Col. (1): Satellite model.
Col. (2): Redshift at which satellite properties were computed.
Col. (3): Look-back time in units of Gyr corresponding to the redshift of Col. (2).
Col. (4): Satellite distance from the center of the host halo in $\rm {kpc}$.  
          Due to the finite spacing of simulation outputs this distance
          differs from $50\kpc$.
Col. (5): Bound satellite mass in units of the mass of the 
          disk, $M_{\rm disk}=3.53 \times 10^{10} \Mo$, used in the
          controlled satellite-disk encounter simulations.
Col. (6): Tidal radius in $\rm {kpc}$.
Col. (7): Maximum circular velocity in ${\rm km s^{-1}}$.
Col. (8): Radius at which the maximum circular velocity occurs in $\rm {kpc}$.
Col. (9): Pericentric radius of the orbit in units of the radial scale length
          of the disk, $R_d = 2.82\kpc$, used in the controlled simulations.
Col. (10): Apocentric radius of the orbit in units of $R_d$.
Col. (11): Angle between the angular momentum of the satellite
           and the {\it initial} angular momentum of the disk in degrees. This
           angle is defined so that  $\theta \sim 90 \degrees$ represents an
           orbit that starts off near the pole of the disk, $\theta < 90
           \degrees$ implies a prograde orbit, and $\theta > 90 \degrees$
           corresponds to a retrograde orbit.
Col. (12): Pericentric radius of the orbit in units of $R_d$ calculated directly from 
           the controlled simulations. This is to be compared with Col. (9).
\label{table:sat_param}
\end{table*}

For each satellite model, we generate an $N$-body realization containing $N_{\rm sat} = 10^6$ 
particles and employ a gravitational softening length of $\epsilon_{\rm sat}=150 \pc$.
This choice enables us to to resolve density structures down to $\sim 1\%$ of the tidal radius of the 
simulated systems. The adopted particle numbers and force resolution are
sufficient to resolve mass loss processes inside a host potential \citep{Kazantzidis_etal04b}.

In what follows, we examine whether the generous allowance for density profile
shapes we considered here can give an accurate representation of the density structure
of cosmological subhalos. Figure~\ref{fig2} presents cumulative mass profiles, $M(r)$, of characteristic 
cosmological satellites S1 and S2 along with representations of their internal structure 
obtained using the density laws discussed above. Specifically, we plot the quantity 
$M(r)/r$, where masses and radii are normalized to the bound mass and tidal
radius of each satellite, respectively. This choice scales out the gross
dependence of the mass profile on radius and shows the acceptability of the fit better.
Figure~\ref{fig2} demonstrates that the satellite models that will be used in 
the numerical experiments of satellite-disk interactions
are initialized according to the internal structure of subhalos found in high-resolution 
cosmological $N$-body simulations. 

Table~\ref{table:sat_param} provides a summary of the infall times, 
orbital parameters, and internal structures of cosmological subhalos 
S1-S6 that we study in the numerical experiments of satellite-disk 
interactions. Columns (2)-(10) list parameters extracted
directly from the cosmological simulation of host halo G$_1$, and used as initial 
conditions in the controlled numerical experiments. We remind the reader that 
the structural properties for each subhalo were measured at the simulation 
output time closest to the first inward crossing of a sphere of radius $50\kpc$ 
(in the rescaled units of the controlled simulations). Several interesting
findings regarding the properties of the simulated satellites can be read
from the entries in this table.

First, the $6$ different satellite models cover the mass range 
$0.21 < M_{\rm sub}/M_{\rm disk} < 0.57$, where $M_{\rm disk} = 3.53 \times
10^{10}\Mo$ is the mass of the stellar disk used in the controlled satellite-disk 
encounter simulations that we describe below in \S~\ref{sub:control_sims}. 
This mass range corresponds to 
$7.4 \times 10^{9} \lesssim M_{\rm sub}/M_{\odot} \lesssim 2 \times 10^{10}$. 
As we stated earlier, this mass regime is considerably larger than the regime
of relevance for the missing galactic satellites problem ($\lesssim 10^{9}
M_{\odot})$. As a comparison,
the Large Magellanic Cloud (LMC) and M32, the most massive
dwarf companions of MW and M31, have an estimated {\it total} present mass 
of $\sim 2 \times 10^{10} M_{\odot}$ \citep[e.g.,][]{Schommer_etal92,Mastropietro_etal05} 
and $2.1 \times 10^{9} M_{\odot}$ \citep{Mateo98}, respectively. The mass of 
satellites S1-S6 corresponds to the upper limit of the mass function
of observed satellites in the Local Group.

Another important characteristic of the infalling subhalos is their 
spatial extent. Specifically, they are very extended, with $r_{\rm tid}
\gtrsim 20 \kpc$, so of comparable size to the disk itself. This fact suggests
that the precise pericenter or angular momentum vector of the 
orbit may not be particularly important in driving the disk response. We
discuss this point in more detail in Paper II. Note that with the exceptions 
of \citet{Font_etal01} and \citet{Gauthier_etal06} all previous studies of 
satellite-disk interactions modeled exclusively the {\it compact}, baryonic
cores of the infalling satellites neglecting the more diffuse dark matter
component. Thus, it remains uncertain whether these earlier investigations
accurately captured the amount of global morphological evolution that
accreting subhalos can induce in cold stellar disks.

Table~\ref{table:sat_param} also shows that the majority of infalling satellites are on fairly eccentric 
orbits with apocenter-to-pericenter ratio of $r_{\rm apo}/r_{\rm peri} > 6$. 
Interestingly, satellites S4 and S6 appear to be on highly radial orbits with 
$r_{\rm apo}/r_{\rm peri} \sim 20$. Moreover, as we stated earlier,
our satellites are consistent with isotropic infall but the sample size is
insufficient to provide significant limits on anisotropic infall. 
Despite this fact we note the existence of an adequate variety of recorded 
orbits ranging from polar (S1,S2) to prograde (S3,S4) to retrograde (S5,S6).

Finally, the orbital pericenters of the infalling systems measured directly in the 
satellite-disk encounter simulations are, in most cases, smaller by a 
factor of $\sim 2$ compared to those calculated from the cosmological
simulation, although there are cases where the former are
larger (S4 and S6). Several reasons may be responsible for these
discrepancies.  First, the presence of the disk in the controlled
simulations provides a larger central potential compared to the dark halo
alone. Secondly, the pericenters measured in the cosmological simulations are 
computed from the the orbit of a test particle in a static NFW potential 
because the spacing of the simulation outputs makes several estimates of pericenters 
directly from orbital tracks rather poor. However, we argue that the {\it global} disk
morphological signatures we report in this study are not sensitive to the
magnitude of this difference in pericenters. The reason is that, as long as 
a central encounter takes place, the subhalos are so extended that their
dynamical effects are felt across the entire disk nearly
independently of the exact value of $r_{\rm peri}$.

\subsection{Primary Disk Galaxy Models}
\label{sub:galaxy_models}

With the notable exception of \citet{Gauthier_etal06}, all previous numerical 
investigations of satellite-disk interactions used approximate schemes for
building $N$-body models of disk galaxies. In our study we employ the 
method of \citet{Widrow_Dubinski05} to construct numerical realizations of 
multi-component disk galaxies consisting of a thin stellar disk, a
central bulge, and an extended dark matter halo. These galaxy models are
derived from explicit DFs for each component and thus represent axisymmetric, 
equilibrium solutions to the coupled collisionless Boltzmann 
and Poisson equations. Because of that, this technique allows us to set up
thin, equilibrium disk galaxies without instabilities and transient effects 
that can be present in other simpler approximate schemes
\citep[e.g.,][]{Hernquist93}. The \citet{Widrow_Dubinski05} galaxy models are 
specified by a large number of parameters that permit detailed modeling of
actual galaxies such as the MW and M31 by fitting to observational data sets. 

For our experiments, we choose their best-fit model MWb, which satisfies a
broad range of observational constraints available for the MW galaxy with
simultaneous fits to the inner and outer rotation curve, bulge velocity
dispersion, vertical force in the solar neighborhood, and total mass at large
radii. The stellar disk never dominates the rotation curve of model
MWb. Direct numerical simulations confirm its stability against bar formation for 
$10$~Gyr. In what follows, we give an overview of the parameters and setup,
but refer the reader to \citet{Widrow_Dubinski05} for a detailed discussion.

The \citet{Widrow_Dubinski05} disk galaxy models comprise an exponential
stellar disk, a Hernquist model bulge \citep{Hernquist90} and an NFW dark
matter halo. The halo density distribution is given by
\be
   \rho_{\rm NFW}(r) = 
      \frac{\rho_h} {\left (r/r_h\right) 
        \left (1 + r/r_h\right )^2} \ ,
\ee
where $\rho_h$ is a characteristic inner density, and $r_h$ denotes the scale radius 
of the profile. For convenience, the inner density $\rho_h$ is expressed in terms of a 
characteristic velocity $\sigma_h$, $\rho_h\equiv\sigma_h^2/4\pi G r_{h}^2$, with $G$ 
being the gravitational constant. The NFW density profile is formally infinite
in extent with a cumulative mass that diverges as $r \rightarrow \infty$. In
order to keep the total mass finite, it is necessary to truncate the profile. 
In practice, this is accomplished by introducing an {\it energy} cutoff which
is described by the halo tidal radius parameter, $\cal E_{\rm h}$ ($0 < \cal
E_{\rm h}$\,$< 1$). This parameter controls the tidal radius of the halo, $R_h$,
which represents the outer edge of the entire system. For model building 
a natural choice for the value of $\cal E_{\rm h}$ 
would be the one for which the resulting $R_h$ becomes roughly equal to 
the cosmologically motivated virial radius, $r_{\rm vir}$. Finally, arbitrary 
amount of rotation controlled by the parameter $\alpha_h$ can be added to the
halo model ($\alpha_h=1/2$ implies no rotation). 

The bulge follows the \citet{Hernquist90} density profile
\be
   \rho_{\rm H}(r) = \frac{\rho_b}{\left (r/a_b \right)
        \left (1 + r/a_b\right )^3} \ ,
   \label{bulge_density}
\ee
where $\rho_b$ and $a_b$ are a characteristic inner density and scale length of the 
bulge, respectively. In analogy to the treatment of the halo, the inner
density of the bulge $\rho_b$ is expressed in terms of a characteristic
velocity dispersion $\sigma_b$, $\rho_b\equiv\sigma_b^2/2\pi G a_b^2$. 
Similarly, the modeling of the stellar bulge is complete once the tidal 
radius parameter, $\cal E_{\rm b}$, which controls the outer edge of the bulge, 
$R_b$, and the rotation parameter, $\alpha_b$, are specified. Finally, the
bulge mass is denoted by $M_b$. 

The stellar disks are assumed to be axisymmetric with density profiles 
$\rho_d = \rho_d(R,z)$ and DFs which are adopted directly from the
self-consistent disk galaxy models of \citet{Kuijken_Dubinski95}. The surface
density profile follows an exponential distribution in cylindrical radius $R$,
while the vertical structure is modeled by constant-thickness self-gravitating 
isothermal sheets \citep{Spitzer42}
\be
   \rho_d(R,z) \propto \exp\left(-\frac{R}{R_d}\right) 
       {\rm sech}^2\left(\frac{z}{z_d}\right) \ ,
   \label{disk_density}
\ee
where $R_d$ and $z_d$ denote the radial scale length and vertical scale height 
of the disk, respectively. We emphasize that in the model formulation the vertical
distribution of disk stars is described by a sech$^2$ function. Note that for $z
\gtrsim z_d$, the sech$^2$ law mimics an exponential profile with scale height 
$h_z \approx z_d/2$. The construction of the stellar disk is also
characterized by a truncation radius, $R_{\rm out}$, and by the parameter, 
$\delta R_{\rm out}$, which controls the sharpness of the truncation. 

The disk phase-space DF is fully determined once the velocity
ellipsoid of the disk is specified. The radial velocity dispersion, 
$\sigma_R(R)$, is assumed to be exponential with $\sigma_R^{2}(R) =
\sigma_{R0}^{2}\exp\left(-R/R_d\right)$, where $\sigma_{R0}$
denotes the central radial velocity dispersion. The disk azimuthal dispersion, 
$\sigma_\phi(R)$, is related to $\sigma_R(R)$ via the epicycle approximation 
\citep{Binney_Tremaine87}, while the vertical velocity dispersion, $\sigma_z$, 
is set by the requirement that the adopted value of the scale height $z_d$ is 
maintained in the total potential of the galaxy model.
In all, the density and phase-space DFs for the three galactic components are 
characterized by $14$ free parameters which may be tuned to fit a wide range
of observational data. Given a choice of these parameters, the
\citet{Widrow_Dubinski05} technique uses an iterative scheme to 
calculate the composite DF and produce equilibrium $N$-body models of disk 
galaxies. 

The simulations reported here use $N_d=10^{6}$ particles in the 
disk, $N_b=5\times10^{5}$ in the bulge, and $N_h=2\times10^{6}$ in the dark
matter halo, and employ a gravitational softening of $\epsilon_d=50 \pc$, 
$\epsilon_b=50 \pc$, and $\epsilon_h=100 \pc$, respectively. Mass and force 
resolution are adequate to resolve the vertical structure of a thin stellar 
disk and minimize artificial disk heating through interactions with the 
massive halo particles.

Physical and numerical parameters for each of the components of the
adopted primary galaxy model MWb are listed in Table~2.
The best-fit value of the solar radius and the total 
circular velocity of the galaxy model there are given by $R_{\odot}
\simeq 8 \kpc$ and $V_c(R_\odot) \approx 234\kms$, respectively.
The initial disk scale height is equal to $z_d = 400\pc$, 
consistent with the observed value for the old, thin stellar
disk of the MW obtained with a variety of methods \citep[e.g.,][]
{Kent_etal91,Dehnen_Binney98,Mendez_Guzman98,Larsen_Humphreys03}.
Observational evidence \citep[e.g.,][]{Quillen_Garnet01,Nordstrom_etal04}
indicates that the scale height of the thin disk of the MW has not changed
significantly over the period modeled in this study ($z \lesssim 1$). Therefore, 
it is reasonable to initialize the stellar disk with such thickness. 

The Toomre stability parameter of the disk is equal to $Q = 2.2$ at $R=2.5 R_d$,
indicating that the model is stable against local non-axisymmetric
instabilities. This suggests that any significant bar growth during
the satellite-disk encounter simulations can be assumed to be tidally
induced by the infalling subhalos. It is also worth emphasizing that
the adopted set of parameters gives $R_h=244.5 \kpc$, a value in
agreement with the nominal virial radius of the MW galaxy as predicted
by CDM models of cosmological structure formation
\citep[e.g.,][]{Klypin_etal02}.


\begin{table}[t!]
\begin{center}
\centering
\caption{Primary Galaxy Model Parameters}
\begin{tabular}{lcc} \hline \hline
\\[1mm]
Component  &  Parameter &  Value 
\\
\\
\hline \hline
\\
Disk:   &                        &                                              \\
        &  $M_{\rm disk}$         & $ 3.53 \times 10^{10}\,\Mo$                  \\
        &  $R_d$                 & $ 2.82\kpc$                                  \\
        &  $z_d$                 & $ 400\pc$                                    \\
        &  $R_{\rm out}$          & $ 30\kpc$                                   \\
        &  $\delta R_{\rm out}$   & $ 1\kpc$                                    \\
        &  $\sigma_{R0}$         & $ 124.4\kms$                                 \\
        &  $Q(2.5R_d)$           & $ 2.2$                                       \\
        &  $R_{\odot}$            & $ 8\kpc$                                    \\
        &  $N_d$                 & $ 10^{6}$                                    \\
        &  $\epsilon_d$          & $ 50\pc$ \\[1mm]
\hline\\[1mm]
Bulge:  &                        &                                              \\
        &  $\cal E_{\rm b}$       & $0.21$                                      \\
        &  $\sigma_b$            & $435.7\kms$                                  \\
        &  $a_b$                 & $0.88\kpc$                                   \\
        &  $\alpha_b$            & $0.5$                                        \\                                      
        &  $M_b$                 & $1.18 \times 10^{10}\,\Mo$                   \\      
        &  $R_b$                 & $3.05\kpc$                                   \\
        &  $N_b$                 & $5\times10^{5}$                              \\
        &  $\epsilon_b$          & $50\pc$ \\[1mm]
\hline\\[1mm]
Halo:   &                        &                                              \\
        &  $\cal E_{\rm h}$       & $0.1$                                       \\      
        &  $\sigma_h$            & $344.7\kms$                                  \\
        &  $r_h$                 & $8.82\kpc$                                   \\
        &  $\alpha_h$            & $0.5$                                        \\
        &  $M_h$                 & $7.35 \times 10^{11}\,\Mo$                   \\      
        &  $R_h$                 & $244.5\kpc$                                  \\
        &  $N_h$                 & $2\times10^{6}$                              \\
        &  $\epsilon_h$          & $100\pc$ \\[1mm]
\hline
\end{tabular}
\end{center}
\label{table:prim_param}
\end{table}
%

\subsection{Description of Satellite-Disk Encounter Simulations}
\label{sub:control_sims}

All satellite-disk encounter simulations were carried out
using PKDGRAV, a multi-stepping, parallel, tree $N$-body code 
\citep{Stadel01}. PKDGRAV uses a spline softening length, such 
that the force is completely Keplerian at twice the quoted softening
lengths, and multi-stepping based on the local acceleration of
particles. We used an adaptive, kick-drift-kick leapfrog integrator
with individual particle time steps $\Delta t_i$ chosen
according to $\Delta t_i \leq \eta (\epsilon_i/\alpha_i)^{1/2}$, where 
$\epsilon_i$ is the gravitational softening length of  the particle, 
$\alpha_i$ is the value of the local acceleration, and $\eta$ is a 
parameter that specifies the size of the individual timesteps and consequently 
the time accuracy of the integration.

For all simulations, we set the base-timestep to be equal to $1\%$ of
the orbital time of the galaxy model at the disk half-mass radius and allowed
individual particle timesteps to be at most a factor of $2^{30}$
smaller. In practice, particle timesteps were found to be at most a 
factor of $2^{12}$ smaller than the base time-step in the specific 
application. The number of timesteps was sufficient to accurately follow 
particle orbits down to the adopted force resolution of the 
simulations and resolve the vertical frequency of disk stars.
The time integration was performed with high enough accuracy
such that the total energy was conserved to better than 
$0.1\%$ in all cases.  

\begin{figure*}[t!]
\centerline{\epsfxsize=6.8in \epsffile{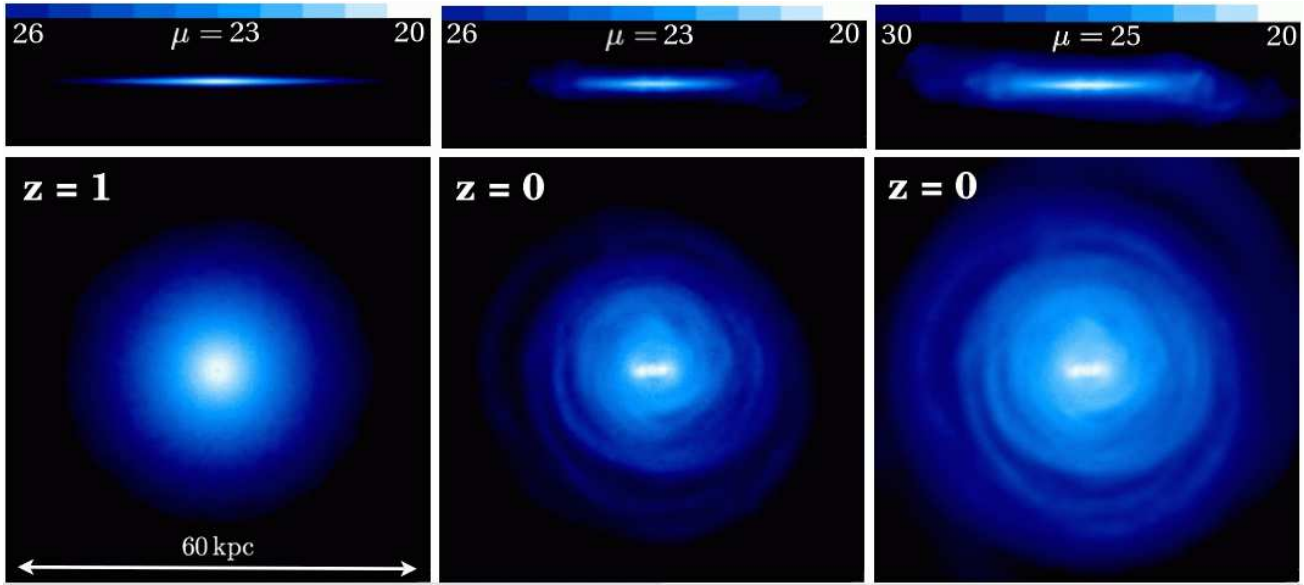}}
\caption{Surface brightness maps of disk stars in the simulated accretion history of 
host halo G$_1$. The edge-on ({\it upper panels}) and face-on ({\it bottom panels}) 
views of the disk are displayed in each frame and the color 
bar in each upper panel indicates the surface brightness limits used to
generate the maps. In constructing these images, a stellar mass-to-light ratio
equal to $M_*/L=3$ is assumed.
Bottom images are $60\kpc$ on a side, while top images measure $18\kpc$ by $60\kpc$. 
The {\it left} panel shows the initial disk assuming that the sequence of satellite-disk 
interactions initiates at $z=1$.  The {\it middle} and {\it right} panels depict the disk 
after the last satellite passage, evolved in isolation for additional $\sim 4$~Gyr, so that 
the evolution of disk stars is followed from $z=1$ to $z=0$. In the left and
middle panels, images are shown to a limit of $\mu = 26$ mag arcsec$^{-2}$,
while the right panel corresponds to a ``deeper'' surface brightness 
threshold of $\mu = 30$ mag arcsec$^{-2}$. Results are presented after
centering the disk to its center of mass and rotating it to a new coordinate
frame defined by the three principal axes of the total disk inertia tensor.
Considerable flaring and a wealth of features that they might falsely be 
identified as tidal streams can be seen in the perturbed disk down to 
$26-30$ mag arcsec$^{-2}$. The existence of non-axisymmetric structures such
as extended outer rings and bars after a significant amount of time subsequent
to the last accretion event confirm their robustness and indicate that
axisymmetry in the disk has been destroyed and is not restored at late times.
\label{fig3}}
\end{figure*}

We modeled satellite impacts S1-S6 as a sequence of encounters.
Starting with subhalo S1 we included subsequent systems at the epoch 
when they were recorded in the cosmological simulation (Table~\ref{table:sat_param}). 
Due to the fact that the simulated infalling substructures were fairly massive and of
comparable size to the disk itself they were not introduced in the simulations
directly, but rather were grown adiabatically in their orbits. This procedure
ensures that the disk does not suffer 
substantial perturbations by the sudden presence of the satellite and change 
in potential at its vicinity \citep{Walker_etal96}. The mass of each satellite
was increased linearly to its final value during a timescale that ranges
between $\sim 150$ and $400$~Myr depending on the subhalo mass. The disk
structure was found to evolve only slightly during the growth period of each satellite. 

Accreting subhalos were removed from the controlled simulations once they
reached their maximum distances from the disk after crossing,
and thus any given satellite was not permitted to complete a second orbit. 
Uniformly, these distances were only slightly smaller compared to the starting 
radii of the orbits. The accretion times of simulated subhalos S1-S6 
(Table~\ref{table:sat_param}) were such that the disk interacted {\it
  simultaneously} with the first two satellites S1 and S2.
We conducted an additional experiment in which subhalo S2 was introduced in the
controlled simulation {\it after} the removal of satellite S1 and confirmed that the disk
exhibited the same global evolution and morphological signatures as in the
standard case. 

In all of the experiments we performed, the satellites lost
$\gtrsim 80 \%$ of their mass after the completion of the first orbit. This
justifies our decision to ignore the computationally costly, but dynamically
insignificant, second (subsequent) crossing event. Note that only the {\it
  self-bound} core of each satellite
was extracted from the simulations and not the unbound material, a significant 
fraction of which remained in the region of the disk. Removing the latter would result
in potential fluctuations that could throw the disk out of equilibrium, and thus interfere
with the interpretation of the results.

In order to save computational time, when time
intervals between subhalo passages were larger than the timescale
needed for the disk to relax after the previous interaction, we introduced
the next satellite in the simulation immediately after the disk had settled 
from the previous encounter. Due to the complexity of the interaction, 
we determine this ``settling'' timescale empirically by monitoring basic properties 
of the disk structure (e.g., surface density, velocity dispersions, thickness) as a 
function of time. When these quantities stop evolving significantly within
radii of interest ($ r \lesssim 6 R_d$), the encounter is deemed
complete. Changes of the order of $5-10\%$ in disk properties were 
considered acceptable. We note that the limiting radius of $6R_d$ in chosen
because it initially contains $\sim 98\%$ of the mass of the disk. Typical
settling timescales correspond to $\sim 100-200$~Myr {\it after} the removal
of the bound core of each orbiting subhalo. This eliminates the
computational expense of simulating the disk during the quiet intervals
between interactions.

Finally, we stress that prior to commencing the satellite-disk encounter
simulations, the primary galaxy model was tilted so that the disk 
angular momentum vector was aligned with the angular momentum of
the host CDM halo G$_1$. This choice is motivated by results of cosmological
simulations suggesting that the angular momenta of galaxies and their host
halos tend to be well aligned \citep[e.g.,][]{Libeskind_etal07}.

\section{Results}
\label{sec:results}

In this section we present results regarding the evolution of an
initially-thin stellar disk subject to a cosmologically-motivated subhalo merger 
history. Recall that we examine the disk response to substructure impacts S1-S6, 
which are designed to mimic a reasonable {\it central} accretion history for a
galaxy-sized halo over the past $\sim 8$~Gyr. The ``final'' disk discussed in the
next sections has experienced the S1-S6 encounters and was 
evolved in isolation for $\sim 4$~Gyr after the last interaction. While this
allows for relaxation after the encounters, it also ensures that all of the
resultant morphological features are long-lived rather than transient.
Consequently, our results are relevant to systems that exhibit no obvious,
ongoing encounters and have been allowed to relax in isolation. 
We also stress that we do not include the bulge component in any of the analysis
presented below. The focus is only on the evolution of the disk material.
We defer any discussion regarding the bulge to future work.

We compute all properties of the disk and show all visualizations of the
disk morphology after centering the disk to its center of mass and
rotating it to a new coordinate frame defined by the three principal axes of
the total disk inertia tensor. The motivation behind performing both 
actions is twofold. As discussed in more detail in Paper II, exchange of
angular momentum between the infalling satellites and the disk tilt the disk
plane substantially over the merging history and cause the disk center of mass
to drift from its initial position at the origin of the coordinate frame. In
the original coordinate frame, rotation in a tilted disk would appear as
vertical motion interfering with the interpretation of the results. 

Finally, we underscore that there is nothing exceptional about host halo G$_1$
as far as the properties of the subhalo populations are concerned. In all
other three galaxy-sized halos we analyzed, substructures of similar numbers,
masses, internal structures, orbital parameters, and accretion times were
identified. Though our simulation program is designed to mimic the activity in 
the inner halo of G$_1$, the similarity of subhalo populations in all four
halos suggests that the results presented next should be regarded as fairly general.

\begin{figure*}[t]
\centerline{\epsfxsize=7.2in \epsffile{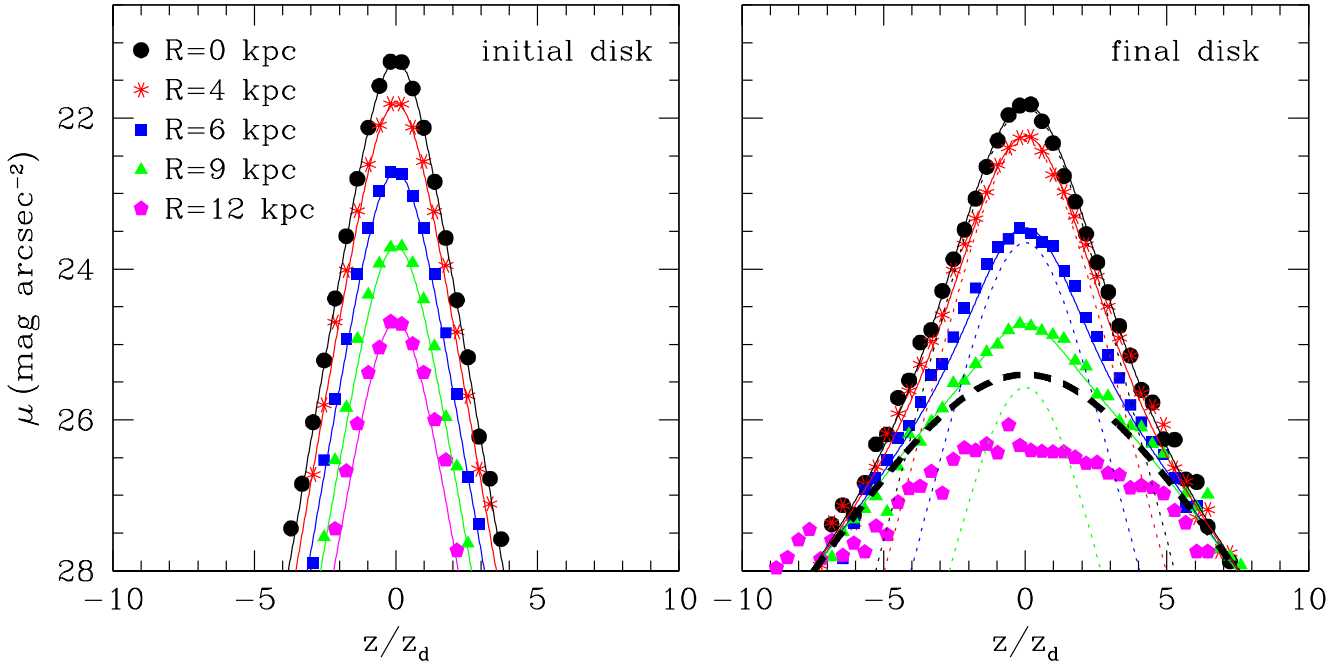}}
\caption{Thin-thick disk decomposition analysis. Surface brightness profiles for disk stars 
as a function of distance above the plane of the disk, $z$. The stellar distribution is observed 
edge-on and results are shown for both the initial ({\it left panel}) and final 
({\it right panel}) disk. The final disk has experienced the S1-S6 encounters and 
was further evolved in isolation for $\sim 4$~Gyr after the last interaction to ensure that is has 
reached a relaxed state. Different symbols correspond to various projected
radii, $|R| = 0, 4, 6, 9, 12\kpc$, averaged between the $\pm |R|$ projections about $R = 0$, 
and within $1\kpc$ strips centered on each of the projected radii.
The {\it solid} lines in the left panel present the analytic sech$^2$ surface brightness profiles of the 
initial thin disk with $z_d = 0.4 \kpc$ and demonstrate a very good agreement with results obtained 
directly from the particle distribution. {\it Solid} lines in the right panel show a ``thin-thick'' 
disk decomposition for the final disk, which consists of a ``thick'' disk
component ({\it dashed line}) with a sech$^2$ scale height of $z_{\rm thick} = 1.6\kpc$ and a
``thin'' disk component ({\it dotted lines}) with a slightly larger scale
height ($z_{\rm thin}=0.6\kpc$) compared to that of the initial thin
disk. The surface brightness profile of the ``thick'' disk is nearly
independent of projected radius for $R<9\kpc$ explaining the existence of only one 
dashed line. At small projected radii ($|R| \lesssim 6\kpc$), the contribution of the thin component to 
the light profile of the stellar disk is dominant for heights $z \lesssim 1\kpc$. As $|R|$ increases, the situation 
is reversed with the thick disk component becoming prevalent. This simple decomposition provides a reasonable 
description of the final disk for $\mu \lesssim 27$ mag arcsec$^{-2}$ and $|R| \lesssim 9\kpc$. 
At fainter magnitudes and beyond $R \sim 9\kpc$, the flare and other diffuse
structures become important and the adopted decomposition fails to provide an
accurate description of the disk structure.
\label{fig4}}
\end{figure*}
\begin{figure*}
\begin{center}
  \includegraphics[height=6cm]{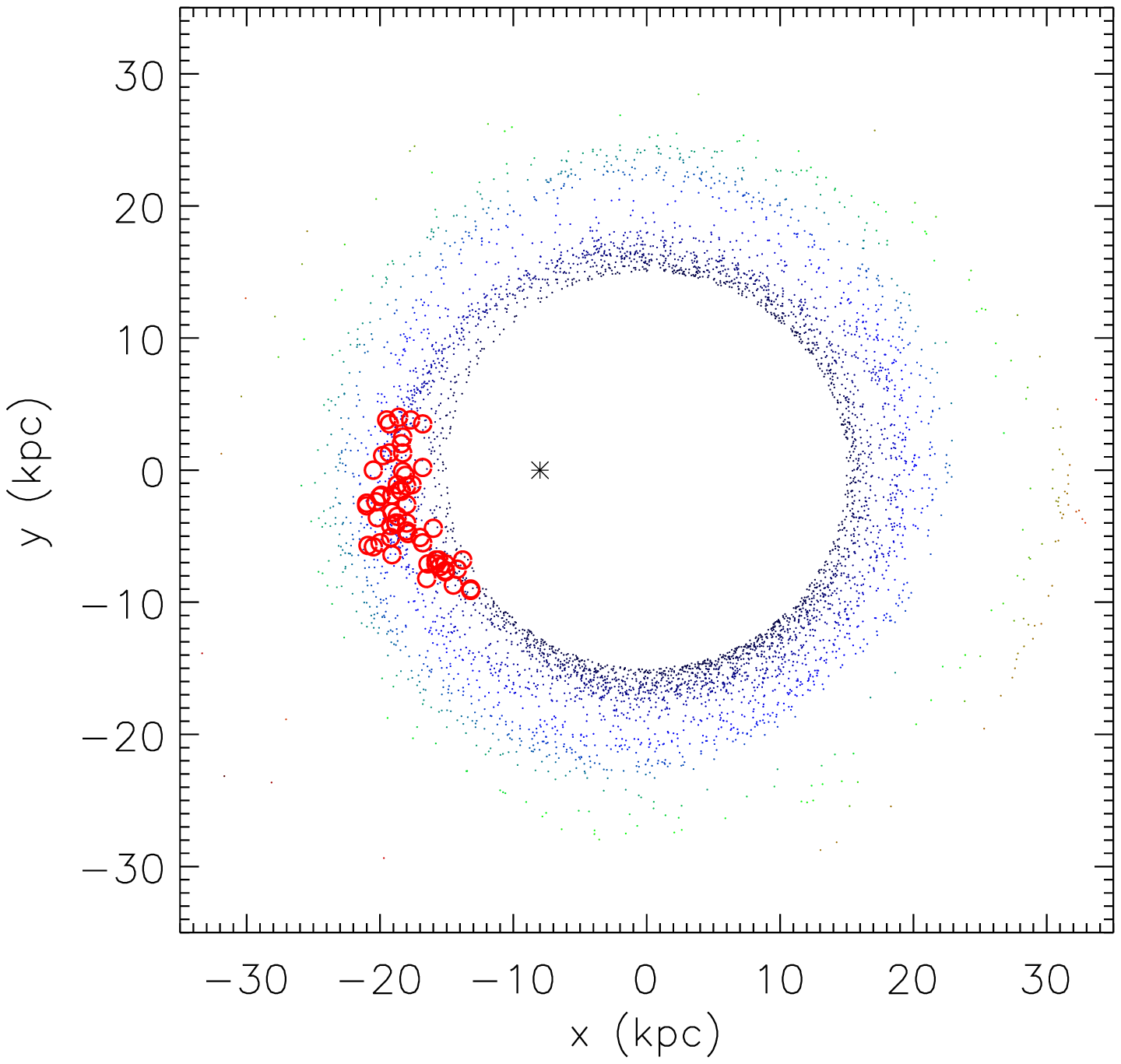}
   \includegraphics[height=6cm]{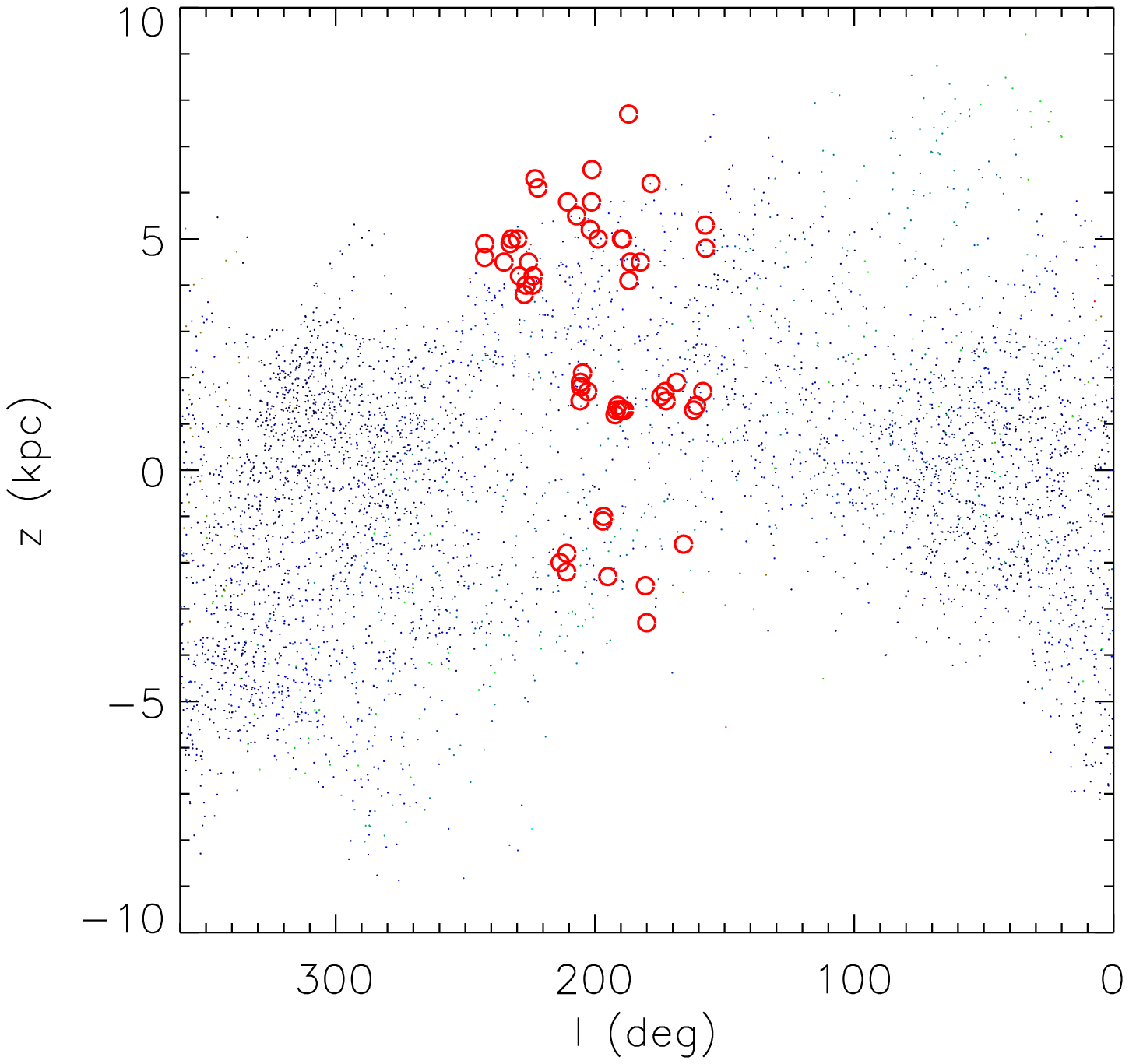} \\
    \includegraphics[height=6cm]{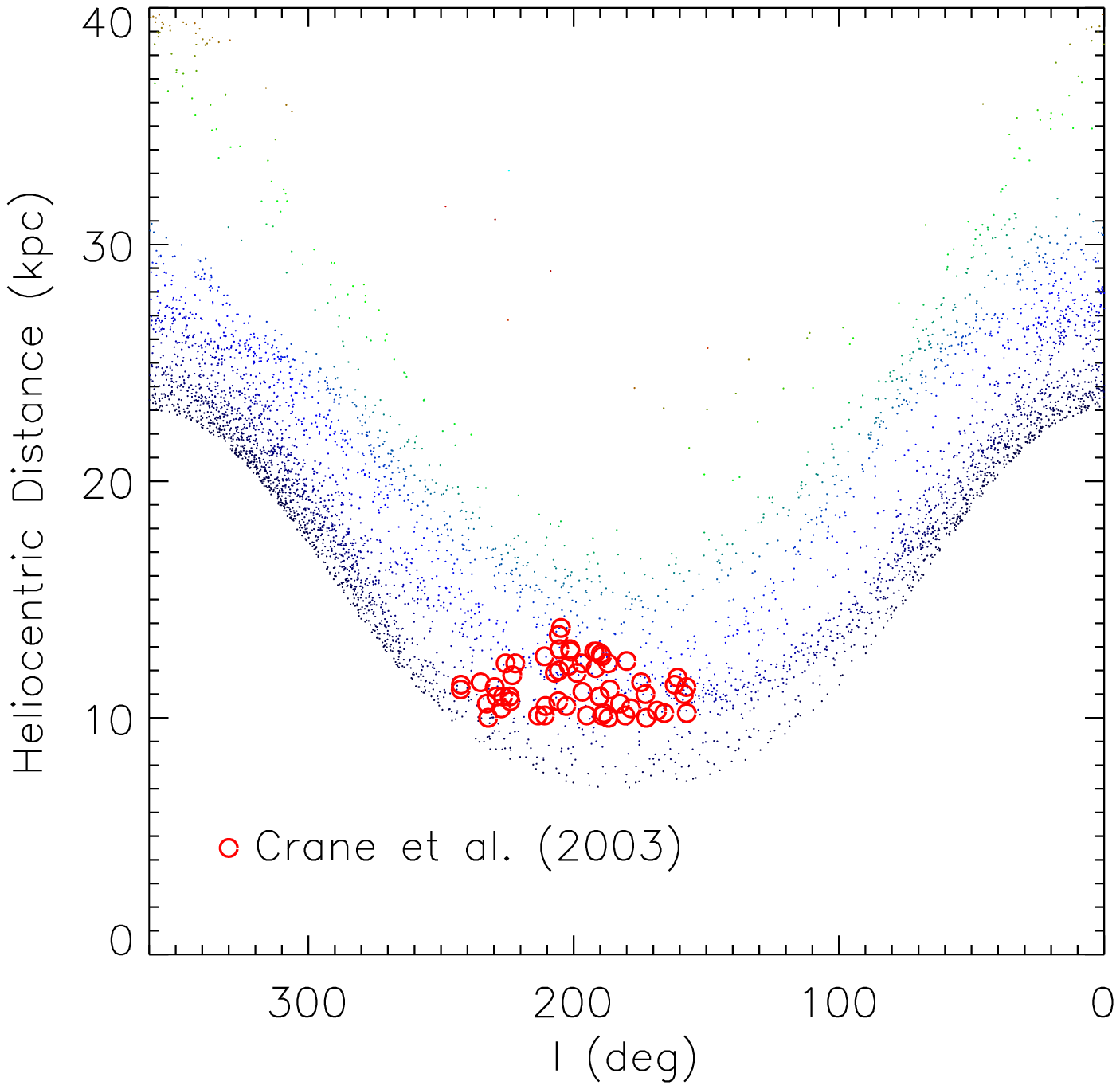}
     \includegraphics[height=6cm]{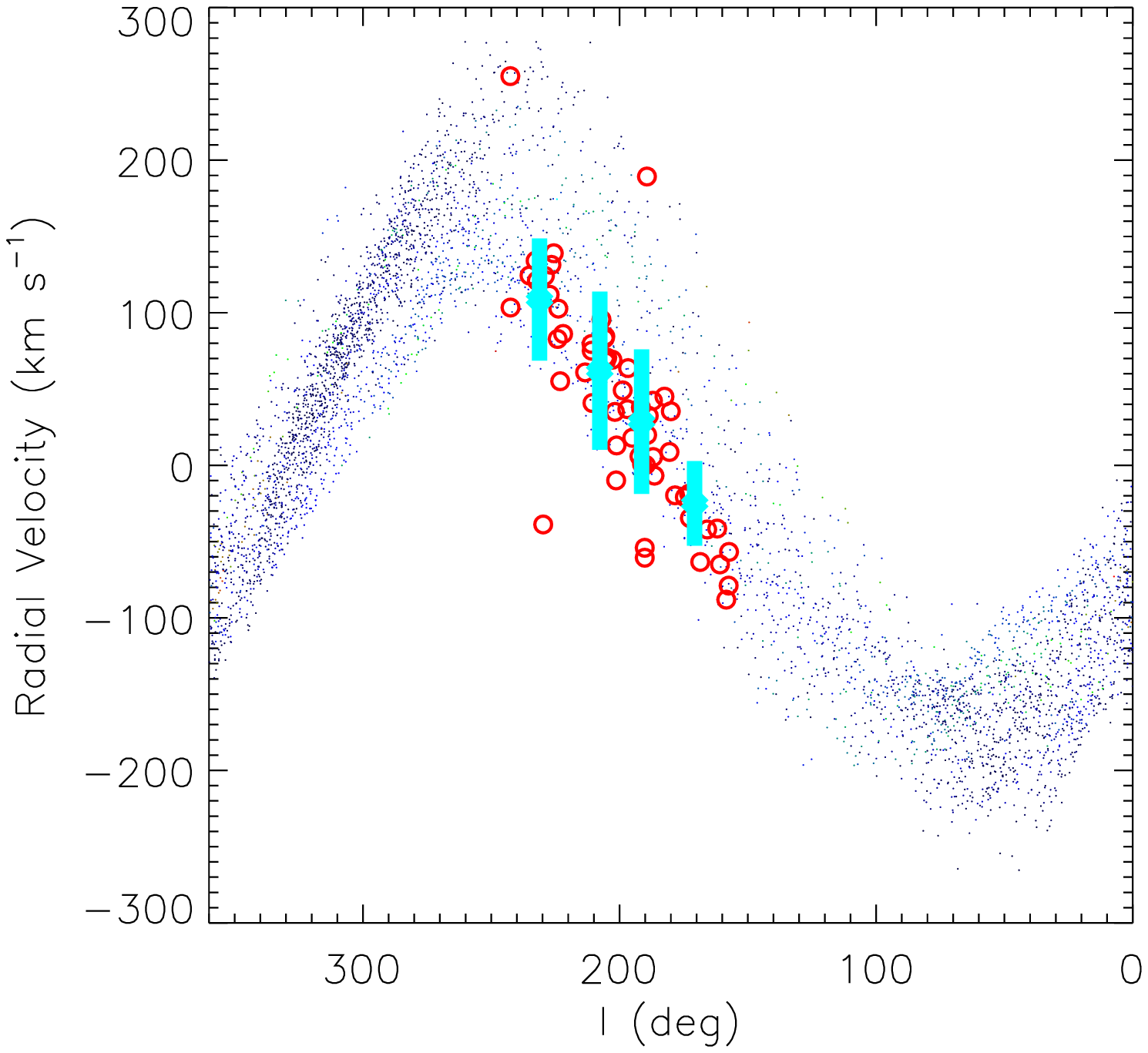}
\end{center}
\caption{Monoceros ring comparison. Circles are data of 
  M giant stars from the \citet{Crane_etal03} study of the Monoceros 
  stream in the direction of the Galactic anticenter. Points in each panel 
  correspond to stars in the outer regions ($R>15\kpc$) of the final 
  disk. These stars are color-coded according to their galacto-centric radius, 
  with dark blue/black at $R=15\kpc$, light-blue at $R\simeq 18\kpc$, 
  green at $R \simeq 22\kpc$, and orange/red at $R \gtrsim 25\kpc$.
  The {\it upper-left panel} shows a face-on view of the disk and the 
  asterisk marks the adopted position of the Sun at $(x,y,z) = (8, 0, 0)\kpc$.
  The {\it upper-right}, {\it lower-left}, and {\it lower-right} panels show vertical  
  distance of stars above the disk plane, $z$, heliocentric distance, and heliocentric
  radial velocity, respectively, as a function of Galactic longitude in degrees, $l$.   
  {\it Diamonds} in the lower-right panel correspond to the median line-of-sight velocities 
  in four bins in longitude $l = [240-210, 210-180, 180-150, 150-120]\degrees$
  for a specific stream-feature identified in the simulations. The associated bars
  reflect the line-of-sight velocity dispersion in each bin. Accretion
  histories of the kind expected in {\LCDM} models produce
  dynamically cold ring-like features around galactic disks that are
  quantitatively similar to the Monoceros ring in the MW.
\label{fig5}}
\end{figure*}
%

\subsection{Global Disk Morphology}
\label{sub:global_morpho}

Figure~\ref{fig3} depicts the evolution of the global structure of a thin stellar disk that
experiences a merging history of the type expected in {\LCDM}.
The left panel shows the initial configuration of disk stars assuming that the
sequence of satellite-disk interactions initiates at $z=1$, while the right and middle panels 
show the final configuration at $z=0$ {\it of the same stars} at two different surface
brightness thresholds. Stellar particles are color-coded by the projected 
{\it surface brightness}. In producing these images, we have 
assumed a stellar mass-to-light ratio of $M_*/L=3$ in the V-band, as would be appropriate
for an old stellar population with colors relevant for thick disks \citep{Dalcanton_Bernstein02}. 
This particular choice produces a peak edge-on central surface brightness of 
$\mu \sim 21$ mag arcsec$^{-2}$, similar to values seen in optical surveys of nearby, edge-on, 
undisturbed disk galaxies \citep{Yoachim_Dalcanton06}.
A different choice for $M_*/L$ would simply result in a dimming or brightening of these images
and subsequent figures accordingly.

Several interesting morphological signatures of satellite-disk interactions are clear from 
these images. First, a wealth of low-surface brightness features have developed both in the 
plane and above the plane of the final disk as a consequence of these disturbances and we 
discuss these in detail in the next sections. Particularly intriguing is the presence of a
conspicuous flare and non-axisymmetric structures such as extended outer rings
and bars. Their existence and robustness indicate that the axisymmetry
of the disk has been destroyed by the encounters with the infalling satellites
and is not restored at late times.

Moreover, as is evident in the edge-on images of the final disk in Figure~\ref{fig3}, 
a high-surface brightness {\it thin} disk component remains after the bombardment 
by CDM substructure. We discuss the implication of this finding in the general
context of disk survivability in a {\LCDM} universe in Section~\ref{sec:discussion}.
It is interesting to ask whether or not the bright in-plane structure
would be recognized as a thin disk component in a standard disk
decomposition analysis. We present such an analysis in Figure~\ref{fig4}.
This figure shows the initial and final surface brightness profiles, $\mu(z)$,
of the stellar disk as a function of distance, $z$, above the disk plane. For
this calculation, the stellar distribution is observed edge-on. 
Symbols correspond to different projected radii, $R$.

\begin{figure*}[t!]
\centerline{\epsfxsize=6.8in \epsffile{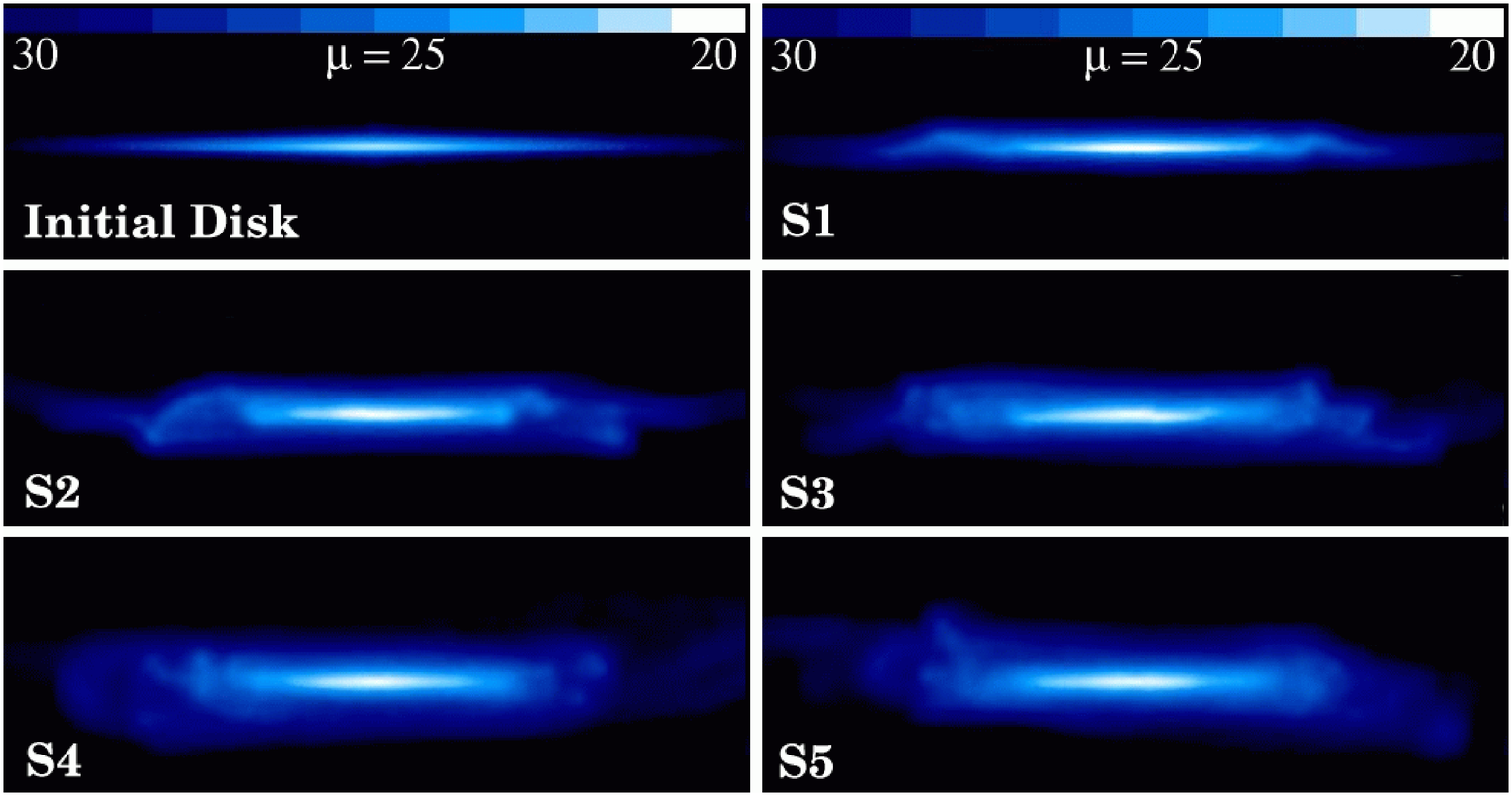}}
\caption{Effect of individual satellites on the global morphological 
evolution of the disk. Each panel shows edge-on surface brightness 
maps of disk stars to a surface brightness limit of $\mu = 30$ mag
arcsec$^{-2}$ and measures $18\kpc$ by $60\kpc$. Labels for individual
satellite passages from S1 to S5 are indicated in the 
{\it lower left-hand} corners of each image and, to aid comparison, the initial
disk is presented in the {\it upper left} panel. Images are constructed 
after the disk has relaxed from each encounter according to the definition in 
\S 2.5. The response to subhalo S6 can be seen in Figure~\ref{fig3}.
A high-surface brightness thin disk component is visible in all images.
The first satellite passage S1 generates a warp in the outer regions of the disk. 
The second encounter which involves the most massive subhalo S2 causes
substantial flaring above the disk plane. The remaining satellites perturb the
disk structure further even though their effect is 
differentially less significant compared to the initial encounters. In broad terms,
most of the morphological signatures have emerged as a result of the
interaction with the most massive subhalo.
\label{fig6}}
\end{figure*}

The left panel shows the analytic sech$^2$ surface brightness profile 
(eq.~[\ref{disk_density}]) for the initial, thin stellar disk with a scale height of 
$z_d = 0.4\kpc$. By construction, there is excellent agreement between the
analytic profiles and results obtained directly from the particle
distribution. As we discussed earlier, Figure~\ref{fig3} demonstrates that the
disk is substantially perturbed by the encounters with the infalling
satellites. This suggests that describing the later stages of 
the disk evolution requires more than the simple, single function that works
well for the initial disk. Indeed, the right panel of Figure~\ref{fig4} presents 
a two-component, ``thin-thick'' disk decomposition for the final disk.
The decomposition consists of a low-surface brightness, ``thick'' disk component with a 
sech$^2$ scale height of $z_{\rm thick} = 1.6\kpc$ and a ``thin'' disk
component with a sech$^2$ scale height of $z_{\rm thin}=0.6\kpc$ which is slightly larger
compared to that of the initial thin disk. 

This simple decomposition provides a reasonably good description of the
final disk down to fairly low surface brightnesses, $\mu \lesssim 27$ mag
arcsec$^{-2}$, and for $|R| \lesssim 9\kpc$. Interestingly, the edge-on thick
disk component is consistent with having a surface brightness profile that is
{\it almost} independent of projected radius for $R \lesssim 9\kpc$.  
At fainter magnitudes, a flare and other diffuse structures become important 
as discussed in \S~\ref{sub:morph}. Beyond $R \sim 9\kpc$, the adopted
decomposition fails to provide an adequate representation of the disk structure. 
At these radii the disk is significantly flared and its structure becomes
strongly irregular. The complexity of the disk structure at $R \gtrsim 9\kpc$
suggests that a more complicated functional form may be needed to describe
the final disk at larger radii.

We note that this decomposition does not constitute a formal fit but merely
provides an acceptable parametrization for the final disk structure. The
analysis presented above simply serves to illustrate that the final disk
cannot be represented by a {\it single} exponential or sech$^2$ 
function. Two components are clearly required, one of which is significantly thicker than 
the ``thin'' one. Deriving a formal fit for the disk vertical structure will not yield 
substantial additional information regarding the global morphological evolution of the disk, 
though it will be explored in the future.

Figure~\ref{fig4} shows a number of interesting results related to the surface 
brightness distributions of the final disk. First, these distributions become
uniformly broader extending at considerably larger vertical distances from the
disk plane compared to those of the initial thin disk. This 
change is caused by the encounters with the infalling subhalos and can be
directly associated with the strong flare seen in the final disk which will be 
discussed in \S~\ref{sub:flaring}. Second, the peak of the surface brightness profiles which 
corresponds to the light distribution in the disk plane ($z=0$) drops systematically at 
all projected radii compared to that of the initial disk. This is attributable to the fact 
that stars are efficiently heated above the disk plane by the merging subhalos.
As a result of this, the number of stars in the plane of the disk decreases with obvious 
consequences for the surface brightness distribution.

As the projected radius increases, the height above the
plane of the disk at which the thick disk component dominates the light 
profile becomes progressively smaller. This is because the surface brightness distribution of 
the thin disk declines exponentially with $R$, while both the scale height, and the 
central (projected) surface brightness of the edge-on thick disk vary much more slowly 
with radius. At small projected radii ($|R| \lesssim 6\kpc$) the thin disk component dominates for 
$z \lesssim 1\kpc$. However, this picture is quickly reversed.
For $|R| = 9\kpc$ ({\it triangles}) the thick disk dominates the light profile of the stellar
distribution for {\it all} heights above the disk plane. 
We note that within $9\kpc$, approximately $\sim 17 \%$ of the final stellar
mass is contained in the thick disk component.

\subsection {Morphological Signatures of Satellite-Disk Encounters}
\label{sub:morph}

The results in the previous section demonstrate that infalling CDM
substructures play a significant role in setting the global morphology of a
stellar disk. The face-on view of the final disk in 
Figure~\ref{fig3} ({\it middle} and {\it right panels}) reveals that satellite 
accretion can also excite strong bars which drive further evolution in the
inner disk regions. The primary disk galaxy model was constructed to be stable
against the formation of a bar. Thus, the observed bar growth should be
regarded as tidally induced by the infalling satellites. 
Interestingly, the edge-on view of the same panels shows the generation of a 
characteristic ``X'' shape in the bright central disk, a finding also reported 
in previous numerical studies of satellite-disk encounters 
\citep{Walker_etal96,Gauthier_etal06}. This noticeable feature is often linked
to secular evolution of galaxies driven by the presence of a bar when it buckles 
as a result of becoming unstable to bending modes
\citep[e.g.,][]{Combes_Sanders81}, and may be associated with the presence of 
peanut-shaped bulges observed in many galaxies \citep[e.g., ][]{Lutticke_etal00}.

Moreover, while the bright ``X''-shaped component is visible at high-surface
brightness, many other interaction-driven signatures appear as low-surface
brightness features. The deep view of the final edge-on 
disk in Figure~\ref{fig3} ({\it right panel}) shows a number of additional
filamentary structures at $\mu \gtrsim 26$ mag arcsec$^{-2}$, and other
complex configurations, that develop above the disk plane. 
These structures bear some resemblance to tidal 
streams, but are in fact disk stars that have been gravitationally excited by 
the subhalo impacts. The final face-on disk in the same panel is also
significantly more structured at faint surface brightness and large radius
compared to the initial disk and is quite reminiscent of the outer 
disk structure seen around M31 \citep{Ibata_etal05}. The same image also
reveals a pronounced ring-like feature at $\mu \sim 26$ mag arcsec$^{-2}$.
The ring is nearly in-plane and is qualitatively similar to the Monoceros
stream known to extend around the MW 
\citep[e.g.,][]{Newberg_etal02,Yanny_etal03,Ibata_etal03}. We emphasize
that this ring is a non-transient feature which survives for a considerable
amount of time after the satellite accretion events. Indeed, though this
structure is produced after the interaction with satellite S2, it remains
apparent at the final simulation output some $\sim 4$~Gyr after the final
impact with subhalo S6.

Though our simulations were neither designed to follow the evolution of nor 
to draw specific conclusions about the MW or any other particular system,
the reminiscence of the resultant rings in our simulations to that of the 
Monoceros ring in the Galaxy is suggestive of the possibility that the latter 
may have been generated via an excitation of an ancient disk's stars. In order to check 
the general validity of this scenario Figure~\ref{fig5} presents a more direct comparison between the 
ring-like structures generated via satellite-disk interactions
and the Monoceros ring feature towards the Galactic anticenter discovered 
by \citep{Newberg_etal02}. The circles in this figure correspond to a kinematic
study of M giant stars by \citet{Crane_etal03}, which followed up
a study by \citet{Rocha-Pinto_etal03} to identify M giants associated with
the ring. Our specific choice to compare to the \citet{Crane_etal03} 
sample is motivated by the fact that these data span the Monoceros stream uniformly 
over nearly $\sim 100\degrees$ in the sky with precise membership criteria and
good velocity determinations. For clarity, we have focused our analysis on the distribution 
of stars at galacto-centric radii larger
than $15\kpc$. These stars are color-coded according to their 
radial position from the center of the disk. Figure~\ref{fig5} shows vertical 
distance above the disk plane ($z$), heliocentric distance, and heliocentric radial 
velocity of the same stars as a function of Galactic longitude, $l$.

It is worth emphasizing that the stars in our simulated data are more finely
sampled compared to those of \citet{Crane_etal03}.  In order to facilitate the   
comparison, in the lower-right panel, we have more precisely  
identified a stream-feature along four bins in longitude $l = [240-210,  
210-180, 180-150, 150-120]\degrees$ and associated cuts in helio-centric distance 
$d_{\rm helio} = [14-12\kpc, 12.5-10\kpc, 12.5-10\kpc, 12.5-10\kpc]$.  
The total mass of this stream-feature is $\sim 10^{8} \Mo$ or $\sim 0.3\% $ 
of the disk mass, $M_{\rm disk}$. Although the total mass of the Monoceros
stream is still a matter of debate, the aforentioned value is consistent with
the mass estimates of \citet{Yanny_etal03}.

The four diamonds correspond to the median line-of-sight velocities and 
longitudes in each of these angular bins. The bars reflect the line-of-sight velocity 
dispersion in each bin, which is approximately $\sigma_{\rm los}\simeq 40\kms$ across
the stream. We emphasize that using the \citet{Crane_etal03} data across the
first three of these bins we measured a line-of-sight velocity dispersion of
$\sigma_{\rm los} \simeq 39 \kms$, which is in excellent agreement with 
the corresponding values associated with the simulated data. 
Note that \citet{Crane_etal03} estimate a line-of-sight
velocity dispersion of $\sigma_{\rm los}= 20 \pm 4\kms$ for the ring, 
which differs significantly
from the aforementioned value of $\sigma_{\rm los} \simeq 39 \kms$.
This is because the former was calculated using a subset of $53$ of the
$58$ stars in their sample, obtained after
a $2.5\sigma$ ($\pm 50 \kms$) rejection threshold was applied
about a third-order polynomial fit to the velocity trend. When we perform
a similar rejection threshold within each of our four angular
bins we find $\sigma_{\rm los} \sim 24.5\kms$ for our simulated ring
feature. Therefore, following the same analysis methods, our results are in
good agreement with the quoted velocity dispersion for Monoceros stream 
from \citet{Crane_etal03}.

The general agreement in the properties of rings (spatial distribution and
kinematics) generated in our simulations with those of the Monoceros ring
structure from the \citet{Crane_etal03} data is encouraging. This agreement 
is particularly noteworthy because our simulation program did not aim to 
reproduce such a feature. Our simulation campaign has explored only one realization of 
a galaxy-sized halo accretion history and we tuned no aspect of the computations to produce 
this agreement. This is a significant point that bears emphasis.  
One may constrast our model of a disk origin for the 
Monoceros stream to previous ones that have attempted to explain this structure via the 
accretion and disruption of a satellite galaxy on a nearly-circular, co-planar orbit
\citep{Penarrubia_etal05}. Neither our model nor the model of 
\citet[][see Figures 2 \& 4 in this reference]{Penarrubia_etal05} can be 
falsified by the extant data; however, our ring was produced 
as an unforeseen byproduct of our simulation campaign while the \citet{Penarrubia_etal05} 
ring was produced as part of a program aimed at tuning a satellite 
accretion event to yield a structure similar to the Monoceros ring.  
As such, a scenario of the kind we propose seems a viable mechanism to 
produce such ring-like structure from a stellar disk.

Given the existence of at least two qualitatively different scenarios
for the formation of structures like the Monoceros ring, it is crucial
to be able to test these formation scenarios and distinguish between them.
In principle, the metallicity of stars in the stream may constrain
the competing models. The stars in an accreted satellite can be
generally expected to have different metallicity from that of the surrounding disk
stars, while in our model the metallicity of the stream should be
comparable to the metallicity of the thick disk stars surrounding the stream.
\footnote{Note that the inner thick disk may have a different metallicity, so the comparison
should be performed with the outer thick disk, which likely formed
during the same heating events that produced the ring. Note also that if a similar ring-like 
feature is generated in the gas distribution, associated star formation and enrichment in the ring itself 
can enhance the metallicities of the stars in the stream.}
Unfortunately, there is no definitive conclusion regarding the metallicity
distribution of the Monoceros ring stars as the metallicity estimates span a 
wide range of values. Similar uncertainties exist for the characterization of the 
outer thick disk.

Indeed, \citet{Yanny_etal03} report mean metallicities for their
sample of Monoceros stream stars of $[{\rm Fe/H}] = -1.6\pm 0.3$, while 
\citet{Crane_etal03} measure a much higher metallicity of $[{\rm Fe/H}] = -0.4\pm 0.3$ 
for the M giants associated with the ring. The most recent measurements from the SDSS
\citep{Ivezic_etal08} indicate a metallicity for the Monoceros ring of $[{\rm Fe/H}] \approx -1$,
which is quite similar to their own determinations for the outer, high-latitude disk,
$[{\rm Fe/H}] \approx -0.9$. Other determinations of the thick disk metallicity distribution 
have shown a dependence on how one differentiates thick disk from halo stars, but the typical
range is $[{\rm Fe/H}] \approx -0.7$ to $-1.8$ \citep{Gilmore_etal95,Chiba_Beers00,Brown_etal08}.
The lower ring metallicities reported by \citet{Yanny_etal03} are
consistent with the lower range of thick-disk metallicities quoted \citep{Brown_etal08},
but would be problematic for our model if the outer thick disk is truly as metal
rich as the canonical thick-disk value, $[{\rm Fe/H}] \approx -0.7$ to $-1$ 
\citep{Gilmore_etal95,Ivezic_etal08}. Taking this canonical thick-disk value, the higher ring 
metallicities estimated by \citet{Crane_etal03} and \citet{Ivezic_etal08} would
seem to be more in line with our scenario. Overall, the use of metallicity measurements as a
robust constraint on our model would require refining the observational measurements for the 
Monoceros ring and securing the metallicity spread in the outer thick disk.  Moreover, precise 
predictions would require numerical simulations with star formation and metal enrichment.

Finally, we note that the so-called TriAnd clump feature \citep{Rocha-Pinto_etal03,Rocha-Pinto_etal04}
is sometimes discussed in association with the Monoceros ring.  It is worth emphasizing that
whether these structures are related remains a matter of debate.
TriAnd is spatially disjoint from the Monoceros ring, with
$d_{\rm helio} \gtrsim 20$ kpc and $l = 100 - 150\degrees$ 
compared to $d_{\rm helio} \sim 12$ kpc for the Monoceros stream. 
Some models suggest that a single disruption event can explain both features \citep[e.g.,][]{Penarrubia_etal05}. 
However, it is also possible that the TriAnd clump
is a disjoint disruption event, or possibly a secondary ring feature, of the type
colored in green in Figure~\ref{fig5}.  More exhaustive comparisons between
data and models will be required to definitively settle these issues.

Following up on the findings reported in Figs.~\ref{fig3} and \ref{fig4},
Figure~\ref{fig6} illustrates the effect of each satellite passage S1-S5 on 
the global morphological evolution of the disk. Images were created 
after the disk has relaxed from the encounter with each infalling subhalo 
according to the definition in \S 2.5. The high-surface brightness thin disk 
component discussed above is visible in all edge-on views of the disk. The 
first satellite passage S1 generates a pronounced warp in the disk beyond some
radius. The second interaction with the most massive subhalo S2 has a dramatic
effect on the disk structure causing substantial flaring above the
disk plane. Subsequent satellite accretion (S3-S5) does not
substantially perturb the disk further. The disk flare is visible at low
surface brightness limits below $26-27$ mag arcsec$^{-2}$ and becomes
particularly prominent after the second impact. This fact suggests
that only a few interactions with satellites with {\LCDM} motivated
internal properties and orbital parameters can generate
the bulk of morphological evolution of a stellar disk. We return to
this point below.

\begin{figure*}[t]
\centerline{\epsfxsize=7.2in \epsffile{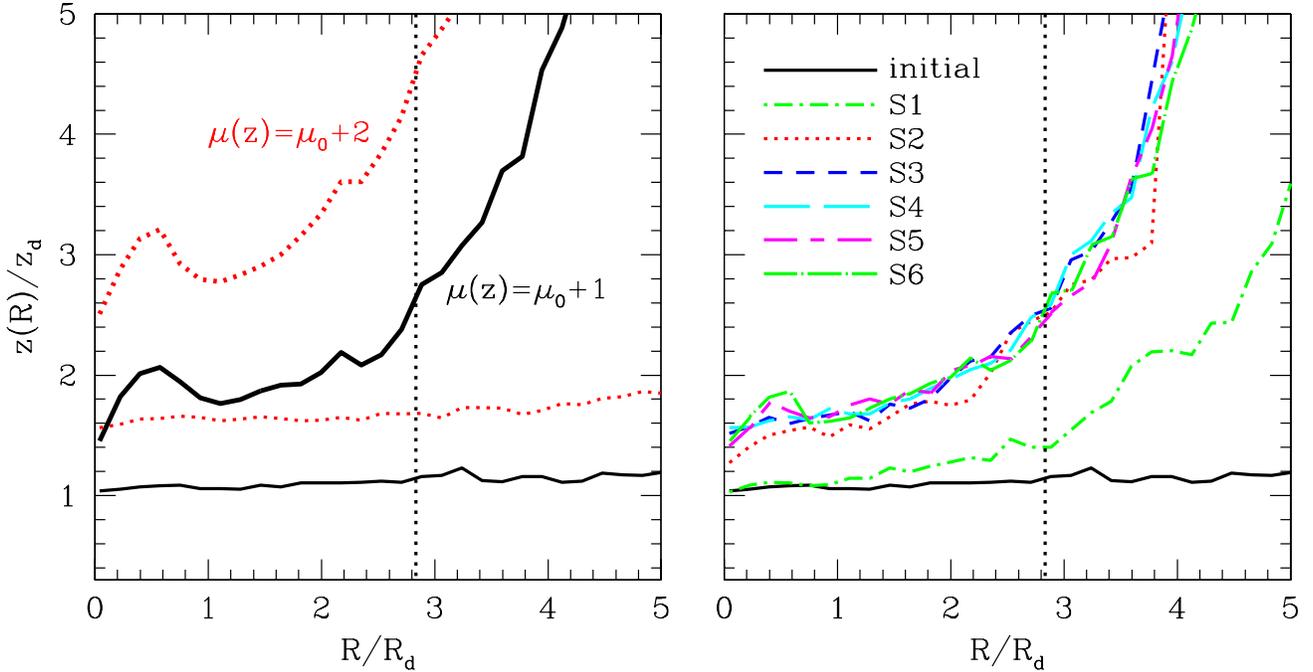}}
\caption{Disk flaring. Scale height profiles, $z(R)$, of the disk 
viewed edge-on as a function of projected radius in units of the disk radial scale 
length, $R_d$. The scale height is defined to be the vertical distance from the 
disk plane where the surface brightness drops by $1$ mag arcsec$^{-2}$
from its maximum along the $z=0$ disk plane, $\mu(z) = \mu_0+1$ where 
$\mu_0 \equiv \mu(z=0)$. Surface brightnesses are averaged about the $z=0$ 
and $R=0$ planes and all profiles are normalized to the initial disk scale height, $z_d$.
{\it Left:} Scale height profiles for the initial ({\it thin lines}) and final  
({\it thick lines}) disk. {\it Solid} lines show results obtained using the above 
definition of scale height, while {\it dotted} lines correspond to scale
heights defined to be the vertical position above $z=0$ where the edge-on
surface brightness falls off from the central value by $2$ mag arcsec$^{-2}$. 
The initial disk is constructed with a constant scale height 
explaining why the corresponding curves are flat. The vertical dotted line indicates 
the location of the solar radius, $R_\odot$. A conspicuous flare is evident in the final disk
beyond $\sim 2-3 R_d$ as a result of the gravitational interaction with CDM substructure.
{\it Right:} Evolution of the scale height profile of the disk. Various
lines correspond to different satellite passages from S1 to S6 and scale
heights are measured using the standard definition, $\mu(z) = \mu_0+1$.
The encounters with the first two satellites give rise to a distinct flare in the 
outskirts of the disk. The combined effect of the remaining satellite passages
(S3-S6) is much less dramatic indicating that the second, most massive
encounter is responsible for setting the global disk structure. The inner disk
appears much less susceptible to damage by the infalling subhalos owing to
its large binding energy and presence of sinks of the orbital energy of
satellites, such as the bulge component.
\label{fig7}}
\end{figure*}
%

\subsection{Disk Flaring}
\label{sub:flaring}

Among the most striking morphological features in the final edge-on
disk is the pronounced flare, which is particularly visible at low surface
brightness levels (Figures~\ref{fig3} and~\ref{fig6}). A more 
quantitative measure of the flare induced by the accretion events in our
simulations is given in Figure~\ref{fig7}. This figure shows scale height
profiles of the disk, $z(R)$, viewed edge-on. Profiles are normalized to the
initial disk scale height, $z_d$, and are plotted as a function of projected
radius in units of the disk radial scale length, $R_d$. 

The flaring of the disk can be formally described by the increase of the vertical scale height. 
For the purpose of our analysis, we define the scale height at any given radius 
to be the vertical distance above the disk midplane where the edge-on surface brightness 
falls off from its maximum along the $z=0$ disk plane by $\Delta \mu = 1 $ mag arcsec$^{-2}$,
$\mu(z) = \mu_0+1$ where $\mu_0 \equiv \mu(z=0)$. Though disk scale heights
must be formally derived by means
of fitting the particle distribution to an appropriate functional form (e.g.,
exponential or sech$^2$ law), scale heights as quantified here are
well-defined and require no assumptions for their interpretation.

The left panel of Figure~\ref{fig7} shows scale height profiles for the
initial and final disk. Results are presented using
both the definition of scale height introduced above, $\mu(z)=\mu_0+1$, and
that for which scale heights are defined 
to be the vertical distance above $z=0$ where the edge-on surface brightness
falls off from the central maximum value by $2$ mag arcsec$^{-2}$,
$\mu(z)=\mu_0+2$. The final disk shows considerable flaring beyond $\sim 2-3
R_d$ indicating that encounters with CDM substructure can be responsible for
producing notable flares in stellar disks. 
Remarkably, the scale height of the disk near the solar radius has increased
in excess of a factor of $2$ as a result of the interactions with satellites
S1-S6. We checked that the scale height of the same disk galaxy evolved in 
isolation for a timescale equal to that corresponding to the final disk 
($\sim 8$~Gyr) has grown by only about $10\%$ indicating the quality of 
the initial conditions and adequate resolution of the simulations. 

The right panel of Figure~\ref{fig7} shows the {\it evolution} of the scale height 
profile of the disk caused by individual satellite passages. After the
encounter with satellite S1, the inner disk ($R \lesssim 2 R_d$) appears
unaffected by the accretion event despite that fact that S1 is quite massive, 
with an initial mass of $\sim 30\%$ of the disk mass, and is characterized by a 
small pericenter ($r_{\rm peri} \sim 1.2 R_d)$. In contrast, the scale height 
of the outer disk ($R \gtrsim 3.5 R_d$) raises considerably giving rise to a distinct 
flare. The second event, which involves the most massive satellite S2 
($M_{\rm sub} \sim 0.6 M_{\rm disk})$, generates a stronger flare by greatly
increasing the scale height of the outer disk. Though the scale height has 
now grown throughout the entire disk, the inner disk regions still exhibit
substantial resilience to the encounter. Indeed, at $R = R_d$ the scale height
only raises by $\sim 50\%$ compared to approximately a factor of $4$ increase
at $R = 4 R_d$. 

This difference in the robustness of inner and outer disks is attributed to
the larger binding energy of the former and to the presence of a massive,
central bulge that acts as a sink of the orbital energy of the infalling satellites. 
The combined effect of the remaining satellite impacts (S3-S6) on the
structure of the disk is much less dramatic increasing 
the scale height only slightly compared to passage S2. This finding suggests
that subsequent accretion events by already thickened disks induce much
smaller changes in the disk scale height compared to the initial encounters. 
This effect was also noted previously by \citet{Quinn_etal93} using numerical
simulations in which halos were treated as rigid potentials. In broad terms, 
the dynamical and morphological evolution of a galactic disk subject to a 
cosmologically-motivated subhalo merger history are driven by the most massive
accretion event.

Given the fact that the self-gravity of the disk grows weaker as a function of
distance from the center, it is not unexpected that the scale height of the
final disk should increase with radius. In the thin disk approximation, we may consider 
the total vertical energy per unit area of the disk, $e_z = t_z + w_{dd} +
w_{ds}$, where $t_z$ is the disk vertical kinetic 
energy, and $w_{dd}$ and $w_{ds}$ are the disk potential energy densities
associated with the disk self-gravity and disk-spheroid gravity, respectively 
\citep{Toth_Ostriker92,Benson_etal04}. In this formulation it is implicitly
assumed that the spheroid includes the dark matter halo and any bulge 
component. For a thin disk of varying scale height, $z(R)$, we expect
\beqa
w_{dd} & =  & G \, \alpha \, \Sigma_d^2(R) \, z(R) \\
w_{ds} & = & G \, \beta \, \Sigma_d(R)  \, \bar{\rho}_{s}(R)  \, z^2(R).
\eeqa
Here, $\bar{\rho}_s(R)$ is the spheroid mass density
averaged over the disk scale height at radius $R$ and $\Sigma_d(R)$ is the projected 
disk surface density. The constants $\alpha$ and $\beta$ are geometrical 
factors of order unity that will depend on the profile shapes of the
disk and spheroid components, respectively.

Now consider the impact of a dark subhalo with the disk.
Some fraction of the orbital energy of the satellite
($\sim M_{\rm sub} V_{\rm sat}^2$, where $V_{\rm sat}$ is the orbital
velocity), will be delivered as vertical kinetic energy to the disk. The
disk will virialize on a dynamical timescale such that $t_z \simeq 0.5
w_{dd} + w_{ds}$ \citep{Toth_Ostriker92}. When a subhalo passes through the
disk, it frequently triggers global modes (e.g., spiral arms, warps, bars)
which can be very efficient in redistributing its orbital energy throughout 
the disk \citep{Sellwood_etal98}. Moreover, if the infalling satellites are
extended, like in the case of our simulations, most disk stars will also 
directly receive kinetic energy during the interaction. Both facts suggest
that the orbital energy of an accreting subhalo will not be deposited locally
at the point of impact as it was assumed by \citet{Toth_Ostriker92}, but rather
globally across the entire disk.

The simplest assumption for energy deposition is that it is roughly constant in
radius such that $e_z(R) \rightarrow e_z(R) + \Delta e_z$, with $\Delta e_z$ nearly
independent of $R$. Moreover, due to the fact that rotational energy dominates 
over random motions in the plane of the disk, the predominant heating will be
in the vertical direction. In this limit we can assume that the {\it
  projected} radial profiles of the disk and spheroid remain unchanged. The 
global change in vertical energy will demand that the scale height changes as:
\be
\Delta z(R) \propto \frac{\Delta e_z}{ \Sigma_d(R)[\alpha \Sigma_d(R) + 2 z(R)\beta \bar{\rho}_{s}(R) ]}.
\label{eq:hd}
\ee
In the limit where the disk surface density dominates the restoring
force, the disk scale height will increase as $\Delta z(R) \propto  \Sigma_d^{-2}(R)$. 
The presence of additional sinks of the orbital energy of satellites such as a
massive, central bulge or a concentrated dark matter halo will further act to
suppress the increase of scale height at small $R$. Indeed, as demonstrated 
in Paper II, a dense bulge is efficient in reducing the damaging effects of
infalling subhalos on stellar disks.

Note also that once the disk is thickened to drive $z(R)$ to be quite large, we
expect the second term in the denominator of eq.~[\ref{eq:hd}] to prevent any further
vertical puffing of the disk. Generally speaking, this suggests that the most massive of the past
encounters will set the scale height of the old thick disk, which is directly confirmed by 
the findings presented in the right panel of Figure~\ref{fig7}.

\begin{figure}[t]
\centerline{\epsfxsize=3.5in \epsffile{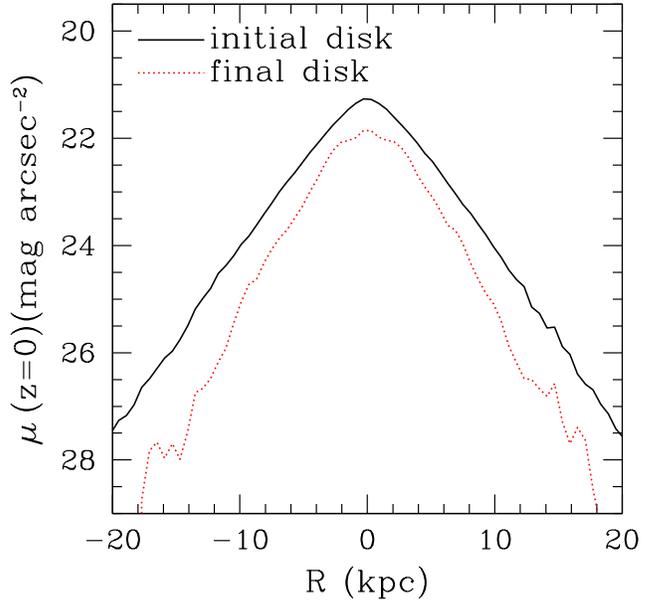}}
\caption{Edge-on surface brightness profiles of the stellar disk 
as a function of projected radius, $R$. The final disk ({\it dotted
  line}) exhibits a steepening in surface brightness distribution beyond 
$R \sim 3.5 R_d$ compared to the initial disk ({\it solid line}). 
The peak of the surface brightness profile of the final disk decreases by 
$\Delta \mu \approx 0.6$ mag arcsec$^{-2}$ from the initial value.
At large distances from the disk center a differential decrease in the surface 
brightness of $\Delta \mu \approx 2$ mag arcsec$^{-2}$ is established. Disk
flaring due to infalling CDM substructures is responsible for generating
breaks in the edge-on surface brightness profiles of galactic disks.
\label{fig8}}
\end{figure}

It is particularly interesting to investigate how the pronounced flare
we predict in our simulations would manifest in the surface brightness
distribution of an edge-on disk. The relevant analysis is presented in Figure~\ref{fig8}. 
This figure shows an edge-on ``in-plane'' ($z=0$) surface brightness 
profile for both initial and final disk as a function of projected radius, 
$R$. The calculation assumes again a stellar mass-to-light ratio of
$M_*/L=3$. A steepening of the exponential slope can be seen in the profile
of the final disk beyond a projected radius of $R \sim 10\kpc$.
The entire surface brightness distribution of the
final disk also decreases as a result of the encounters with the accreting
satellites. Specifically, the surface brightness in
the outskirts ($R \sim 6 R_d$) drops from $\mu \approx 27$ 
to $\mu \approx 29$ mag arcsec$^{-2}$. 

Furthermore, the central surface brightness decreases from the initial value of 
$\mu \sim 21.2$ mag arcsec$^{-2}$ by $\Delta \mu \approx 0.6$ mag arcsec$^{-2}$. This is 
consistent with the decrease in the peak of the surface brightness profiles at $R=0$ between 
initial and final disk seen in Figure~\ref{fig4}. As we discussed earlier,
this drop as well as the overall decrease of the brightness distribution can
be explained by the fact that infalling satellites heat stars efficiently
above the disk plane.

A low-surface brightness phenomenon seen in many disk galaxies is the tendency
for exponential disks to be truncated at faint magnitudes \citep{Kruit79,Kruit_Searle81,Dejong_etal07b}.
Such breaks were originally discovered in edge-on systems and appear to be
less readily observed in face-on galaxies because of the line-of-sight
projection and the associated lower surface brightness. 
It is important to emphasize that even though the behavior of the final
edge-on disk depicted in Figure~\ref{fig8} resembles observed disk
truncations, the steepening of the exponential slope disappears when the disk
is viewed face-on. This suggests that the 3D stellar distribution does not
actually truncate. Thus, the steepening of the surface brightness profile $\mu(z=0,R)$
reported here does not constitute a separate physical effect. 
On the contrary, it is a consequence of the flare. 

In the outskirts, the binding energy of the
disk is low and stars are efficiently heated above the disk plane by the infalling 
satellites. As a result, at large distances from the center, very little
material remains in the disk plane which gives rise to a distinct drop-off in
the surface brightness beyond $R \sim 3.5 R_d$. On the other hand, the larger
self-gravity of the inner, exponential disk and the presence of a massive 
bulge that acts as a sink of the satellites' orbital energy increase the
robustness of the inner disk to accretion events and prohibit the
development of such a break in the exponential slope at smaller radii.

%

\section{Discussion}
\label{sec:discussion}

In what follows we discuss the implications of our findings 
and caveats of the present study. 

\subsection{Implications}
\label{sec:implications}

We have demonstrated that encounters with CDM substructure generate a
wealth of morphological signatures in disk galaxies. These include
conspicuous flares, bars, low-surface brightness ring-like 
features in the outskirts of the disk, faint filamentary structures above the
disk plane, and a complex vertical morphology that resembles the commonly
adopted thin-thick disk profiles used in the analysis of disk galaxies. 
\citet{Font_etal01} and \citet{Gauthier_etal06} performed similar numerical 
investigations of the gravitational interaction between multi-component disk 
galaxies and infalling subhalos. Both of these investigations considered the 
population of surviving substructures at $z=0$ and both studies found
negligible effect on the global structure of the disk.

In particular, the aforementioned studies showed only mild 
evolution of the disk scale height as a function of radius due to the
satellite population. In contrast, the present study reports substantial disk 
flaring stemming from the gravitational interaction with infalling CDM
subhalos. As Figure~\ref{fig7} indicates, the flaring we find 
is associated with an increase of the disk scale height by more than a factor 
of $2$ near the solar radius. The primary reason for this discrepancy is that 
we have followed the formation {\it history} of the host halo since $z \sim
1$, and consequently we have considered impacts of satellites onto the disk 
that are significantly more massive that those included in the numerical
experiments of \citet{Font_etal01} and \citet{Gauthier_etal06}.

This difference is critical because massive subhalos on highly eccentric orbits
at early epochs become preferentially disrupted or suffer substantial mass
loss during their orbital evolution, and so they are
more likely to be absent from the $z=0$ population 
\citep[e.g.,][]{Zentner_Bullock03,Kravtsov_etal04,Gao_etal04}.
To capture these relatively short-lived objects, it is necessary to trace the
interaction history of the central halo with nearby, massive systems as we
have done here. Interestingly, orbital evolution models of the LMC and SMC
using 3D velocities constrained by recent proper motion
measurements suggest that these massive Galactic companions may currently be
on their first passage about the MW \citep{Besla_etal07}. If confirmed, this
is in line with the idea that massive objects would be present today only if they 
were recently-accreted. 

As discussed in the introduction, it is tempting to associate the faint 
structures resulting from the satellite-disk interactions (Figures~\ref{fig3} 
and~\ref{fig6}) with the recently-discovered complex disk
phenomena seen in and around the MW and M31, and in other nearby and distant disk
galaxies. Stellar counts in the MW from 2MASS and SDSS \citep{Momany_etal06,Bell_etal07}
identify stellar streams, many of which may be associated with recent merger
or fly-by activity from Local Group galaxies, or other perturbations in the galactic
disk. These studies also find a great deal more structure at radii
beyond $R_{\rm GC}\sim10$\,kpc \citep{Bell_etal07}, as well as a many-fold increase in the 
scale height of the disk, compared to that at the solar circle 
\citep{Momany_etal06}. 

Figure~\ref{fig6} also illustrates that encounters with
infalling subhalos can generate warps in the outer disk. Warp excitation by
tidal interactions with small companion galaxies constitutes an attractive 
mechanism. Indeed, it has recently been argued that the origin of the Galactic
warp can be attributed to the interaction of the MW disk with the LMC
\citep{Weinberg_Blitz06}. In more distant galaxies, extremely deep imaging has long revealed that 
multiple-component, thick-disk, and more extended diffuse structures are not limited to MW and M31
extending the applicability of our results beyond the Local Group 
\citep[e.g.,][]{Sackett_etal04,Lequeux_etal98,Abe_etal99,
Dalcanton_Bernstein02,Zibetti_Ferguson04,Dalcanton_etal05,Yoachim_Dalcanton06,Dejong_etal07a}.

The ring-like features shown in Figure~\ref{fig3} are
particularly intriguing in light of the Monoceros ring structure of the MW. 
In Figure~\ref{fig5}, we presented a comparison between 
the ring-like structures in our simulations and the Monoceros ring and 
demonstrated that there is a striking agreement in a number of properties
including the R and z spatial distributions, tight velocity widths and line of 
sight velocity dispersions. To this end, the present study suggests
that dynamically cold rings of this kind are generated naturally as a result
of impacts with halo substructure, which can excite perturbations in {\it disk stars}. This phenomenon
is different from the well-known ``shell'' structures that arise in stars that are {\it accreted}
from dynamically cold systems that interact with larger galaxies 
\citep[e.g.,][]{Hernquist_Quinn88,Helmi_etal03}. In our simulations, the ring
is caused by a perturbation within an initially cold disk. 

It is important to emphasize that our simulation set was not designed to produce 
a ring-like structure, nor any of the other morphological features observed in the remnant 
stellar disk. Our campaign was designed to mimic the general process of satellite infall in the 
prevailing cosmology and as such, we expect our results to be general consequences of this 
process as well. In comparison to models that attempt to explain the Monoceros stream and extended M31 disk structure
using minor merger events that directly deposit stars, our scenario requires no special orbital
configuration for the infalling satellites. In order to explain the origin of the aforementioned 
features, accreted-star models seem to require fairly circular, prograde, or
(at least) low-inclination orbits \citep[e.g.,][]{Helmi_etal03,Penarrubia_etal05}. 
In studies of {\it dissipationless} simulations, orbits of this type are quite rare for
subhalos that penetrate deep into the host halo and interact with disks 
\citep[e.g.,][]{Ghigna_etal98,Knebe_etal04,Zentner_etal05a,Zentner_etal05b,Benson05}. 
The subhalo orbital eccentricity distribution presented in the right panel
of Figure~\ref{fig1} confirms this conclusion. 

More circular orbits may be motivated by numerical experiments of satellite decay in
disk-dominated or flattened systems that have reported rapid circularization of the orbit
\citep[e.g.,][]{Quinn_Goodman86,Penarrubia_etal02,Meza_etal05}. 
However, because dynamical friction depends sensitively on the mass ratio
of the interacting systems, among other things, the mass of the satellite must
be a non-negligible fraction of the disk mass for such mechanism to be
viable, and in the very least subhalo orbits will not be driven to nearly circular {\it prior} to an encounter 
with the disk. The extended disk components and morphological features we report in the present 
study arise naturally in the context of a typical {\LCDM} accretion history. 

While ring features in the simulated disks capture many of the properties of the
Monoceros stream, there are a number of observational constraints that cannot be
directly addressed with our simulations and could potentially falsify our proposed 
model. A promising way to distinguish between the two different scenarios for the 
formation of the Monoceros ring is to establish that the stream contains stellar 
populations that are not expected to pre-exist in the outer disk. 
Interestingly, there have been indications that some globular clusters may be 
associated with the Monoceros stream \citep[e.g.,][]{Crane_etal03}. While this association
is still a matter of debate, the existence of such globular clusters 
would point to an external origin for the Monoceros structure.

In principle, the metallicity distribution of stars in the stream can also 
be used to constrain the competing models. Indeed, in our scenario the metallicity 
of the stream should be comparable to that of the thick disk stars surrounding 
the stream. If the outer thick disk is truly as metal
rich as the canonical thick-disk value, $[{\rm Fe/H}] \approx -0.7$ 
to $-1$ \citep{Gilmore_etal95,Ivezic_etal08}, the higher ring 
metallicities estimated by \citet{Crane_etal03} and \citet{Ivezic_etal08}
would be consistent with our scenario. Taking this canonical thick-disk value,
the lower ring metallicities reported by \citet{Yanny_etal03} would seem to be 
problematic for our model. Overall, in order to use metallicity as a reliable 
constraint on our model, we would have to refine observational measurements for the 
Monoceros ring, secure the metallicity spread in the outer thick disk, and 
perform numerical simulations with star formation and metal enrichment.

In the context of distinctive observational signatures produced in stellar
disks by accreting subhalos, an intriguing result is reported in
Figure~\ref{fig7}. This plot demonstrates
that the scale height of the simulated disk increases with radius as a result 
of the gravitational interactions with infalling satellites. 
Indeed, disk flaring appears to be a natural result of encounters with CDM
substructure and one of the most important observational consequences of the
present study. Many late-type galaxies (including the MW and M31) exhibit a great deal of
non-smooth ``structure'' in their stellar distributions, often with clear
flaring at large galacto-centric radii.  

Actual disk-flaring, such as that described in this paper, is seen in the MW
in both the stellar disk \citep{Lopez_etal02,Momany_etal06} and atomic hydrogen 
layer \citep[e.g.,][]{Merrifield92,Nakanishi_Sofue03}.
Flaring is also observed in external galaxies seen edge-on, in their stellar light
\citep[e.g.,][]{DeGrijs_Peletier97,Narayan_Jog02} (though more-rarely
in lower-mass galaxies; \citealt{DeGrijs_Peletier97,Seth_etal05}), as
well as in $H_{\scriptsize I}$ gas \citep[e.g.,][]{Brinks_Burton84,0lling96,Matthews_Wood03}.
Extending the search for disk flaring in external systems would provide a new
observational test for the {\LCDM} paradigm and its competitors on non-linear,
galactic scales, and thus offer unique opportunities to constrain cosmological models.

The structure and kinematics of galactic disks have long been recognized as a key 
constraint on structure formation models. Our results have intriguing implications for tests of 
the nature of dark matter through detailed observations of galactic
structure. Alternatives to the standard CDM paradigm such
as models with warm dark matter or decaying dark matter
\citep[e.g.,][]{Hogan_Dalcanton00,Avila-Reese_etal01,Cembranos_etal05,Kaplinghat05,Strigari_etal06},
or non-standard power spectra \citep[e.g.,][]{Kamionkowski_Liddle00} predict
substantially different accretion histories for 
galaxy-sized dark matter halos as well as different satellite abundances and structural
properties \citep{Zentner_Bullock03}. 

We have assessed the sensitivity of disk flaring to the density structure of
the infalling systems by adopting satellite models with radically different density 
profiles, but requiring that the total bound mass of each object be exactly the same. 
Specifically, we repeated the
first two satellite passages S1 and S2 modeling the infalling subhalos using a constant
density core ($\rho(r) \propto r^{-\gamma}$, where $\gamma=0$) and adopting
the same outer, $\beta$, and intermediate, $\alpha$, slopes of their CDM
counterparts. The size of the density core was chosen to be approximately
equal to $10\%$ of the tidal radius of each subhalo ($\gtrsim 2\kpc$). 

We find that cored satellites of equal bound mass produce much smaller amounts
of disk flaring at the solar radius and at larger radii, and smaller overall
disk morphological evolution compared to the standard CDM cosmological satellites 
whose density distribution is well described by a cusp slope of $\rho(r)
\propto r^{-1}$. This is not unexpected since the former are more susceptible to mass 
loss and will, on average, approach and penetrate the disk with smaller
amounts of orbital energy available to be converted into random motions of
disk stars. These findings suggest that differences in the properties of the
subhalo populations should be imprinted on the fine structure of disk galaxies. 

Such differences can be used to provide fundamental constraints on structure
formation theories by distinguishing between competing cosmological
models. Indeed, specifically-designed surveys such as SEGUE and RAVE and 
planned revolutionary instruments like SIM and GAIA will be able to quantify 
the structure and kinematic properties of the MW disk. Moreover, future
surveys like SEGUE-2, PanSTARRS, and LSST will provide ever more precise
star-by-star maps of the Galaxy and multi-object spectrographs on thirty-meter
class telescopes will enable star count maps for many galaxies in the local
volume, similar to the maps being made for M31 and M33. Confronting the
resulting data sets with theoretical predictions such as those of the present 
study will allow to test whether the detailed structure of galactic disks is 
as perturbed as predicted by the CDM paradigm.

We expect disk morphological evolution in galaxy-sized dark matter halos, $M_{\rm vir}
\sim 10^{12} \Mo$, to be driven by the few most massive satellite accretion 
events $M_{\rm sub} \sim  10^{10-11} \Mo \sim M_{\rm disk}$.
Indeed, mergers of this size dominate the mass buildup in galaxy-sized CDM
halos \citep[e.g.,][]{Zentner_Bullock03,Purcell_etal07}. In addition,
massive accretion events of this kind should 
have been relatively common since $z \sim 1$. We have showed this
directly using our four well-resolved cosmological $N$-body halos in
\S~2.2. This result is corroborated by \citet{Stewart_etal07}
who used thousands of halos extracted from a {\LCDM}
$N$-body simulation ($\sigma_8 = 0.9$) to show that $\sim 80 \%$ ($\sim 70
\%$) of MW-sized objects should have accreted at least one subhalo with mass
comparable to that of the disk in our controlled simulations, 
$M_{\rm sub} > 0.02 M_{\rm vir}$ ($> 0.05 M_{\rm vir}$) 
since $z\sim 1$. Moreover, $\sim 60 \%$ should have accreted a system
with more than twice that mass over the same period ($> 0.1 M_{\rm vir}$). 
Of course, the importance of these accretion events for 
both driving and altering disk morphologies will depend on the orbital
evolution of the satellites once accreted into the host halo virial
radius. However, given that dynamical friction will drive massive subhalos 
towards the central regions of the host halo within a few Gyr, these results
serve as general motivation that significant central encounters should be fairly 
commonplace in the history of galaxy-sized halos.

The above discussion is particularly relevant in the context of the survival
of thin, stellar disks to hierarchical satellite
accretion. Figures~\ref{fig3}, ~\ref{fig4} and ~\ref{fig6} illustrate that while
infalling satellites substantially perturb stellar disks affecting their
overall morphology, a significant thin disk component can survive bombardment
by CDM substructure that is an appreciable fraction of the disk mass.
Though most MW-sized galaxies are expected to have accreted an object at least
as massive as the disk itself since $z\sim 1$ (Figure~\ref{fig1}) we have
explicitly ignored these violent interactions in the present study in an attempt to be 
conservative. This is because our main goal was to examine the formation of
distinct morphological features about a well-preserved disk galaxy similar to
the MW and M31. An important future study will be to model the impact of a
satellite as massive as (or more massive than) the disk itself in order to test the 
robustness of disks in this general framework. We intend to examine this
issue fully in a forthcoming work. The results presented here provide some
encouragement that such an experiment may not prove ruinous to the survival of 
thin disks in a {\LCDM} universe.

Of specific concern for hierarchical models of structure formation is the prevalence 
of {\it old} ($\sim 10$~Gyr) stars in the thin disk \citep[e.g.,][]{Wyse01}.
\citet{Quillen_Garnet01} showed that the age-velocity dispersion
relation in the solar neighborhood for disk stars was relatively flat 
between $\sim 2$ and $\sim 10$~Gyr, with a distinct jump in the velocity dispersion for
older stars, as perhaps would be expected if thick disk formation were
driven by a single ancient event. However, utilizing a sample $\sim 75$ times
as large, \citet{Nordstrom_etal04} found a continuous increase 
of stellar velocity dispersion with age, and argued that ongoing heating could 
provide an explanation. \citet{Seabroke_Gilmore07} used the same data set and
similarly concluded that vertical disk heating does not saturate at early 
times. Regardless, the age distribution  of stars in the MW suggests that a
significant fraction of the thin disk ($\sigma_z \lesssim 20 \kms$)
was in place by $z \sim 1$.

It is tempting to argue that the existence of an old, thin stellar disk 
would imply an absence of satellite impacts over a period that corresponds 
to at least the age of the oldest stars in the thin disk. If an impact occurs 
and destroys the disk, then a thin disk component must be 
generated later by secondary processes such as gas inflow. 
This would suggest that thick disks originating during 
damaging encounters with substructures must be invariably older than the 
oldest stars in the thin disk, making thin and 
thick disks distinct in age as well as scale height. While this is true for
the MW and other external galaxies, our 
results are consistent with a picture in which a thick disk arises naturally 
in the context of a typical {\LCDM} accretion history while at the same time 
a substantial thin disk component survives the violent gravitational
encounters with CDM substructure. Our results thus indicate that the absence
of thick disks in significant fraction of disk galaxies, rather than
ubiquitous thick disks, would represent a challenge for the CDM model. 

The findings of the present study have implications for the 
formation of thick disks which constitute a ubiquitous component of 
disk galaxies \citep[e.g.,][]{Dalcanton_Bernstein02,Yoachim_Dalcanton06}. 
Several formation mechanisms have been proposed to explain their origin and
properties with a growing body of theoretical and observational 
evidence favoring a merger-driven scenario. There are two qualitatitavely 
different mechanisms that are associated with this formation scenario. In the first,
thick disk stars form initially in the thin disk and then are dynamically
heated to large scale heights by encounters with orbiting satellites 
\citep[e.g.,][]{Quinn_etal93,Walker_etal96,Robin_etal96,Velazquez_White99,Chen_etal01}. 
In the second, thick disk stars form in external galaxies and subsequently are 
deposited by accretion events at large scale heights 
\citep[e.g.,][]{Statler88,Abadi_etal03,Yoachim_Dalcanton05}. 

The results presented in this paper suggest that at least part of a 
galaxy's thick disk component may plausibly originate from the gravitational 
interaction between an existing thin disk and infalling satellites with 
mass functions, density structures, and orbital distributions
of the kind expected in the {\LCDM} paradigm of structure formation. 
CDM substructure increases the scale height of stellar disks 
(Figures~\ref{fig4} and~\ref{fig7}) and should be regarded as being at least 
partially responsible for the origin of thick disks. Indeed, this conclusion is 
supported by observational studies of the vertical distribution of stars in
the MW \citep{Chen_etal01} and star count data from a number of Galactic
sample fields \citep{Robin_etal96}. Photometric constraints in combination
with kinematic data from these investigations favor the mechanism in which the
thick disk of the MW formed through the heating
of a preexisting thin disk, with the heating mechanism being the merging of 
a satellite galaxy. Additional supporting proof comes from recent HST 
studies of resolved stellar populations in nearby edge-on disks
finding evidence for continuous vertical heating \citep{Seth_etal05} and chemical
abundance analysis of F and G dwarf stars in the Galactic thin and thick disks
\citep{Bensby_etal05}.

In contrast to the previous investigations, other ``thin-thick'' disk kinematic studies
favor models in which the thick disk forms from direct accretion of stars from 
infalling satellites. For example, \citet{Yoachim_Dalcanton05} presented 
GMOS kinematic measurements of edge-on galaxies FGC 1415 and FGC 227 and 
found that both thick disks were rotating much slower compared to 
the thin ones. In the case of FGC 227, the thick disk even showed weak signs of 
counterrotation. As argued by these authors, the detection of very
slowly rotating or counterrotating thick disks support an accretion origin 
for thick disk stars. While it is sufficiently unlikely that 
the thin disk forms from subsequent accretion of gas with angular 
momentum opposite to that of the original disk, disks heated by satellites 
do exhibit a vertical gradient in rotational velocity \citep{Hayashi_Chiba06}
which is consistent with what is inferred for the thick disk of the 
MW \citep{Allende_etal06}. Interestingly, analysis of the mean azimuthal
velocity of disk stars at the solar radius in our simulations also reveals a
vertical gradient in rotational velocity of $\sim -20\kms \kpc^{-1}$ between
$1$ and $3$~kpc from the disk plane, which is in good agreement with the
results of \citep{Hayashi_Chiba06} and \citep{Allende_etal06}.

In addition to slower rotation, MW thick disk stars exhibit larger
velocity dispersions compared to stars in the thin disk
\citep[e.g.,][]{Chiba_Beers00}. This is relevant as our simulations do reveal
that the stellar disk becomes substantially heated in all
three directions by the encounters with CDM substructure. Indeed, as we show 
in Paper II, the disk velocity ellipsoid at the solar radius, 
$R_\odot=8\kpc$, increases from $(\sigma_R, \sigma_\phi, \sigma_z) = (31, 24, 17)\kms$ to 
$(\sigma_R,\sigma_\phi,\sigma_z) \simeq (61, 49, 31)\kms$. For reference, the velocity ellipsoid 
of the thick disk of the MW is estimated to be $\sim (46, 50, 35)\kms$ by \citet{Chiba_Beers00}
and $\sim (63,39,39)\kms$ by \citet{Soubiran_etal03}. However, it is important
not to over-interpret this comparison because the final dispersions are calculated 
considering all stars instead of using only those that belong to the thick
disk component according to the decomposition analysis of \S~3.1.

Chemically, thick disk stars in the MW are more metal-poor than stars in the thin disk 
\citep[e.g.,][]{Chiba_Beers00}, are significantly enhanced in 
$\alpha$-elements compared to thin disk stars of similar iron 
abundances which suggests an extended star formation history
\citep[e.g.,][]{Bensby_etal05}, and exhibit no vertical metallicity 
gradient \citep{Bensby_etal05,Allende_etal06}. Most of the above 
properties have been also confirmed in external systems 
\citep[e.g.,][]{Seth_etal05}. Obviously, our dissipationless 
simulations can neither verify nor disprove any of the aforementioned 
trends. However, relevant in this context are the collisionless simulations 
of \citet{Hayashi_Chiba06} reporting that satellite-disk interactions
constitute an efficient mixing mechanism capable of erasing any metallity
gradient present in the original thin disk.

In summary, there is no definitive observational or theoretical evidence 
to conclusively rule out either of the merger-driven scenarios for the 
origin of thick disks. To this end, the existence of very slowly rotating or 
counterrotating thick disks in a significant fraction of disk 
galaxies would be problematic for the ``vertical heating'' model.
Most likely, both mechanisms do operate at a different degree
in forming the thick disk of a galaxy. This is further corroborated by the
results of \citet{Abadi_etal03} reporting that only $\sim 60\% $ of the thick
disk in their galaxy consisted of the tidal debris of disrupted systems.
Our study demonstrates that thick disks with realistic properties can result 
from the dynamical heating of pre-existing thin disks by CDM substructure. 
More detailed modeling of the thick disk properties would require the inclusion
of gasdynamics, star formation, and metal enrichment.

\subsection{Caveats}
\label{sec:caveats}

Certainly our approach is not without drawbacks. The most evident one is that the present study
used only four {\it dissipationless} galaxy-sized dark matter halos in order to derive
``typical'' orbits and merger histories, and the controlled simulations of
satellite-disk interactions were based on the accretion history of just one of
these host halos. It will be important to explore a range of accretion histories and orbital
distributions of infalling objects using a larger sample of halos from
cosmological $N$-body simulations. Moreover, since the orbital evolution of
substructure once accreted into host halos will be influenced by baryonic 
processes, a self-consistent, hydrodynamical simulation of disk galaxy
formation would be required to fully refine the conclusions presented here.
Unfortunately, hydrodynamical simulations that produce statistical samples of
realistic disk galaxies at $z=0$ do not yet exist.

Despite the aforementioned limitations, several facts suggest that our study has 
captured to at least a certain extent the amount of global morphological
transformation that CDM satellites can cause to stellar disks.
First, all four of the host halos we explored showed similar numbers of 
substantial central mergers, and their accretion histories are typical of
systems in this mass range \citep[e.g.,][]{Wechsler_etal02}. Second, we used a 
{\it conservative} subset of one of the merger histories to seed the controlled
satellite-disk encounter simulations, and neglected interactions with extremely massive 
subhalos that could prove ruinous to disk survival. If anything, the morphological
response of most disks should be more pronounced compared to that we find
here. Third, our numerical experiments took into account the effects of a realistic baryonic
component on the orbital evolution of the infalling satellites by including a
live central galaxy. Finally, most of the observational signatures on stellar disks
(Figures~\ref{fig3},~\ref{fig4},~\ref{fig6},~\ref{fig7}, and~\ref{fig8})
were generated by the single most massive encounter
($M_{\rm sub} \sim 0.6 M_{\rm disk}$, $r_{\rm tid} \sim 20 \kpc$, and 
$r_{\rm peri} \lesssim 10\kpc$). We identify accretion events of this kind in all four
of the host halo histories studied and, as discussed above, it is reasonable
to expect that most MW-sized halos should have accreted numerous such
systems since $z\sim 1$.

A second caveat is that we have neglected stellar components in the infalling
satellites modeling the latter as pure dark matter subhalos. Thus, the present
study is inadequate to address the contribution of satellite stars to the
formation of the thick disk and elucidate their ability in driving existing or 
creating their own morphological signatures. Notwithstanding this limitation,
the present work does demonstrate that dark matter subhalos can be capable of 
generating a wealth of morphological features in the simulated disks that
resemble those seen in real galaxies. A third drawback is related to the fact
that our models track only stars that were in place in a disk at $z\sim 1$,
and therefore apply most directly to {\it old} disk stars. We have also
adopted a primary disk galaxy model motivated by the present-day structure of 
the MW. A more complete investigation would have to include the ongoing
formation of the disk. 

Furthermore, we have only addressed the gravitational interaction of 
disk galaxies and dark matter substructures in the collisionless regime, a
choice with obvious limitations. Spiral galaxies contain atomic and molecular
gas which can absorb and subsequently radiate away 
part of the orbital energy deposited by the sinking satellites. Owing to this 
property, the heated gas component will eventually resettle to form a new thin 
disk consisting of a younger stellar population 
compared to that of the thick disk. Modeling the evolution of a dissipative component 
in the primary galaxy would be required to determine the extent to which the
presence of gas can alter the dynamical effects of substructure on stellar
disks. Given that the gas fraction in massive disk galaxies at $z\sim 0$ is 
typically low ($\sim 10 \%$ of the stellar disk mass; \citealt{Read_Trentham05}), 
the role of a dissipative component in stabilizing the disks against the
violent gravitational encounters with satellites may not be important, but
this will depend sensitively on whether the gas
content was higher in the past. Indeed, high gas-fractions at early times may
be required to explain the formation of disk galaxies in a hierarchical
universe \citep{Robertson_etal06}.

Finally, the present-day structure of galactic disks originates 
from a complex interplay of effects and a full explanation requires detailed knowledge 
of their star formation history and chemical evolution, amongst others. 
In fact, the thin and thick disk stars of the MW represent distinct populations not only
kinematically but also chemically \citep[e.g.,][]{Wyse_Gilmore95,Bensby_etal05}.
Adding star formation as a further ingredient will provide the opportunity to
track ongoing star formation in an ever re-forming thin disk and refine the
conclusions presented here. All of these effects will be discussed in forthcoming works.

\section{Summary}
\label{sec:summary}

Using high-resolution, fully self-consistent dissipationless $N$-body 
simulations we have examined the cumulative effect of substructure 
impacts onto thin disk galaxies in the context of the {\LCDM} paradigm of 
structure formation. The main goals of the present study have been 
to (1) demonstrate that interactions with CDM substructure {\it do} 
lead to morphological signatures and substantial changes in the structure 
of galactic disks and (2) study generic features of the evolution of 
disk galaxies subject to a {\LCDM}-motivated satellite accretion history.

Our simulation campaign utilizes cosmological 
simulations of the formation of galaxy-sized dark matter halos to derive the 
properties of subhalo populations and controlled numerical experiments of 
consecutive satellite encounters with an initially-thin disk 
galaxy. The present study extends and expands upon past numerical 
investigations into the interaction between disk galaxies and merging satellites in 
at least three major respects. 

First, we incorporate a model in which the infalling satellite populations are
truly representative of those accreted and possibly destroyed in the past,
instead of the $z=0$ surviving substructure present in a CDM halo. Second, 
the primary disk galaxy models are fully self-consistent and flexible enough to permit 
detailed modeling of actual galaxies. Third, individual satellites are
initialized according to the internal structure of subhalos culled from
high-resolution cosmological simulations. All of these elements inherent in
the modeling ensure that the numerical experiments of satellite-disk
encounters were performed with a high degree of realism.

In contrast to what can be inferred from statistics of the present-day
substructure populations in galaxy-sized CDM halos, encounters between massive 
subhalos ($ M_{\rm sub} \gtrsim 0.2 M_{\rm disk}$) and disks should be
common occurrences in standard {\LCDM} since $z \sim 1$. Most of the
difference is caused by the fact that satellites that pass closest to the halo 
center at early times are precisely those that are most likely to suffer significant
mass loss or become tidally disrupted. As subhalos on highly eccentric orbits
lose mass, the fraction of massive satellites with small orbital pericenters
declines with redshift so that few remain by $z=0$. The specific merger history we studied
involved $6$ accretion events of satellites with masses, orbital pericenters, and tidal radii of 
$7.4 \times 10^{9} \lesssim M_{\rm sub}/M_{\odot} \lesssim 2 \times 10^{10}$,
$r_{\rm peri} \lesssim 20\kpc$, and $r_{\rm tid} \gtrsim 20 \kpc$,
respectively, over a $\sim 8$~Gyr period. The morphological impact of these
events on an initially-thin ($z_d = 0.4\kpc$) disk galaxy ($M_{\rm disk} \approx 3.5 \times
10^{10} \Mo$) included the following
signatures, many of which resemble features seen in the MW and M31, and in 
other disk galaxies.

\begin{itemize}

\item The development of a ``thick'' disk component and the survival of a 
significant ``thin'' disk component, as characterized by standard
``thin-thick'' disk decomposition analysis. 

\item The growth of a strong bar.

\item The formation of a pronounced flare at intermediate and large radii, 
particularly visible at low surface brightness levels.

\item The production of long-lived, low-surface brightness, dynamically
cold ring-like features in the outskirts of the disk similar
to observed coherent stellar structures, such as the Monoceros 
ring in the MW.

\item The generation of faint filamentary structures that develop above the 
disk plane and (spuriously) resemble tidal streams in configuration space.

\end{itemize}

Finally, the results presented in this study of a ``typical" halo merger
history highlight the potential difficulty in 
explaining the fraction of observed {\it featureless} disk galaxies (e.g., those with extremely
thin discs, non-flared and/or non-warped and/or non-barred disk galaxies etc) in the context
of the CDM model of structure formation. Our findings also raise questions
regarding whether the existence of pure exponential discs or ``antitruncated'' disks which exhibit
an excess of surface brightness at very large radii \citep[e.g.,][]{Pohlen_etal07} 
can be explained in a model where mergers are as common as predicted in {\LCDM}.

A relevant intriguing issue concerns the survivability of bulgeless disk
galaxies.  To this end, we have performed numerical
experiments of interactions between satellites and bulgeless disk galaxies and found
that the latter experience substantially more damage compared to their
counterparts with bulges (see also \citealt{Velazquez_White99} for a similar
conclusion). In this context, the existence of a large number of very thin 
bulgeless disk galaxies would be difficult to reconcile with {\LCDM}. 
Interestingly, all systems in the sample of bulgeless galaxies of
\citet{Yoachim_Dalcanton06} have pronounced thick disks and there are no signs of 
companion systems in the vicinity of the prototype thin bulgeless disk 
galaxy M33. Directly relevant to this issue are recent, complementary studies
of edge-on disk galaxies using the {\it Sloan Digital Sky Survey} database
which revealed a significant fraction of ``super-thin'' bulgeless disks with 
much larger axial ratios compared to typical disk galaxies
\citep{Kautsch_etal06}. On the other hand, most MW-sized halos are expected to 
have accreted at least one subhalo as massive as the disk in our controlled 
simulations since $z\sim 1$ (see Figure~\ref{fig1} and \citealt{Stewart_etal07}).

The present work tackles the focused problem of identifying generic features in 
disk galaxies that arise in {\it typical} {\LCDM} merger histories.
To this end, we have demonstrated that gravitational interactions between existing thin
galactic disks and CDM substructure should be at least partially responsible 
for the formation of thick disks, disk flares, bars, low-surface brightness 
ring-like configurations and faint filamentary structures around disk galaxies. 
The fact that many of these features appear ubiquitous in real
galaxies is encouraging for the {\LCDM} paradigm of structure formation.
For example, bars are present in about $70\%$ of disk galaxies \citep[e.g.,][]
{Knapen99,Eskridge_etal00} and the occurence of thick disks is even
more frequent \citep[e.g.,][]{Dalcanton_Bernstein02,Yoachim_Dalcanton06}.
Detailed comparisons with the observed populations of disk galaxies would 
require an extensive series of numerical experiments to fully sample the 
statistical variation in halo accretion histories 
predicted in {\LCDM} as well as an equally careful statistical accounting of 
the fraction of galactic disks in the Universe that are indeed featureless. \\

\acknowledgments

The authors are grateful to Andrew Benson, Jeffrey Crane, Annette Ferguson, Andreea Font, 
Zeljko Ivezi{\'c}, Kathryn Johnston, Lucio Mayer, Ben Moore, Jorge Pe\~{n}arrubia, Tom Quinn,
Helio Rocha-Pinto, Steven Snell, Joachim Stadel, and Octavio Valenzuela for
many stimulating discussions, and Jeffrey Crane for making available in
electronic format data from his sample of M giants in the Monoceros stream. 
SK would like to thank Frank van den Bosch for communicating unpublished
results, and John Dubinski and 
Larry Widrow for kindly making available the software used to set up the 
primary galaxy model. SK, JSB, and AVK acknowledge the Aspen Center for 
Physics for hosting the summer workshop ``Deconstructing the Local Group -- 
Dissecting Galaxy Formation in our Own Background'' where some of this work 
was completed. SK is also grateful to the Research Center for Astronomy and 
Applied Mathematics at the Academy of Athens for their hospitality during a visit when
the final stages of this work were completed. SK is supported by a Kavli
Institute for Particle Astrophysics and Cosmology (KIPAC) Postdoctoral
Fellowship at Stanford University.   JSB is supported by NSF grants
AST 05-07916 and AST 06-07377. ARZ is funded by the University 
of Pittsburgh. AVK is supported by the NSF grants AST-0239759 and AST-0507596, and by KICP.
The work of LAM was carried out at the Jet Propulsion Laboratory, California
Institute of Technology under a contract with NASA. The numerical simulations were
performed on the zBox supercomputer at The University of Z\"urich and on the
Cosmos cluster at the Jet Propulsion Laboratory. This research made use of the 
NASA Astrophysics Data System.

\bibliography{disk1}

\begin{thebibliography}{171}
\expandafter\ifx\csname natexlab\endcsname\relax\def\natexlab#1{#1}\fi

\bibitem[{{Abadi} {et~al.}(2006){Abadi}, {Navarro}, \&
  {Steinmetz}}]{Abadi_etal06}
{Abadi}, M.~G., {Navarro}, J.~F., \& {Steinmetz}, M. 2006, \mnras, 365, 747

\bibitem[{{Abadi} {et~al.}(2003){Abadi}, {Navarro}, {Steinmetz}, \&
  {Eke}}]{Abadi_etal03}
{Abadi}, M.~G., {Navarro}, J.~F., {Steinmetz}, M., \& {Eke}, V.~R. 2003, \apj,
  591, 499

\bibitem[{{Abe} {et~al.}(1999)}]{Abe_etal99}
{Abe}, F. {et~al.} 1999, \aj, 118, 261

\bibitem[{{Allende Prieto} {et~al.}(2006){Allende Prieto}, {Beers}, {Wilhelm},
  {Newberg}, {Rockosi}, {Yanny}, \& {Lee}}]{Allende_etal06}
{Allende Prieto}, C., {Beers}, T.~C., {Wilhelm}, R., {Newberg}, H.~J.,
  {Rockosi}, C.~M., {Yanny}, B., \& {Lee}, Y.~S. 2006, \apj, 636, 804

\bibitem[{{Ardi} {et~al.}(2003){Ardi}, {Tsuchiya}, \& {Burkert}}]{Ardi_etal03}
{Ardi}, E., {Tsuchiya}, T., \& {Burkert}, A. 2003, \apj, 596, 204

\bibitem[{{Avila-Reese} {et~al.}(2001){Avila-Reese}, {Col{\'{\i}}n},
  {Valenzuela}, {D'Onghia}, \& {Firmani}}]{Avila-Reese_etal01}
{Avila-Reese}, V., {Col{\'{\i}}n}, P., {Valenzuela}, O., {D'Onghia}, E., \&
  {Firmani}, C. 2001, \apj, 559, 516

\bibitem[{{Barth}(2007)}]{Barth07}
{Barth}, A.~J. 2007, \aj, 133, 1085

\bibitem[{{Bell} {et~al.}(2007)}]{Bell_etal07}
{Bell}, E.~F. {et~al.} 2007, ApJ submitted (astro-ph/0706.0004)

\bibitem[{{Belokurov} {et~al.}(2006)}]{Belokurov_etal06}
{Belokurov}, V. {et~al.} 2006, \apjl, 642, L137

\bibitem[{{Bensby} {et~al.}(2005){Bensby}, {Feltzing}, {Lundstr{\"o}m}, \&
  {Ilyin}}]{Bensby_etal05}
{Bensby}, T., {Feltzing}, S., {Lundstr{\"o}m}, I., \& {Ilyin}, I. 2005, \aap,
  433, 185

\bibitem[{{Benson}(2005)}]{Benson05}
{Benson}, A.~J. 2005, \mnras, 358, 551

\bibitem[{{Benson} {et~al.}(2004){Benson}, {Lacey}, {Frenk}, {Baugh}, \&
  {Cole}}]{Benson_etal04}
{Benson}, A.~J., {Lacey}, C.~G., {Frenk}, C.~S., {Baugh}, C.~M., \& {Cole}, S.
  2004, \mnras, 351, 1215

\bibitem[{{Besla} {et~al.}(2007){Besla}, {Kallivayalil}, {Hernquist},
  {Robertson}, {Cox}, {van der Marel}, \& {Alcock}}]{Besla_etal07}
{Besla}, G., {Kallivayalil}, N., {Hernquist}, L., {Robertson}, B., {Cox},
  T.~J., {van der Marel}, R.~P., \& {Alcock}, C. 2007, \apj, 668, 949

\bibitem[{{Binney} \& {Tremaine}(1987)}]{Binney_Tremaine87}
{Binney}, J. \& {Tremaine}, S. 1987, {Galactic dynamics} (Princeton, NJ,
  Princeton University Press, 1987, 747 p.)

\bibitem[{{Blumenthal} {et~al.}(1984){Blumenthal}, {Faber}, {Primack}, \&
  {Rees}}]{Blumenthal_etal84}
{Blumenthal}, G.~R., {Faber}, S.~M., {Primack}, J.~R., \& {Rees}, M.~J. 1984,
  \nat, 311, 517

\bibitem[{{Brinks} \& {Burton}(1984)}]{Brinks_Burton84}
{Brinks}, E. \& {Burton}, W.~B. 1984, \aap, 141, 195

\bibitem[{{Brook} {et~al.}(2004){Brook}, {Kawata}, {Gibson}, \&
  {Flynn}}]{Brook_etal04}
{Brook}, C.~B., {Kawata}, D., {Gibson}, B.~K., \& {Flynn}, C. 2004, \mnras,
  349, 52

\bibitem[{{Brown} {et~al.}(2006)}]{Brown_etal06}
{Brown}, T.~M. {et~al.} 2006, \apj, 652, 323

\bibitem[{{Brown} {et~al.}(2008){Brown}, {Beers}, {Wilhelm}, {Allende Prieto},
  {Geller}, {Kenyon}, \& {Kurtz}}]{Brown_etal08}
{Brown}, W.~R., {Beers}, T.~C., {Wilhelm}, R., {Allende Prieto}, C., {Geller},
  M.~J., {Kenyon}, S.~J., \& {Kurtz}, M.~J. 2008, \aj, 135, 564

\bibitem[{{Bullock} \& {Johnston}(2005)}]{Bullock_Johnston05}
{Bullock}, J.~S. \& {Johnston}, K.~V. 2005, \apj, 635, 931

\bibitem[{{Bullock} {et~al.}(2001){Bullock}, {Kravtsov}, \&
  {Weinberg}}]{Bullock_etal01}
{Bullock}, J.~S., {Kravtsov}, A.~V., \& {Weinberg}, D.~H. 2001, \apj, 548, 33

\bibitem[{{Cembranos} {et~al.}(2005){Cembranos}, {Feng}, {Rajaraman}, \&
  {Takayama}}]{Cembranos_etal05}
{Cembranos}, J.~A., {Feng}, J.~L., {Rajaraman}, A., \& {Takayama}, F. 2005,
  Physical Review Letters, 95, 181301

\bibitem[{{Chen} {et~al.}(2001)}]{Chen_etal01}
{Chen}, B. {et~al.} 2001, \apj, 553, 184

\bibitem[{{Chiba} \& {Beers}(2000)}]{Chiba_Beers00}
{Chiba}, M. \& {Beers}, T.~C. 2000, \aj, 119, 2843

\bibitem[{{Cole} \& {Lacey}(1996)}]{Cole_Lacey96}
{Cole}, S. \& {Lacey}, C. 1996, \mnras, 281, 716

\bibitem[{{Combes} \& {Sanders}(1981)}]{Combes_Sanders81}
{Combes}, F. \& {Sanders}, R.~H. 1981, \aap, 96, 164

\bibitem[{{Conn} {et~al.}(2005){Conn}, {Martin}, {Lewis}, {Ibata},
  {Bellazzini}, \& {Irwin}}]{Conn_etal05}
{Conn}, B.~C., {Martin}, N.~F., {Lewis}, G.~F., {Ibata}, R.~A., {Bellazzini},
  M., \& {Irwin}, M.~J. 2005, \mnras, 364, L13

\bibitem[{{Crane} {et~al.}(2003){Crane}, {Majewski}, {Rocha-Pinto},
  {Frinchaboy}, {Skrutskie}, \& {Law}}]{Crane_etal03}
{Crane}, J.~D., {Majewski}, S.~R., {Rocha-Pinto}, H.~J., {Frinchaboy}, P.~M.,
  {Skrutskie}, M.~F., \& {Law}, D.~R. 2003, \apjl, 594, L119

\bibitem[{{Dalcanton} \& {Bernstein}(2002)}]{Dalcanton_Bernstein02}
{Dalcanton}, J.~J. \& {Bernstein}, R.~A. 2002, \aj, 124, 1328

\bibitem[{{Dalcanton} {et~al.}(2005){Dalcanton}, {Seth}, \&
  {Yoachim}}]{Dalcanton_etal05}
{Dalcanton}, J.~J., {Seth}, A., \& {Yoachim}, P. 2005, in Proceedings of
  "Island Universes: Structure and Evolution of Disk Galaxies", Terschelling,
  the Netherlands (astro-ph/0509700)

\bibitem[{{de Grijs} \& {Peletier}(1997)}]{DeGrijs_Peletier97}
{de Grijs}, R. \& {Peletier}, R.~F. 1997, \aap, 320, L21

\bibitem[{{de Jong} {et~al.}(2007{\natexlab{a}})}]{Dejong_etal07a}
{de Jong}, R.~S. {et~al.} 2007{\natexlab{a}}, in the proceedings of IAUS 241:
  "Stellar Populations as Building Blocks of Galaxies" (astro-ph/0702168)

\bibitem[{{de Jong} {et~al.}(2007{\natexlab{b}})}]{Dejong_etal07b}
---. 2007{\natexlab{b}}, \apjl, 667, L49

\bibitem[{{Dehnen} \& {Binney}(1998)}]{Dehnen_Binney98}
{Dehnen}, W. \& {Binney}, J. 1998, \mnras, 294, 429

\bibitem[{{Eskridge} {et~al.}(2000)}]{Eskridge_etal00}
{Eskridge}, P.~B. {et~al.} 2000, \aj, 119, 536

\bibitem[{{Faria} {et~al.}(2007)}]{Faria_etal07}
{Faria}, D. {et~al.} 2007, \aj, 133, 1275

\bibitem[{{Ferguson}(2007)}]{Ferguson07}
{Ferguson}, A. 2007, in Proceedings of "From Stars to Galaxies: Building the
  Pieces to Build up the Universe", eds. Vallenari et al, ASP Conf Series
  (astro-ph/0702224)

\bibitem[{{Ferguson} {et~al.}(2005)}]{Ferguson_etal05}
{Ferguson}, A.~M.~N. {et~al.} 2005, \apjl, 622, L109

\bibitem[{{Font} {et~al.}(2001){Font}, {Navarro}, {Stadel}, \&
  {Quinn}}]{Font_etal01}
{Font}, A.~S., {Navarro}, J.~F., {Stadel}, J., \& {Quinn}, T. 2001, \apjl, 563,
  L1

\bibitem[{{Gao} {et~al.}(2004){Gao}, {White}, {Jenkins}, {Stoehr}, \&
  {Springel}}]{Gao_etal04}
{Gao}, L., {White}, S.~D.~M., {Jenkins}, A., {Stoehr}, F., \& {Springel}, V.
  2004, \mnras, 355, 819

\bibitem[{{Gauthier} {et~al.}(2006){Gauthier}, {Dubinski}, \&
  {Widrow}}]{Gauthier_etal06}
{Gauthier}, J.-R., {Dubinski}, J., \& {Widrow}, L.~M. 2006, \apj, 653, 1180

\bibitem[{{Ghigna} {et~al.}(1998){Ghigna}, {Moore}, {Governato}, {Lake},
  {Quinn}, \& {Stadel}}]{Ghigna_etal98}
{Ghigna}, S., {Moore}, B., {Governato}, F., {Lake}, G., {Quinn}, T., \&
  {Stadel}, J. 1998, \mnras, 300, 146

\bibitem[{{Gilmore} {et~al.}(1995){Gilmore}, {Wyse}, \&
  {Jones}}]{Gilmore_etal95}
{Gilmore}, G., {Wyse}, R.~F.~G., \& {Jones}, J.~B. 1995, \aj, 109, 1095

\bibitem[{{Gilmore} {et~al.}(2002){Gilmore}, {Wyse}, \&
  {Norris}}]{Gilmore_etal02}
{Gilmore}, G., {Wyse}, R.~F.~G., \& {Norris}, J.~E. 2002, \apjl, 574, L39

\bibitem[{{Gnedin} \& {Kravtsov}(2006)}]{Gnedin_Kravtsov06}
{Gnedin}, N.~Y. \& {Kravtsov}, A.~V. 2006, \apj, 645, 1054

\bibitem[{{Gordon} {et~al.}(2006)}]{Gordon_etal06}
{Gordon}, K.~D. {et~al.} 2006, \apjl, 638, L87

\bibitem[{{Governato} {et~al.}(2004)}]{Governato_etal04}
{Governato}, F. {et~al.} 2004, \apj, 607, 688

\bibitem[{{Governato} {et~al.}(2007)}]{Governato_etal07}
---. 2007, \mnras, 374, 1479

\bibitem[{{Grillmair}(2006{\natexlab{a}})}]{Grillmair06a}
{Grillmair}, C.~J. 2006{\natexlab{a}}, \apjl, 645, L37

\bibitem[{{Grillmair}(2006{\natexlab{b}})}]{Grillmair06b}
---. 2006{\natexlab{b}}, \apjl, 651, L29

\bibitem[{{Grillmair} \& {Dionatos}(2006)}]{Grillmair_Dionatos06}
{Grillmair}, C.~J. \& {Dionatos}, O. 2006, \apjl, 643, L17

\bibitem[{{Hayashi} \& {Chiba}(2006)}]{Hayashi_Chiba06}
{Hayashi}, H. \& {Chiba}, M. 2006, \pasj, 58, 835

\bibitem[{{Helmi} {et~al.}(2003){Helmi}, {Navarro}, {Meza}, {Steinmetz}, \&
  {Eke}}]{Helmi_etal03}
{Helmi}, A., {Navarro}, J.~F., {Meza}, A., {Steinmetz}, M., \& {Eke}, V.~R.
  2003, \apjl, 592, L25

\bibitem[{{Hernquist}(1990)}]{Hernquist90}
{Hernquist}, L. 1990, \apj, 356, 359

\bibitem[{{Hernquist}(1993)}]{Hernquist93}
---. 1993, \apjs, 86, 389

\bibitem[{{Hernquist} \& {Quinn}(1988)}]{Hernquist_Quinn88}
{Hernquist}, L. \& {Quinn}, P.~J. 1988, \apj, 331, 682

\bibitem[{{Hogan} \& {Dalcanton}(2000)}]{Hogan_Dalcanton00}
{Hogan}, C.~J. \& {Dalcanton}, J.~J. 2000, PRD, 62, 063511

\bibitem[{{Huang} \& {Carlberg}(1997)}]{Huang_Carlberg97}
{Huang}, S. \& {Carlberg}, R.~G. 1997, \apj, 480, 503

\bibitem[{{Ibata} {et~al.}(2005){Ibata}, {Chapman}, {Ferguson}, {Lewis},
  {Irwin}, \& {Tanvir}}]{Ibata_etal05}
{Ibata}, R., {Chapman}, S., {Ferguson}, A.~M.~N., {Lewis}, G., {Irwin}, M., \&
  {Tanvir}, N. 2005, \apj, 634, 287

\bibitem[{{Ibata} {et~al.}(2001{\natexlab{a}}){Ibata}, {Irwin}, {Lewis},
  {Ferguson}, \& {Tanvir}}]{Ibata_etal01c}
{Ibata}, R., {Irwin}, M., {Lewis}, G., {Ferguson}, A.~M.~N., \& {Tanvir}, N.
  2001{\natexlab{a}}, \nat, 412, 49

\bibitem[{{Ibata} {et~al.}(2001{\natexlab{b}}){Ibata}, {Irwin}, {Lewis}, \&
  {Stolte}}]{Ibata_etal01b}
{Ibata}, R., {Irwin}, M., {Lewis}, G.~F., \& {Stolte}, A. 2001{\natexlab{b}},
  \apjl, 547, L133

\bibitem[{{Ibata} {et~al.}(2001{\natexlab{c}}){Ibata}, {Lewis}, {Irwin},
  {Totten}, \& {Quinn}}]{Ibata_etal01a}
{Ibata}, R., {Lewis}, G.~F., {Irwin}, M., {Totten}, E., \& {Quinn}, T.
  2001{\natexlab{c}}, \apj, 551, 294

\bibitem[{{Ibata} {et~al.}(2007){Ibata}, {Martin}, {Irwin}, {Chapman},
  {Ferguson}, {Lewis}, \& {McConnachie}}]{Ibata_etal07}
{Ibata}, R., {Martin}, N.~F., {Irwin}, M., {Chapman}, S., {Ferguson}, A.~M.~N.,
  {Lewis}, G.~F., \& {McConnachie}, A.~W. 2007, ApJ submitted
  (astro-ph/0704.1318)

\bibitem[{{Ibata} {et~al.}(1994){Ibata}, {Gilmore}, \& {Irwin}}]{Ibata_etal94}
{Ibata}, R.~A., {Gilmore}, G., \& {Irwin}, M.~J. 1994, \nat, 370, 194

\bibitem[{{Ibata} {et~al.}(2003){Ibata}, {Irwin}, {Lewis}, {Ferguson}, \&
  {Tanvir}}]{Ibata_etal03}
{Ibata}, R.~A., {Irwin}, M.~J., {Lewis}, G.~F., {Ferguson}, A.~M.~N., \&
  {Tanvir}, N. 2003, \mnras, 340, L21

\bibitem[{{Ivezi{\'c}} {et~al.}(2000)}]{Ivezic_etal00}
{Ivezi{\'c}}, {\v Z}. {et~al.} 2000, \aj, 120, 963

\bibitem[{{Ivezi{\'c}} {et~al.}(2008)}]{Ivezic_etal08}
---. 2008, ApJ accepted (astro-ph/0804.3850)

\bibitem[{{Johnston} {et~al.}(1995){Johnston}, {Spergel}, \&
  {Hernquist}}]{Johnston_etal95}
{Johnston}, K.~V., {Spergel}, D.~N., \& {Hernquist}, L. 1995, \apj, 451, 598

\bibitem[{{Juric} {et~al.}(2008)}]{Juric_etal08}
{Juric}, M. {et~al.} 2008, \apj, 673, 864

\bibitem[{{Kalirai} {et~al.}(2006){Kalirai}, {Guhathakurta}, {Gilbert},
  {Reitzel}, {Majewski}, {Rich}, \& {Cooper}}]{Kalirai_etal06}
{Kalirai}, J.~S., {Guhathakurta}, P., {Gilbert}, K.~M., {Reitzel}, D.~B.,
  {Majewski}, S.~R., {Rich}, R.~M., \& {Cooper}, M.~C. 2006, \apj, 641, 268

\bibitem[{{Kamionkowski} \& {Liddle}(2000)}]{Kamionkowski_Liddle00}
{Kamionkowski}, M. \& {Liddle}, A.~R. 2000, Physical Review Letters, 84, 4525

\bibitem[{{Kaplinghat}(2005)}]{Kaplinghat05}
{Kaplinghat}, M. 2005, \prd, 72, 063510

\bibitem[{{Kautsch} {et~al.}(2006){Kautsch}, {Grebel}, {Barazza}, \&
  {Gallagher}}]{Kautsch_etal06}
{Kautsch}, S.~J., {Grebel}, E.~K., {Barazza}, F.~D., \& {Gallagher}, III, J.~S.
  2006, \aap, 445, 765

\bibitem[{{Kazantzidis} {et~al.}(2004{\natexlab{a}}){Kazantzidis}, {Magorrian},
  \& {Moore}}]{Kazantzidis_etal04a}
{Kazantzidis}, S., {Magorrian}, J., \& {Moore}, B. 2004{\natexlab{a}}, \apj,
  601, 37

\bibitem[{{Kazantzidis} {et~al.}(2004{\natexlab{b}}){Kazantzidis}, {Mayer},
  {Mastropietro}, {Diemand}, {Stadel}, \& {Moore}}]{Kazantzidis_etal04b}
{Kazantzidis}, S., {Mayer}, L., {Mastropietro}, C., {Diemand}, J., {Stadel},
  J., \& {Moore}, B. 2004{\natexlab{b}}, \apj, 608, 663

\bibitem[{{Kazantzidis} {et~al.}(2005){Kazantzidis}, {Zentner}, \&
  {Nagai}}]{Kazantzidis_etal06}
{Kazantzidis}, S., {Zentner}, A.~R., \& {Nagai}, D. 2005, in Proceedings of
  "Mass Profiles and Shapes of Cosmological Structures", eds. Mammon et al, EAS
  Publications Series (astro-ph/0508114)

\bibitem[{{Kent} {et~al.}(1991){Kent}, {Dame}, \& {Fazio}}]{Kent_etal91}
{Kent}, S.~M., {Dame}, T.~M., \& {Fazio}, G. 1991, \apj, 378, 131

\bibitem[{{Klypin} {et~al.}(1999){Klypin}, {Kravtsov}, {Valenzuela}, \&
  {Prada}}]{Klypin_etal99}
{Klypin}, A., {Kravtsov}, A.~V., {Valenzuela}, O., \& {Prada}, F. 1999, \apj,
  522, 82

\bibitem[{{Klypin} {et~al.}(2002){Klypin}, {Zhao}, \&
  {Somerville}}]{Klypin_etal02}
{Klypin}, A., {Zhao}, H., \& {Somerville}, R.~S. 2002, \apj, 573, 597

\bibitem[{{Klypin} {et~al.}(2001){Klypin}, {Kravtsov}, {Bullock}, \&
  {Primack}}]{Klypin_etal01}
{Klypin}, A.~A., {Kravtsov}, A.~V., {Bullock}, J.~S., \& {Primack}, J.~R. 2001,
  \apj, 554, 903

\bibitem[{{Knapen}(1999)}]{Knapen99}
{Knapen}, J.~H. 1999, in Astronomical Society of the Pacific Conference Series,
  Vol. 187, The Evolution of Galaxies on Cosmological Timescales, ed. J.~E.
  {Beckman} \& T.~J. {Mahoney}, 72--87

\bibitem[{{Knebe} {et~al.}(2004){Knebe}, {Gill}, {Gibson}, {Lewis}, {Ibata}, \&
  {Dopita}}]{Knebe_etal04}
{Knebe}, A., {Gill}, S.~P.~D., {Gibson}, B.~K., {Lewis}, G.~F., {Ibata}, R.~A.,
  \& {Dopita}, M.~A. 2004, \apj, 603, 7

\bibitem[{{Kravtsov}(1999)}]{Kravtsov99}
{Kravtsov}, A.~V. 1999, PhD thesis, New Mexico State University

\bibitem[{{Kravtsov} {et~al.}(2004){Kravtsov}, {Gnedin}, \&
  {Klypin}}]{Kravtsov_etal04}
{Kravtsov}, A.~V., {Gnedin}, O.~Y., \& {Klypin}, A.~A. 2004, \apj, 609, 482

\bibitem[{{Kravtsov} {et~al.}(1998){Kravtsov}, {Klypin}, {Bullock}, \&
  {Primack}}]{Kravtsov_etal98}
{Kravtsov}, A.~V., {Klypin}, A.~A., {Bullock}, J.~S., \& {Primack}, J.~R. 1998,
  \apj, 502, 48

\bibitem[{{Kravtsov} {et~al.}(1997){Kravtsov}, {Klypin}, \&
  {Khokhlov}}]{Kravtsov_etal97}
{Kravtsov}, A.~V., {Klypin}, A.~A., \& {Khokhlov}, A.~M. 1997, \apjs, 111, 73

\bibitem[{{Kuijken} \& {Dubinski}(1995)}]{Kuijken_Dubinski95}
{Kuijken}, K. \& {Dubinski}, J. 1995, \mnras, 277, 1341

\bibitem[{{Larsen} \& {Humphreys}(2003)}]{Larsen_Humphreys03}
{Larsen}, J.~A. \& {Humphreys}, R.~M. 2003, \aj, 125, 1958

\bibitem[{{Lequeux} {et~al.}(1998){Lequeux}, {Combes}, {Dantel-Fort},
  {Cuillandre}, {Fort}, \& {Mellier}}]{Lequeux_etal98}
{Lequeux}, J., {Combes}, F., {Dantel-Fort}, M., {Cuillandre}, J.-C., {Fort},
  B., \& {Mellier}, Y. 1998, \aap, 334, L9

\bibitem[{{Libeskind} {et~al.}(2007){Libeskind}, {Cole}, {Frenk}, {Okamoto}, \&
  {Jenkins}}]{Libeskind_etal07}
{Libeskind}, N.~I., {Cole}, S., {Frenk}, C.~S., {Okamoto}, T., \& {Jenkins}, A.
  2007, \mnras, 374, 16

\bibitem[{{L{\'o}pez-Corredoira} {et~al.}(2002){L{\'o}pez-Corredoira},
  {Cabrera-Lavers}, {Garz{\'o}n}, \& {Hammersley}}]{Lopez_etal02}
{L{\'o}pez-Corredoira}, M., {Cabrera-Lavers}, A., {Garz{\'o}n}, F., \&
  {Hammersley}, P.~L. 2002, \aap, 394, 883

\bibitem[{{L{\"u}tticke} {et~al.}(2000){L{\"u}tticke}, {Dettmar}, \&
  {Pohlen}}]{Lutticke_etal00}
{L{\"u}tticke}, R., {Dettmar}, R.-J., \& {Pohlen}, M. 2000, \aaps, 145, 405

\bibitem[{{Majewski} {et~al.}(2003){Majewski}, {Skrutskie}, {Weinberg}, \&
  {Ostheimer}}]{Majewski_etal03}
{Majewski}, S.~R., {Skrutskie}, M.~F., {Weinberg}, M.~D., \& {Ostheimer}, J.~C.
  2003, \apj, 599, 1082

\bibitem[{{Martin} {et~al.}(2004){Martin}, {Ibata}, {Bellazzini}, {Irwin},
  {Lewis}, \& {Dehnen}}]{Martin_etal04}
{Martin}, N.~F., {Ibata}, R.~A., {Bellazzini}, M., {Irwin}, M.~J., {Lewis},
  G.~F., \& {Dehnen}, W. 2004, \mnras, 348, 12

\bibitem[{{Mart{\'{\i}}nez-Delgado} {et~al.}(2005){Mart{\'{\i}}nez-Delgado},
  {Butler}, {Rix}, {Franco}, {Pe{\~n}arrubia}, {Alfaro}, \&
  {Dinescu}}]{Mdelgado_etal05}
{Mart{\'{\i}}nez-Delgado}, D., {Butler}, D.~J., {Rix}, H.-W., {Franco}, V.~I.,
  {Pe{\~n}arrubia}, J., {Alfaro}, E.~J., \& {Dinescu}, D.~I. 2005, \apj, 633,
  205

\bibitem[{{Mastropietro} {et~al.}(2005){Mastropietro}, {Moore}, {Mayer},
  {Wadsley}, \& {Stadel}}]{Mastropietro_etal05}
{Mastropietro}, C., {Moore}, B., {Mayer}, L., {Wadsley}, J., \& {Stadel}, J.
  2005, \mnras, 363, 509

\bibitem[{{Mateo}(1998)}]{Mateo98}
{Mateo}, M.~L. 1998, \araa, 36, 435

\bibitem[{{Matthews} \& {Wood}(2003)}]{Matthews_Wood03}
{Matthews}, L.~D. \& {Wood}, K. 2003, \apj, 593, 721

\bibitem[{{Mayer} {et~al.}(2007){Mayer}, {Kazantzidis}, {Mastropietro}, \&
  {Wadsley}}]{Mayer_etal07}
{Mayer}, L., {Kazantzidis}, S., {Mastropietro}, C., \& {Wadsley}, J. 2007,
  \nat, 445, 738

\bibitem[{{Mendez} \& {Guzman}(1998)}]{Mendez_Guzman98}
{Mendez}, R.~A. \& {Guzman}, R. 1998, \aap, 333, 106

\bibitem[{{Merrifield}(1992)}]{Merrifield92}
{Merrifield}, M.~R. 1992, \aj, 103, 1552

\bibitem[{{Meza} {et~al.}(2005){Meza}, {Navarro}, {Abadi}, \&
  {Steinmetz}}]{Meza_etal05}
{Meza}, A., {Navarro}, J.~F., {Abadi}, M.~G., \& {Steinmetz}, M. 2005, \mnras,
  359, 93

\bibitem[{{Moitinho} {et~al.}(2006){Moitinho}, {V{\'a}zquez}, {Carraro},
  {Baume}, {Giorgi}, \& {Lyra}}]{Moitinho_etal06}
{Moitinho}, A., {V{\'a}zquez}, R.~A., {Carraro}, G., {Baume}, G., {Giorgi},
  E.~E., \& {Lyra}, W. 2006, \mnras, 368, L77

\bibitem[{{Momany} {et~al.}(2006){Momany}, {Zaggia}, {Gilmore}, {Piotto},
  {Carraro}, {Bedin}, \& {de Angeli}}]{Momany_etal06}
{Momany}, Y., {Zaggia}, S., {Gilmore}, G., {Piotto}, G., {Carraro}, G.,
  {Bedin}, L.~R., \& {de Angeli}, F. 2006, \aap, 451, 515

\bibitem[{{Momany} {et~al.}(2004){Momany}, {Zaggia}, {Bonifacio}, {Piotto}, {De
  Angeli}, {Bedin}, \& {Carraro}}]{Momany_etal04}
{Momany}, Y., {Zaggia}, S.~R., {Bonifacio}, P., {Piotto}, G., {De Angeli}, F.,
  {Bedin}, L.~R., \& {Carraro}, G. 2004, \aap, 421, L29

\bibitem[{{Moore} {et~al.}(1999){Moore}, {Ghigna}, {Governato}, {Lake},
  {Quinn}, {Stadel}, \& {Tozzi}}]{Moore_etal99}
{Moore}, B., {Ghigna}, S., {Governato}, F., {Lake}, G., {Quinn}, T., {Stadel},
  J., \& {Tozzi}, P. 1999, \apjl, 524, L19

\bibitem[{{Moore} {et~al.}(2004){Moore}, {Kazantzidis}, {Diemand}, \&
  {Stadel}}]{Moore_etal04}
{Moore}, B., {Kazantzidis}, S., {Diemand}, J., \& {Stadel}, J. 2004, \mnras,
  354, 522

\bibitem[{{Nakanishi} \& {Sofue}(2003)}]{Nakanishi_Sofue03}
{Nakanishi}, H. \& {Sofue}, Y. 2003, \pasj, 55, 191

\bibitem[{{Narayan} \& {Jog}(2002)}]{Narayan_Jog02}
{Narayan}, C.~A. \& {Jog}, C.~J. 2002, \aap, 390, L35

\bibitem[{{Navarro} {et~al.}(1995){Navarro}, {Frenk}, \&
  {White}}]{Navarro_etal95}
{Navarro}, J.~F., {Frenk}, C.~S., \& {White}, S.~D.~M. 1995, \mnras, 275, 56

\bibitem[{{Navarro} {et~al.}(1996){Navarro}, {Frenk}, \&
  {White}}]{Navarro_etal96}
---. 1996, \apj, 462, 563

\bibitem[{{Newberg} {et~al.}(2006){Newberg}, {Mayeur}, {Yanny}, \& {the SEGUE
  collaboration}}]{Newberg_etal06}
{Newberg}, H.~J., {Mayeur}, P.~A., {Yanny}, B., \& {the SEGUE collaboration}.
  2006, Memorie della Societa Astronomica Italiana, 77, 1049

\bibitem[{{Newberg} {et~al.}(2002)}]{Newberg_etal02}
{Newberg}, H.~J. {et~al.} 2002, \apj, 569, 245

\bibitem[{{Nordstr{\"o}m} {et~al.}(2004)}]{Nordstrom_etal04}
{Nordstr{\"o}m}, B. {et~al.} 2004, \aap, 418, 989

\bibitem[{{Olling}(1996)}]{0lling96}
{Olling}, R.~P. 1996, \aj, 112, 457

\bibitem[{{Pe{\~n}arrubia} {et~al.}(2002){Pe{\~n}arrubia}, {Kroupa}, \&
  {Boily}}]{Penarrubia_etal02}
{Pe{\~n}arrubia}, J., {Kroupa}, P., \& {Boily}, C.~M. 2002, \mnras, 333, 779

\bibitem[{{Pe{\~n}arrubia} {et~al.}(2006){Pe{\~n}arrubia}, {McConnachie}, \&
  {Babul}}]{Penarrubia_etal06}
{Pe{\~n}arrubia}, J., {McConnachie}, A., \& {Babul}, A. 2006, \apjl, 650, L33

\bibitem[{{Pe{\~n}arrubia} {et~al.}(2005)}]{Penarrubia_etal05}
{Pe{\~n}arrubia}, J. {et~al.} 2005, \apj, 626, 128

\bibitem[{{Pohlen} {et~al.}(2007){Pohlen}, {Zaroubi}, {Peletier}, \&
  {Dettmar}}]{Pohlen_etal07}
{Pohlen}, M., {Zaroubi}, S., {Peletier}, R.~F., \& {Dettmar}, R.-J. 2007,
  \mnras, 378, 594

\bibitem[{{Prada} {et~al.}(2006){Prada}, {Klypin}, {Simonneau},
  {Betancort-Rijo}, {Patiri}, {Gottl{\"o}ber}, \&
  {Sanchez-Conde}}]{Prada_etal06}
{Prada}, F., {Klypin}, A.~A., {Simonneau}, E., {Betancort-Rijo}, J., {Patiri},
  S., {Gottl{\"o}ber}, S., \& {Sanchez-Conde}, M.~A. 2006, \apj, 645, 1001

\bibitem[{{Purcell} {et~al.}(2007){Purcell}, {Bullock}, \&
  {Zentner}}]{Purcell_etal07}
{Purcell}, C.~W., {Bullock}, J.~S., \& {Zentner}, A.~R. 2007, \apj, 666, 20

\bibitem[{{Quillen} \& {Garnett}(2000)}]{Quillen_Garnet01}
{Quillen}, A.~C. \& {Garnett}, D.~R. 2000, ApJ submitted (astro-ph/0004210)

\bibitem[{{Quinn} \& {Goodman}(1986)}]{Quinn_Goodman86}
{Quinn}, P.~J. \& {Goodman}, J. 1986, \apj, 309, 472

\bibitem[{{Quinn} {et~al.}(1993){Quinn}, {Hernquist}, \&
  {Fullagar}}]{Quinn_etal93}
{Quinn}, P.~J., {Hernquist}, L., \& {Fullagar}, D.~P. 1993, \apj, 403, 74

\bibitem[{{Read} {et~al.}(2008){Read}, {Lake}, {Agertz}, \&
  {Debattista}}]{Read_etal08}
{Read}, J.~I., {Lake}, G., {Agertz}, O., \& {Debattista}, V.~P. 2008, MNRAS
  accepted (astro-ph/0803.2714)

\bibitem[{{Read} \& {Trentham}(2005)}]{Read_Trentham05}
{Read}, J.~I. \& {Trentham}, N. 2005, Royal Society of London Philosophical
  Transactions Series A, 363, 2693

\bibitem[{{Robertson} {et~al.}(2006){Robertson}, {Bullock}, {Cox}, {Di Matteo},
  {Hernquist}, {Springel}, \& {Yoshida}}]{Robertson_etal06}
{Robertson}, B., {Bullock}, J.~S., {Cox}, T.~J., {Di Matteo}, T., {Hernquist},
  L., {Springel}, V., \& {Yoshida}, N. 2006, \apj, 645, 986

\bibitem[{{Robertson} {et~al.}(2004){Robertson}, {Yoshida}, {Springel}, \&
  {Hernquist}}]{Robertson_etal04}
{Robertson}, B., {Yoshida}, N., {Springel}, V., \& {Hernquist}, L. 2004, \apj,
  606, 32

\bibitem[{{Robin} {et~al.}(1996){Robin}, {Haywood}, {Creze}, {Ojha}, \&
  {Bienayme}}]{Robin_etal96}
{Robin}, A.~C., {Haywood}, M., {Creze}, M., {Ojha}, D.~K., \& {Bienayme}, O.
  1996, \aap, 305, 125

\bibitem[{{Rocha-Pinto} {et~al.}(2003){Rocha-Pinto}, {Majewski}, {Skrutskie},
  \& {Crane}}]{Rocha-Pinto_etal03}
{Rocha-Pinto}, H.~J., {Majewski}, S.~R., {Skrutskie}, M.~F., \& {Crane}, J.~D.
  2003, \apjl, 594, L115

\bibitem[{{Rocha-Pinto} {et~al.}(2004){Rocha-Pinto}, {Majewski}, {Skrutskie},
  {Crane}, \& {Patterson}}]{Rocha-Pinto_etal04}
{Rocha-Pinto}, H.~J., {Majewski}, S.~R., {Skrutskie}, M.~F., {Crane}, J.~D., \&
  {Patterson}, R.~J. 2004, \apj, 615, 732

\bibitem[{{Rocha-Pinto} {et~al.}(2006){Rocha-Pinto}, {Majewski}, {Skrutskie},
  {Patterson}, {Nakanishi}, {Mu{\~n}oz}, \& {Sofue}}]{Rocha-Pinto_etal06}
{Rocha-Pinto}, H.~J., {Majewski}, S.~R., {Skrutskie}, M.~F., {Patterson},
  R.~J., {Nakanishi}, H., {Mu{\~n}oz}, R.~R., \& {Sofue}, Y. 2006, \apjl, 640,
  L147

\bibitem[{{Sackett} {et~al.}(1994){Sackett}, {Morrison}, {Harding}, \&
  {Boroson}}]{Sackett_etal04}
{Sackett}, P.~D., {Morrison}, H.~L., {Harding}, P., \& {Boroson}, T.~A. 1994,
  \nat, 370, 441

\bibitem[{{Schommer} {et~al.}(1992){Schommer}, {Suntzeff}, {Olszewski}, \&
  {Harris}}]{Schommer_etal92}
{Schommer}, R.~A., {Suntzeff}, N.~B., {Olszewski}, E.~W., \& {Harris}, H.~C.
  1992, \aj, 103, 447

\bibitem[{{Seabroke} \& {Gilmore}(2007)}]{Seabroke_Gilmore07}
{Seabroke}, G.~M. \& {Gilmore}, G. 2007, \mnras, 380, 1348

\bibitem[{{Sellwood} {et~al.}(1998){Sellwood}, {Nelson}, \&
  {Tremaine}}]{Sellwood_etal98}
{Sellwood}, J.~A., {Nelson}, R.~W., \& {Tremaine}, S. 1998, \apj, 506, 590

\bibitem[{{Seth} {et~al.}(2005){Seth}, {Dalcanton}, \& {de Jong}}]{Seth_etal05}
{Seth}, A.~C., {Dalcanton}, J.~J., \& {de Jong}, R.~S. 2005, \aj, 130, 1574

\bibitem[{{Siegel} {et~al.}(2002){Siegel}, {Majewski}, {Reid}, \&
  {Thompson}}]{Siegel_etal02}
{Siegel}, M.~H., {Majewski}, S.~R., {Reid}, I.~N., \& {Thompson}, I.~B. 2002,
  \apj, 578, 151

\bibitem[{{Sommer-Larsen} \& {Dolgov}(2001)}]{Sommer-Larsen_Dolgov01}
{Sommer-Larsen}, J. \& {Dolgov}, A. 2001, \apj, 551, 608

\bibitem[{{Sommer-Larsen} {et~al.}(2003){Sommer-Larsen}, {G{\"o}tz}, \&
  {Portinari}}]{Sommer_Larsen_etal03}
{Sommer-Larsen}, J., {G{\"o}tz}, M., \& {Portinari}, L. 2003, \apj, 596, 47

\bibitem[{{Soubiran} {et~al.}(2003){Soubiran}, {Bienaym{\'e}}, \&
  {Siebert}}]{Soubiran_etal03}
{Soubiran}, C., {Bienaym{\'e}}, O., \& {Siebert}, A. 2003, \aap, 398, 141

\bibitem[{{Spitzer}(1942)}]{Spitzer42}
{Spitzer}, L.~J. 1942, \apj, 95, 329

\bibitem[{{Stadel}(2001)}]{Stadel01}
{Stadel}, J.~G. 2001, Ph.D.~Thesis, Univ. of Washington

\bibitem[{{Statler}(1988)}]{Statler88}
{Statler}, T.~S. 1988, \apj, 331, 71

\bibitem[{{Stewart} {et~al.}(2007){Stewart}, {Bullock}, {Wechsler}, {Maller},
  \& {Zentner}}]{Stewart_etal07}
{Stewart}, K.~R., {Bullock}, J.~S., {Wechsler}, R.~H., {Maller}, A.~H., \&
  {Zentner}, A.~R. 2007, ApJ submitted (astro-ph/0711.5027)

\bibitem[{{Stoehr} {et~al.}(2002){Stoehr}, {White}, {Tormen}, \&
  {Springel}}]{Stoehr_etal02}
{Stoehr}, F., {White}, S.~D.~M., {Tormen}, G., \& {Springel}, V. 2002, \mnras,
  335, L84

\bibitem[{{Strigari} {et~al.}(2007{\natexlab{a}}){Strigari}, {Bullock},
  {Kaplinghat}, {Diemand}, {Kuhlen}, \& {Madau}}]{Strigari_etal07}
{Strigari}, L.~E., {Bullock}, J.~S., {Kaplinghat}, M., {Diemand}, J., {Kuhlen},
  M., \& {Madau}, P. 2007{\natexlab{a}}, ApJ accepted (astro-ph/0704.1817)

\bibitem[{{Strigari} {et~al.}(2007{\natexlab{b}}){Strigari}, {Kaplinghat}, \&
  {Bullock}}]{Strigari_etal06}
{Strigari}, L.~E., {Kaplinghat}, M., \& {Bullock}, J.~S. 2007{\natexlab{b}},
  \prd, 75, 061303

\bibitem[{{Thacker} \& {Couchman}(2001)}]{Thacker_Couchman01}
{Thacker}, R.~J. \& {Couchman}, H.~M.~P. 2001, \apjl, 555, L17

\bibitem[{{Toth} \& {Ostriker}(1992)}]{Toth_Ostriker92}
{Toth}, G. \& {Ostriker}, J.~P. 1992, \apj, 389, 5

\bibitem[{{van der Kruit}(1979)}]{Kruit79}
{van der Kruit}, P.~C. 1979, \aaps, 38, 15

\bibitem[{{van der Kruit} \& {Searle}(1981)}]{Kruit_Searle81}
{van der Kruit}, P.~C. \& {Searle}, L. 1981, \aap, 95, 105

\bibitem[{{Velazquez} \& {White}(1999)}]{Velazquez_White99}
{Velazquez}, H. \& {White}, S.~D.~M. 1999, \mnras, 304, 254

\bibitem[{{Villalobos} \& {Helmi}(2008)}]{Villalobos_Helmi08}
{Villalobos}, {\'A}. \& {Helmi}, A. 2008, MNRAS submitted (astro-ph/0803.2323)

\bibitem[{{Walker} {et~al.}(1996){Walker}, {Mihos}, \&
  {Hernquist}}]{Walker_etal96}
{Walker}, I.~R., {Mihos}, J.~C., \& {Hernquist}, L. 1996, \apj, 460, 121

\bibitem[{{Wechsler} {et~al.}(2002){Wechsler}, {Bullock}, {Primack},
  {Kravtsov}, \& {Dekel}}]{Wechsler_etal02}
{Wechsler}, R.~H., {Bullock}, J.~S., {Primack}, J.~R., {Kravtsov}, A.~V., \&
  {Dekel}, A. 2002, \apj, 568, 52

\bibitem[{{Weil} {et~al.}(1998){Weil}, {Eke}, \& {Efstathiou}}]{Weil_etal98}
{Weil}, M.~L., {Eke}, V.~R., \& {Efstathiou}, G. 1998, \mnras, 300, 773

\bibitem[{{Weinberg} \& {Blitz}(2006)}]{Weinberg_Blitz06}
{Weinberg}, M.~D. \& {Blitz}, L. 2006, \apjl, 641, L33

\bibitem[{{White}(2000)}]{White00}
{White}, S. 2000, in KITP: Blackboard Lunch Series

\bibitem[{{Widrow} \& {Dubinski}(2005)}]{Widrow_Dubinski05}
{Widrow}, L.~M. \& {Dubinski}, J. 2005, \apj, 631, 838

\bibitem[{{Wyse}(2001)}]{Wyse01}
{Wyse}, R.~F.~G. 2001, in ASP Conf. Ser. 230: Galaxy Disks and Disk Galaxies,
  ed. J.~G. {Funes} \& E.~M. {Corsini}, 71--80

\bibitem[{{Wyse} \& {Gilmore}(1995)}]{Wyse_Gilmore95}
{Wyse}, R.~F.~G. \& {Gilmore}, G. 1995, \aj, 110, 2771

\bibitem[{{Yanny} {et~al.}(2000)}]{Yanny_etal00}
{Yanny}, B. {et~al.} 2000, \apj, 540, 825

\bibitem[{{Yanny} {et~al.}(2003)}]{Yanny_etal03}
---. 2003, \apj, 588, 824

\bibitem[{{Yoachim} \& {Dalcanton}(2005)}]{Yoachim_Dalcanton05}
{Yoachim}, P. \& {Dalcanton}, J.~J. 2005, \apj, 624, 701

\bibitem[{{Yoachim} \& {Dalcanton}(2006)}]{Yoachim_Dalcanton06}
---. 2006, \aj, 131, 226

\bibitem[{{Zentner} {et~al.}(2005{\natexlab{a}}){Zentner}, {Berlind},
  {Bullock}, {Kravtsov}, \& {Wechsler}}]{Zentner_etal05a}
{Zentner}, A.~R., {Berlind}, A.~A., {Bullock}, J.~S., {Kravtsov}, A.~V., \&
  {Wechsler}, R.~H. 2005{\natexlab{a}}, \apj, 624, 505

\bibitem[{{Zentner} \& {Bullock}(2003)}]{Zentner_Bullock03}
{Zentner}, A.~R. \& {Bullock}, J.~S. 2003, \apj, 598, 49

\bibitem[{{Zentner} {et~al.}(2005{\natexlab{b}}){Zentner}, {Kravtsov},
  {Gnedin}, \& {Klypin}}]{Zentner_etal05b}
{Zentner}, A.~R., {Kravtsov}, A.~V., {Gnedin}, O.~Y., \& {Klypin}, A.~A.
  2005{\natexlab{b}}, \apj, 629, 219

\bibitem[{{Zhao}(1996)}]{Zhao96}
{Zhao}, H. 1996, \mnras, 278, 488

\bibitem[{{Zibetti} \& {Ferguson}(2004)}]{Zibetti_Ferguson04}
{Zibetti}, S. \& {Ferguson}, A.~M.~N. 2004, \mnras, 352, L6

\end{thebibliography}

\end{document}